\documentclass[aps,prc,twoside,twocolumn,nofootinbib,10pt,showpacs,floatfix]{revtex4-1}
\usepackage{amsmath,amssymb}
\usepackage{graphicx,bm}
\usepackage{slashed}
\usepackage{epstopdf}
\usepackage{ulem} 
\usepackage[usenames]{color}
\usepackage{float}
\usepackage{hyperref}
\usepackage{subcaption}
\usepackage{rotating}
\usepackage{color,xcolor}
\usepackage{multirow}
\usepackage{dcolumn}
\usepackage{overpic}
\usepackage{lipsum}
\usepackage{booktabs}
\usepackage{makecell}
\usepackage{diagbox}
\usepackage{array}
\newcolumntype{|}{!{\vline}}

\renewcommand\sout{\bgroup \color{red} \ULdepth=-.5ex \ULset}

\newsavebox{\tablebox}
\begin{document}
\title{$D^{(*)}\bar{B}^{(*)}$ dynamics in chiral effective field theory}

\author{Zhe Liu$^{1,2,5}$}\email{zhliu20@lzu.edu.cn}
\author{Hao Xu$^{4}$}\email{xuh2020@nwnu.edu.cn}
\author{Xiang Liu$^{1,2,3,5}$}\email{xiangliu@lzu.edu.cn}
\affiliation{
$^1$School of Physical Science and Technology, Lanzhou University, Lanzhou 730000, China\\
$^2$Lanzhou Center for Theoretical Physics,
Key Laboratory of Theoretical Physics of Gansu Province,
Key Laboratory of Quantum Theory and Applications of MoE,
Gansu Provincial Research Center for Basic Disciplines of Quantum Physics, Lanzhou University, Lanzhou 730000, China\\
$^3$MoE Frontiers Science Center for Rare Isotopes, Lanzhou University, Lanzhou 730000, China\\
$^4$Institute of Theoretical Physics, College of Physics and Electronic Engineering,
Northwest Normal University, Lanzhou 730070, China\\
$^5$Research Center for Hadron and CSR Physics, Lanzhou University and Institute of Modern Physics of CAS, Lanzhou 730000, China}

\date{\today}
\begin{abstract}

In this work, we systematically study the interactions of the $S$-wave $D^{(*)}\bar{B}^{(*)}$ systems within the framework of chiral effective field theory in heavy hadron formalism. We calculate the $D^{(*)}\bar{B}^{(*)}$ effective potentials up to next-to-leading order, explore the bound state formations, and investigate the $D^{(*)}\bar{B}^{(*)}$ scattering properties such as scattering rate, scattering length, and effective range. Our results show that all $I=1$ $D^{(*)}\bar{B}^{(*)}$ potentials are repulsive, preventing the formation of bound states, while the $I=0$ potentials are generally attractive. Specifically, we get two important observations: first, the shallow bound state is more likely to exist in the $D\bar{B}[I(J^{P})=0(0^{+})]$ system than in the $D\bar{B}^{*}[I(J^{P})=0(1^{+})]$ system; second, $D^{*}\bar{B}^{*}[I(J^{P})=0(0^{+})]$ and $D^{*}\bar{B}^{*}[I(J^{P})=0(1^{+})]$ systems possess relatively large binding energies and positive scattering lengths, which suggests strong bound state formations in these channels. So the attractions in the $D^{*}\bar{B}^{*}[I=0]$ systems are deeper than those in the $D\bar{B}^{(*)}[I=0]$ systems, thus we strongly recommend the future experiment to search for the $D^{*}\bar{B}^{*}[I=0]$ tetraquark systems. We also investigate the coupled-channel effects on the $J=0, 1$ systems and conclude that the inclusion of the coupled channels introduces small but visible influences. In addition, we also investigate the dependencies of the $D\bar{B}^{(*)}$ binding energies on the contact low-energy coupling constants.

\end{abstract}

\maketitle
\thispagestyle{empty} %

\section{Introduction}
\label{Sec: Introduction}

The study of exotic hadrons has become a vibrant frontier in particle physics, offering profound insights into the nonperturbative regime of strong interactions. In recent years, LHCb Collaboration has made groundbreaking discoveries that have significantly advanced our understanding of multiquark states. In 2020, LHCb reported the observation of two resonances, $X_{0}(2900)$ and $X_{1}(2900)$, in the $D^{-}K^{+}$ invariant mass spectrum of the $B^{+} \to D^{+}D^{-}K^{+}$ decay process \cite{LHCb:2020bls, LHCb:2020pxc}. These states, with quantum numbers $J^{P}=0^{+}$ and $J^{P}=1^{+}$, respectively, were identified as the first tetraquark candidates with the exotic quark content $ud\bar{s}\bar{c}$. This discovery was followed in 2021 by the observation of the double-charm tetraquark state $T^{+}_{cc}$ in the $D^{0}D^{0}\pi^{+}$ invariant mass spectrum \cite{LHCb:2021vvq, LHCb:2021auc}. Shortly thereafter, LHCb reported two additional open heavy-flavor tetraquark candidates, $T_{c\bar{s}0}^{a}(2900)^{++}$ and $T_{c\bar{s}0}^{a}(2900)^{0}$, in the $D_{s}^{+}\pi^{\pm}$ final states of the $B^{+}\to D^{-}D^{+}_{s}\pi^{+}$ and $B^{0}\to \bar{D}^{0}D^{+}_{s}\pi^{+}$ decays \cite{LHCb:2022sfr, LHCb:2022lzp}. These states are believed to have the quark content $c\bar{s}u\bar{d}$.

These experimental discoveries have spurred extensive theoretical investigations into the nature of open heavy-flavor multiquark states. Various approaches, including QCD sum rules, lattice QCD, and effective field theories, have been employed to explore their properties and structures (see Refs.~\cite{Cui:2006mp, Detmold:2007wk, Carlucci:2007um, Navarra:2007yw, Yang:2009zzp, Molina:2010tx, Carames:2011zz, Du:2012wp, Hyodo:2012pm, Ohkoda:2012hv, Vijande:2013qr, Ikeda:2013vwa, Bicudo:2015kna, Bicudo:2015vta, Peters:2015tra, Peters:2016isf, Francis:2016hui, Bicudo:2017szl, Chen:2017rhl, Azizi:2018mte, Bicudo:2019mny, Wang:2019xzt, Lu:2020qmp, He:2020jna, Cheng:2020nho, Albuquerque:2020ugi, Guo:2021mja, Chen:2022asf} for comprehensive reviews). These studies have deepened our understanding of the exotic hadron spectrum and highlighted the importance of investigating the interactions of dimeson systems, particularly those involving doubly heavy quarks.

Doubly heavy dimeson systems have attracted significant theoretical interest due to their potential to form stable or quasistable molecular states. For instance, Li \textit{et al.} explored the possibility of deuteronlike molecular states in $D^{(*)}D^{(*)}$, $\bar{B}^{(*)}\bar{B}^{(*)}$, and $D^{(*)}\bar{B}^{(*)}$ systems, identifying several promising candidates \cite{Li:2012ss}. QCD sum rules have been applied to study exotic open-flavor tetraquark states such as $bc\bar{q}\bar{q}$, $bc\bar{s}\bar{s}$, $qc\bar{q}\bar{b}$, and $sc\bar{s}\bar{b}$, revealing that some of these states lie below the $D^{(*)}B^{(*)}$ and $D_{s}^{(*)}B_{s}^{(*)}$ thresholds, making them susceptible to decay via fall-apart mechanisms \cite{Chen:2013aba}. The color-magnetic interaction model has also been used to calculate mass splittings for $qq\bar{Q}\bar{Q}$ tetraquark states \cite{Luo:2017eub}, while the molecular picture has been employed to investigate the binding energies of $T_{cc}^{+}$ and its bottom and strange partners \cite{Deng:2021gnb}. Additionally, the complex scaling method has been utilized to study doubly heavy tetraquark bound and resonant states \cite{Wu:2024zbx}.

The stability of doubly heavy tetraquarks, particularly those involving bottom and charm quarks, remains a topic of active debate. While some studies suggest that states such as $bb\bar{u}\bar{d}$, $bb\bar{u}\bar{s}$, and $bb\bar{d}\bar{s}$ are stable against strong decay \cite{Eichten:2017ffp}, others argue that bottom-charm tetraquarks may not be stable \cite{Silvestre-Brac:1993zem, Ebert:2007rn, Park:2018wjk, Lu:2020rog, Braaten:2020nwp, Kim:2022mpa, Song:2023izj}. The production mechanisms and detection prospects for these states at colliders have also been explored, with estimates suggesting significant production cross sections for double-bottom and mixed bottom-charm tetraquarks at the LHC \cite{Ali:2018xfq}.

Recent lattice QCD studies have provided further insights into the properties of doubly heavy tetraquarks. Meinel \textit{et al.} found no evidence for QCD-stable $\bar{b}\bar{c}ud$ tetraquark states \cite{Meinel:2022lzo}, while Alexandrou \textit{et al.} and Padmanath \textit{et al.} reported shallow bound states for isoscalar $\bar{b}\bar{c}ud$ systems \cite{Alexandrou:2023cqg, Padmanath:2023rdu}. These findings, along with the identification of subthreshold poles in the $D\bar{B}$ scattering amplitude \cite{Radhakrishnan:2024ihu}, significantly advanced our understanding of the weak decay behaviors of $D\bar{B}$ molecules \cite{Liu:2024lmv}.

In this work, we investigate the interactions of $S$-wave $D\bar{B}$, $D\bar{B}^{*}$, and $D^{*}\bar{B}^{*}$ systems using chiral effective field theory (ChEFT) in the heavy hadron formalism. ChEFT, a low-energy effective theory of QCD, has proven to be a powerful tool for studying hadronic interactions, particularly in the context of nucleon-nucleon systems \cite{Machleidt:2011zz, Hammer:2019poc}. By incorporating heavy-quark and chiral symmetries, ChEFT provides a robust framework for analyzing the interactions of heavy mesons. Previous studies have applied ChEFT to systems such as $D^{(*)}D^{(*)}$ \cite{Xu:2017tsr, Wang:2022jop}, $\bar{B}^{(*)}\bar{B}^{(*)}$ \cite{Wang:2018atz, Abreu:2022sra}, and hidden-charm systems \cite{Xu:2021vsi}, as well as to the interactions of heavy pentaquark molecular states \cite{Meng:2019ilv, Wang:2019ato, Wang:2022ztm}. These investigations have demonstrated the effectiveness of ChEFT in describing the low-energy dynamics of heavy hadrons.

This paper is organized as follows: In Sec.~\ref{Sec: Lagrangian}, we introduce the chiral effective Lagrangians with $SU(2)$ flavor symmetry used in our calculations. In Sec.~\ref{Sec: Potentials}, we present the scattering amplitudes and derive the effective meson-meson potentials. In Sec.~\ref{Sec: Results}, we analyze the numerical results, examining the behaviors of the effective potentials and searching for possible bound states by solving the Schr\"odinger equation. We also examine the coupled-channel effects. We discuss the scattering $T$-matrix and extract physical quantities such as the scattering rate, scattering length, and effective range. Finally, in Sec.~\ref{Sec: Summary}, we summarize our findings and discuss their implications for future studies of doubly heavy tetraquark systems.


\section{Effective Lagrangians}
\label{Sec: Lagrangian}	

Within the framework of ChEFT in the heavy hadron formalism, the low-energy $D^{(*)}$-$\bar{B}^{(*)}$ scattering amplitudes are expanded order by order in terms of a small parameter $\epsilon = q / \Lambda_{\chi}$, where $q$ represents the momentum of a Goldstone boson, the mass difference between $D^{*}$ and $D$ mesons (or $\bar{B}^{*}$ and $\bar{B}$ mesons), or the residual momentum of a heavy meson. The parameter $\Lambda_{\chi}$ denotes the chiral symmetry broken scale or the mass of the heavy mesons. 

\subsection{Lagrangians at the leading order}

At leading order (LO), the Lagrangian describing the LO contact $D^{(*)}\bar{B}^{(*)}$ interaction is:
\begin{align}
	 \mathcal{L}_{4H}^{(0)}=
	 &D_{a}\mathrm{Tr}[H\gamma_{\mu}\bar{H}]\mathrm{Tr}[H\gamma^{\mu}\bar{H}] \notag \\
	 &+D_{b}\mathrm{Tr}[H\gamma_{\mu}\gamma_{5}\bar{H}]\mathrm{Tr}[H\gamma^{\mu}\gamma_{5}\bar{H}] \notag \\
	 &+E_{a}\mathrm{Tr}[H\gamma_{\mu}\tau^{a}\bar{H}]\mathrm{Tr}[H\gamma^{\mu}\tau_{a}\bar{H}] \notag \\
	 &+E_{b}\mathrm{Tr}[H\gamma_{\mu}\gamma_{5}\tau^{a}\bar{H}]\mathrm{Tr}[H\gamma^{\mu}\gamma_{5}\tau_{a}\bar{H}],  \label{L1}
\end{align}	 
where $D_a$, $D_b$, $E_a$, and $E_b$ are independent low-energy coupling
constants (LECs), and $\tau^\alpha$ represents the Pauli matrix in isospin space. The heavy meson doublet $(D, D^{*})$ or $(\bar{B}, \bar{B}^{*})$ is described by the $H$ field in the heavy hadron formalism \cite{Xu:2017tsr, Xu:2021vsi, Wang:2018atz}:
\begin{align}
	H& =\frac{1+\slashed{v}}{2}(P^{*}_{\mu}\gamma^{\mu}+iP\gamma_{5}),\notag \\
	\bar{H}& =\gamma^0 H^\dagger \gamma^0=(P_\mu^{*\dagger}\gamma^\mu+i P^\dagger \gamma_{5})\frac{1+\slashed{v}}{2}, \notag \\
	P & =(D^{0},D^{+})\quad \rm{or} \quad (\it{B}^{-},\bar{\it{B}}^{\rm{0}}), \notag \\
	P^{*}_{\mu}&=(D^{*0},D^{*+})_{\mu}\quad \rm{or} \quad (\it{B}^{*-},\bar{\it{B}}^{*\rm{0}})_{\mu},   
\end{align}
where $v=(1,0,0,0)$ is the four-velocity of the heavy mesons. For the $D^{*}\bar{B}^{*}$ system, the scattering amplitudes receive contributions from the one-pion-exchange interactions at LO. The Lagrangian for the LO $D^{*}D^{*}\pi$ vertex or $\bar{B}^{*}\bar{B}^{*}\pi$ vertex is given by \cite{Xu:2017tsr, Xu:2021vsi, Wang:2018atz}
\begin{align}
    \mathcal{L}_{H\phi}^{(1)}=
    &-\langle(iv\cdot\partial H)\bar{H}\rangle+\langle H v\cdot\Gamma\bar{H}\rangle+g\langle H\slashed{u}\gamma_{5}\bar{H}\rangle \notag \\ 
    &-\frac{1}{8}\Delta\langle H\sigma^{\mu\nu}\bar{H}\sigma_{\mu\nu}\rangle,   \label{L2}
\end{align}

\begin{align}
    \Gamma_\mu=&\frac{i}{2}[\xi^\dagger,\partial_\mu\xi],  \quad u_\mu=\frac{i}{2}\{\xi^\dagger,\partial_\mu\xi\},
\end{align}
where $\Delta$ in the last term represents the mass splitting between $D$ and $D^{*}$ (or $\bar{B}$ and $\bar{B}^{*}$), $\Gamma_{\mu}$ is the chiral connection, $u_{\mu}$ is the axial vector current, and $\xi=\exp(i\phi/2f)$. Here, $f$ is the bare pion decay constant, and the pion field $\phi$ is 
\begin{align}
	\phi= \begin{pmatrix}
	\pi^{0}   & \sqrt{2} \pi^{+}  \\
	\sqrt{2} \pi^{-}  &  -\pi^{0}
	\end{pmatrix} .
\end{align}
	
\subsection{Lagrangians at next-to-leading order}
	
At next-to-leading order (NLO), the scattering amplitudes receive contributions from one-loop corrections to the LO contact interaction, one-loop corrections to LO one-pion-exchange (OPE), and the two-pion-exchange (TPE) terms. To renormalize these loop contributions, we employ the contact Lagrangians at NLO as follows:

\begin{align}
    \mathcal{L}_{4H}^{(2,h)} = &  
    D_a^h\mathrm{Tr}[H\gamma_\mu\bar{H}]\mathrm{Tr}[H\gamma^\mu\bar{H}]\mathrm{Tr}(\chi_+)  \notag  \\
	&+D_b^h\mathrm{Tr}[H\gamma_\mu\gamma_5\bar{H}]\mathrm{Tr}[H\gamma^\mu\gamma_5\bar{H}]\mathrm{Tr}(\chi_+) \notag  \\
	&+E_a^h\mathrm{Tr}[H\gamma_\mu\tau^a\bar{H}]\mathrm{Tr}[H\gamma^\mu\tau_a\bar{H}]\mathrm{Tr}(\chi_+)  \notag  \\
	&+E_{b}^{h}\mathrm{Tr}[H\gamma_{\mu}\gamma_{5}\tau^{a}\bar{H}]\mathrm{Tr}[H\gamma^{\mu}\gamma_{5}\tau_{a}\bar{H}]\mathrm{Tr}(\chi_{+}),  \label{L3}
\end{align}

\begin{align}
    \mathcal{L}_{4H}^{(2,v)} = &
    \{D_{a1}^{v}\mathrm{Tr}[(v\cdot DH)\gamma_{\mu}(v\cdot D\bar{H})]\mathrm{Tr}[H\gamma^{\mu}\bar{H}]  \notag  \\
	&+D_{a2}^v\mathrm{Tr}[(v\cdot DH)\gamma_\mu\bar{H}]\mathrm{Tr}[(v\cdot DH)\gamma^\mu\bar{H}] \notag  \\
    & +D_{a3}^v\mathrm{Tr}[(v\cdot DH)\gamma_\mu\bar{H}]\mathrm{Tr}[H\gamma^\mu(v\cdot D\bar{H})]  \notag  \\
	&+D_{a4}^v\mathrm{Tr}[((v\cdot D)^2H)\gamma_\mu\bar{H}]\mathrm{Tr}[H\gamma^\mu\bar{H}] \notag  \\
    &+D_{b1}^v\operatorname{Tr}[(v\cdot DH)\gamma_\mu\gamma_5(v\cdot D\bar{H})]\operatorname{Tr}[H\gamma^\mu\gamma_5\bar{H}]+\cdots \notag  \\
    &+E_{a1}^v\mathrm{Tr}[(v\cdot DH)\gamma_\mu\tau^a(v\cdot D\bar{H})]\mathrm{Tr}[H\gamma^\mu\tau_a\bar{H}]+\cdots \notag  \\
    &+E_{b1}^v\mathrm{Tr}[(v\cdot DH)\gamma_\mu\gamma_5\tau^a(v\cdot D\bar{H})]\mathrm{Tr}[H\gamma^\mu\gamma_5\tau_a\bar{H}]  \notag  \\
	&+\cdots\}+\mathrm{H.c.}, \label{L4}
\end{align}

\begin{align}
    \mathcal{L}_{4H}^{(2,q)} =
    &\{D_{1}^{q}\mathrm{Tr}[(D^{\mu}H)\gamma_{\mu}\gamma_{5}(D^{\nu}\bar{H})]\mathrm{Tr}[H\gamma_{\nu}\gamma_{5}\bar{H}]  \notag  \\
	&+D_2^q\mathrm{Tr}[(D^\mu H)\gamma_\mu\gamma_5\bar{H}]\mathrm{Tr}[(D^\nu H)\gamma_\nu\gamma_5\bar{H}] \notag  \\
    &+D_3^q\mathrm{Tr}[(D^\mu H)\gamma_\mu\gamma_5\bar{H}]\mathrm{Tr}[H\gamma_\nu\gamma_5(D^\nu\bar{H})]  \notag  \\
	&+D_4^q\mathrm{Tr}[(D^\mu D^\nu H)\gamma_\mu\gamma_5\bar{H}]\mathrm{Tr}[H\gamma_\nu\gamma_5\bar{H}] \notag  \\
    &+E_1^q\mathrm{Tr}[(D^\mu H)\gamma_\mu\gamma_5\tau^a(D^\nu\bar{H})]\mathrm{Tr}[H\gamma_\nu\gamma_5\tau_a\bar{H}]  \notag  \\
	&+ \cdots\}+\mathrm{H.c.},  \label{L5}
\end{align}

where
\begin{align}
    \tilde{\chi}_{\pm}=\chi_{\pm}-\frac12\mathrm{Tr}[\chi_{\pm}],~
    \chi_{\pm}=\xi^\dagger\chi\xi^\dagger\pm\xi\chi\xi,~ 
    \chi=m_\pi^2.
\end{align}

The low-energy constants appearing in the above equations are split into finite and infinite parts. The infinite parts are used to cancel the divergences of the loop contributions at NLO. The finite parts also contribute, but are not determined here due to the lack of data input. Therefore, we just discard them.

After the scattering amplitudes for the $D^{(*)}\bar{B}^{(*)}$ systems are calculated using the above Lagrangians, the effective potentials in momentum space can be obtained via the relation,
\begin{align}
     \mathcal{V}(q)=-\frac{\mathcal{M}(q)}{4},
\end{align}
where the factor $-1/4$ comes from the Breit approximation. To understand the interactions and further investigate whether the $D^{*}\bar{B}^{(*)}$ systems can form stable molecular states, we need the potentials in coordinate space. Fourier transformations of the momentum-space potentials allow for their derivation,
\begin{align}
     \mathcal{V}(r)=\int\frac{d^3\mathbf{q}}{(2\pi)^3}\mathcal{V}(\mathbf{q})e^{-i\mathbf{q}\cdot\mathbf{r}}\mathcal{F}(\mathbf{q}),
\end{align}
where $\mathcal{F}(\mathbf{q})$ is a regulator function. To avoid ultraviolet (UV) divergence in the integral, we use a Gaussian regulator function $\mathcal{F}(\mathbf{q}) = \exp(-\mathbf{q}^{2n}/\Lambda^{2n})$ and an adequately large $n$ is adopted so that the powers generated by the regulator are beyond the order at which our calculation is performed so that it does not affect the accuracy at the given order \cite{Xu:2017tsr, Wang:2018atz, Xu:2021vsi, Entem:2003ft, Entem:2017gor}. Therefore, we chose $2n > \epsilon$ for the contact terms at order $\epsilon$. Here, we take $n=2$. Based on the potentials in coordinate space, we can solve the Schrödinger equations to search for the possible bound states. 

Simultaneously, according to the effective potentials in momentum space, we can also solve the Lippmann-Schwinger equations to calculate the partial-wave two-body scattering amplitudes $T$,
\begin{align}
    T_{l}(k,k') & =V_{l}(k,k') \notag \\ 
    & \quad+\int \frac{q^{2}dq}{(2\pi)^{3}}\mathcal{V}_{l}(k,q)\mathcal{G}(E,q)T_{l}(q,k') ,\label{LSE}
\end{align}
where $E=p^{2}/(2\mu)$ is the energy and $\mu$ is the reduced mass. The Green's function $\mathcal{G}(E,q)$ is given by
\begin{align}
    \mathcal{G}(E,q)=\frac{1}{E-(q^{2}/2\mu)+i\epsilon}.
\end{align}

The relation between the $T$ matrix and the phase shift $\delta$ is
\begin{align}
    T_{l}=e^{i\delta}\rm{sin}\delta_{\it{l}}. \label{phase}
\end{align}

Considering the $S$-wave, in the low-energy limit the $k\rm{cot}\delta_{0}(k)$ can be expanded in powers of $k^{2}$,
\begin{align}
    k\rm{cot}\delta_{0}(\it{k}) = -\frac{\rm{1}}{a}+\frac{\rm{1}}{\rm{2}}r_{\rm{0}}k^{\rm{2}}+..., \label{ERE}
\end{align}
where $a$ is the scattering length and $r_0$ is the effective range. The ellipsis represents higher-order terms in $k$. The $S$-wave scattering cross section is proportional to the phase shift,
\begin{align}
    \sigma_{0}(k) = \frac{4\pi}{k^{2}}\rm{sin}^{2}\delta_{0}(\it{k}). \label{sigma}
\end{align}

To facilitate a comparison with lattice QCD results \cite{Alexandrou:2023cqg}, we will also compute the scattering rate $k\sigma(k)$.

\section{effective potentials of the $D^{(*)}\bar{B}^{(*)}$ systems}
\label{Sec: Potentials}	
\subsection{$D\bar{B}$ systems}
    
\begin{figure*}[ht]
     \centering
     \includegraphics[width=0.7\textwidth]{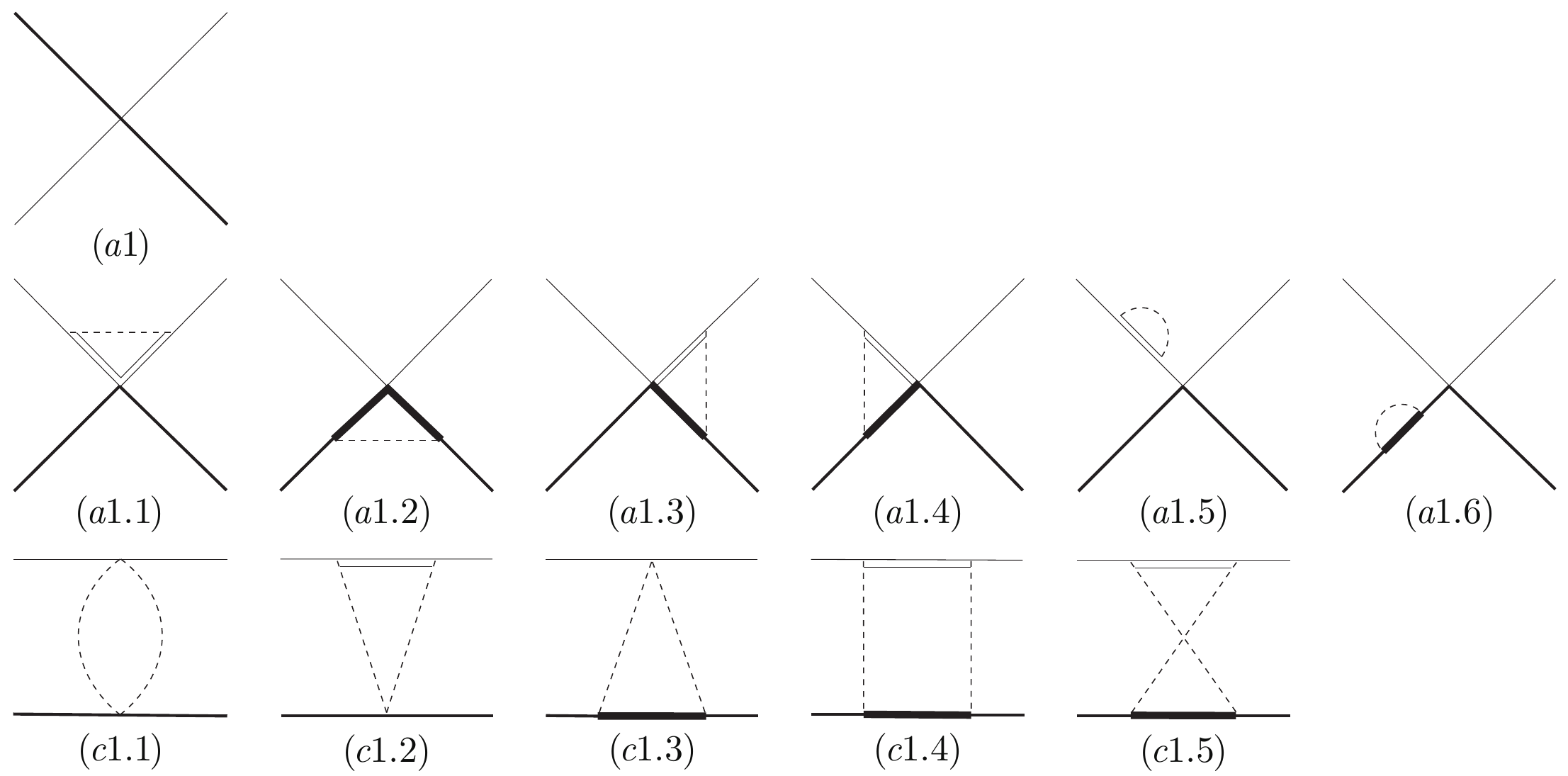}
     \captionsetup{justification=raggedright, singlelinecheck=false}
     \caption{LO contact ($a1$), NLO contact ($a1.1-a1.6$) and NLO TPE ($c1.1-c1.5$) diagrams of the process $D\bar{B} \to D\bar{B}$. The thin solid, double-thin solid, solid, thick solid, and dashed lines stand for the $\bar{B}$, $\bar{B}^{*}$, $D$, $D^{*}$, and a pion, respectively.}\label{BD1}
\end{figure*}

As shown in Fig.~\ref{BD1}, at LO, there is only a tree-level contact diagram for the scattering process $D(p_1)\bar{B}(p_2) \to D(p_3)\bar{B}(p_4)$. Using the Lagrangians in Eq.~(\ref{L1}), the amplitudes can be written as follows:
\begin{align}
    \mathcal{M}_{a1}^{(0)}& =4(D_a+E_a),
\end{align}
 for isospin $I=1$, and for isospin $I=0$  
\begin{align}
    \mathcal{M}_{a1}^{(0)}& =4(D_a-3E_a).
\end{align}

Next, we consider NLO contributions illustrated in Fig.~\ref{BD1}. There are two types of NLO diagrams. The diagrams in Fig.~\ref{BD1} represent one-loop corrections to the LO contact interaction, and the corresponding amplitudes are given by, 
\begin{align}
    \mathcal{M}_{a1.1}^{(2)}& =-4A(d-1)\frac{g_{2}^{2}}{f^{2}}J_{22}^{g}, \label{a1.1} 
\end{align}

\begin{align}
    \mathcal{M}_{a1.2}^{(2)}& =-4A(d-1)\frac{g_{1}^{2}}{f^{2}}J_{22}^{g}, \label{a1.2} 
\end{align}

\begin{align}
    \mathcal{M}_{a1.3}^{(2)}& =4A(d-1)\frac{g_{1}g_{2}}{f^{2}}J_{22}^{h}, \label{a1.3} 
\end{align}

\begin{align}
    \mathcal{M}_{a1.4}^{(2)}& =4A(d-1)\frac{g_{1}g_{2}}{f^{2}}J_{22}^{h}, \label{a1.4} 
\end{align}

\begin{align}
    \mathcal{M}_{a1.5}^{(2)}& =-\frac{3}{2} A(d-1)\frac{g_{1}^{2}}{f^{2}}\partial_{\omega}J_{22}^{b}, \label{a1.5} 
\end{align}

\begin{align}
    \mathcal{M}_{a1.6}^{(2)}& =-\frac{3}{2} A(d-1)\frac{g_{1}^{2}}{f^{2}}\partial_{\omega}J_{22}^{b}.\label{a1.6}
\end{align}
Here, $d$ represents the space-time dimension, and the coefficient $A$ depends on each diagram and isospin $I$, which is listed in Table~\ref{1}. The coupling constants $g_1$ and $g_2$ refer to the bare coupling constants for the $DD^*\pi$ and $\bar{B}\bar{B}^*\pi$ vertices, respectively.

Using the Lagrangians in Eq.~(\ref{L2}), the amplitudes for the TPE diagrams in Fig.~\ref{BD1} can be written as 
\begin{align}
    \mathcal{M}_{c1.1}^{(2)}=& -4\frac{1}{f^{4}}[A_1(q_{0}^{2}J_{0}^{F}+J_{22}^F)\notag \\
    &-A_{15}(q_{0}^{2}J_{11}^{F}+q_0^2J_{21}^F+J_{22}^F) \notag \\
    &-A_{51}(q_{0}^{2}J_{11}^{F}+q_0^2J_{21}^F+J_{22}^F) \notag \\
    &+A_{5}(q_{0}^{2}J_{0}^{F}+2q_{0}^{2}J_{11}^{F}+ q_0^2J_{21}^F+J_{22}^F)], \label{t1.1} 
\end{align}

\begin{align}
    \mathcal{M}_{c1.2}^{(2)}=& -4i\frac{g_{2}^{2}}{f^{4}}\{A_{1}[q_{0} \vec{q}^{2} J_{22}^{T}+\vec{q}^{2} J_{24}^{T}-(d-1)q_{0} J_{31}^{T} \notag \\
    & +q_{0} \vec{q}^{2} J_{32}^{T} +\vec{q}^{2} J_{33}^{T}-(d-1) J_{34}^{T}] \notag \\
    & -A_{5}[q_{0} \vec{q}^{2} J_{11}^{T}-(d-1)q_{0} J_{21}^{T}  \notag  \\
    &+2q_{0} \vec{q}^{2} J_{22}^{T}+\vec{q}^{2} J_{24}^{T}-(d-1)q_{0} J_{31}^{T} \notag \\
    & +q_{0} \vec{q}^{2} J_{32}^{T}+\vec{q}^{2} J_{33}^{T} -(d-1) J_{34}^{T}]\}, \label{t1.2} 
\end{align}

\begin{align}
    \mathcal{M}_{c1.3}^{(2)}=&-4i\frac{g_{2}^{2}}{f^{4}}\{ A_{1}[q_{0} \vec{q}^{2} J_{22}^{T}+\vec{q}^{2} J_{24}^{T}-(d-1)q_{0} J_{31}^{T}  \notag \\
    & + q_{0} \vec{q}^{2} J_{32}^{T} + \vec{q}^{2} J_{33}^{T}-(d-1) J_{34}^{T}] \notag \\
    & -A_{5}[q_{0} \vec{q}^{2} J_{11}^{T}-(d-1)q_{0} J_{21}^{T} \notag  \\
    &+2q_{0} \vec{q}^{2} J_{22}^{T}+\vec{q}^{2} J_{24}^{T}-(d-1)q_{0} J_{31}^{T} \notag \\
    & +q_{0} \vec{q}^{2} J_{32}^{T}+\vec{q}^{2} J_{33}^{T} -(d-1) J_{34}^{T}]\}, \label{t1.3} 
\end{align}

\begin{align}
    \mathcal{M}_{c1.4}^{(2)}=& -4\frac{g_{1}^{2}g_{2}^{2}}{f^{4}} A[-\vec{q}^{2} J_{21}^{B}+\vec{q}^{4} J_{22}^{B}-2(d+1)\vec{q}^{2} J_{31}^{B} \notag \\
    & +2 \vec{q}^{4} J_{32}^{B} + (d^{2}-1) J_{41}^{B} \notag \\
    & -2(d+1)\vec{q}^{2} J_{42}^{B}+\vec{q}^{4} J_{43}^{B}], \label{t1.4}
\end{align}

\begin{align}
    \mathcal{M}_{c1.5}^{(2)}=& -4\frac{g_{1}^{2}g_{2}^{2}}{f^{4}} A[-\vec{q}^{2} J_{21}^{R}+\vec{q}^{4} J_{22}^{R} -2(d+1)\vec{q}^{2} J_{31}^{R} \notag \\
    & +2 \vec{q}^{4} J_{32}^{R} +(d^{2}-1) J_{41}^{R} \notag \\
    & -2(d+1)\vec{q}^{2} J_{42}^{R}+\vec{q}^{4} J_{43}^{R}], \label{t1.5}      
\end{align}
 
where coefficients $A_{1}$, $A_{15}$, $A_{51}$ and $A_{5}$ for each amplitude can be found in Table \ref{2}. 
     
In the above amplitudes (\ref{a1.1})-(\ref{t1.5}), the loop functions $J^{a/b}_{ij}(m,\omega)$, $J^{g/h}_{ij}(m,\omega_{1},\omega_{2})$, $J^{F}_{ij}(m_{1},m_{2},q)$, $J^{T/S}_{ij}(m_{1},m_{2},\omega,q)$, and $J^{B/R}_{ij}(m_{1},m_{2},\omega_{1},\omega_{2},q)$ are abbreviated as $J^{a/b}_{ij}$, $J^{g/h}_{ij}$, $J^{F}_{ij}$, $J^{T/S}_{ij}$, and $J^{B/R}_{ij}$, respectively. The variables $m, m_{1}$, and $m_{2}$ denote the pion mass, while the mass-splitting-dependent variables $\omega_{1}$ and $\omega_{2}$ for every amplitude are collected in Tables \ref{1} and \ref{2}. Here, we define $\delta_{1}=M_{D^{*}}-M_{D}$ and $\delta_{2}=M_{\bar{B}^{*}}-M_{\bar{B}}$ as the $D^{*}$-$D$ and $\bar{B}^{*}$-$\bar{B}$ mass differences, respectively. We list the definitions of the loop functions in Appendix~\ref{SecAppB}, and the calculation procedures of these loop functions follow the Refs.~\cite{Xu:2017tsr, Xu:2021vsi}.

\renewcommand{\arraystretch}{1.4}
\begin{center}
    \begin{table}[htbp]
    	\centering
            \captionsetup{justification=raggedright, singlelinecheck=false}
    	\caption{The coefficients appearing in the contact amplitudes [Eqs.(~\ref{a1.1})-(\ref{a1.6})].}\label{1}
    	\setlength{\tabcolsep}{2.8mm}{ 
    		\begin{tabular}{cccccc}
    			\toprule[1pt]
    			& { $I=1$  } & { $I=0$ } &    & \\
    			&  $A$     &    $A$    &   $\omega_1$   & $\omega_2$ \\
    			\midrule[1pt]
    			$A_{a1.1}$ & $0$  & $0$  & $-\delta_{2}$ & $-\delta_{2}$ \\
    			$A_{a1.2}$ & $0$  & $0$  & $-\delta_{1}$ & $-\delta_{1}$ \\
    			$A_{a1.3}$ & $\frac{1}{4}(D_{b}+E_{b})$  & $\frac{1}{4}(-3D_{b}+9E_{b})$  & $-\delta_{1}$ & $-\delta_{2}$ \\
    			$A_{a1.4}$ & $\frac{1}{4}(D_{b}+E_{b})$  & $\frac{1}{4}(-3D_{b}+9E_{b})$  & $-\delta_{1}$ & $-\delta_{2}$ \\
    			$A_{a1.5}$ & $D_{a}+E_{a}$  & $D_{a}-3E_{a}$  & $-\delta_{1}$ & $0$ \\
    			$A_{a1.6}$ & $D_{a}+E_{a}$  & $D_{a}-3E_{a}$  & $-\delta_{2}$ & $0$ \\
    			\bottomrule[1pt]
    	\end{tabular}}
    \end{table}
\end{center}

\renewcommand{\arraystretch}{1.4}
\begin{center}
    \begin{table*}[!htbp]
     \centering
     \captionsetup{justification=raggedright, singlelinecheck=false}
     \caption{The coefficients appearing in the TPE amplitudes [Eqs.~(\ref{t1.1})-(\ref{t1.5})]. Note that we have $A_{51}=A_{15}$.}\label{2}
    	\setlength{\tabcolsep}{7mm}{
    	\begin{tabular}{ccrcrrccc}	
    		\toprule[1pt]
    			&  \multicolumn{3}{c}{ $I=1$ } &\multicolumn{3}{c}{ $I=0$ } &       & \\
    			\cline{2-4} \cline{5-7}
    			&  $A_1$   &  $A_5$   &  $A_{15}$   &   $A_1$   &  $A_5$   &  $A_{15}$   &  
                    $\omega_1$   & $\omega_2$ \\
    			\midrule[1pt]
    				$A_{c1.1}$ & $\frac{1}{16} $  & $\frac{1}{16} $ & $-\frac{1}{16}$ & $-\frac{3}{16}$ & $-\frac{3}{16}$  & $\frac{3}{16}$ & $0$       & $0$ \\
    				$A_{c1.2}$ & $\frac{i}{8}$  & $-\frac{i}{8}$ & $0$              & $-\frac{3i}{8}$ & $\frac{3i}{8}$  & $0$ & $-\delta_{2}$  &  $0$ \\
    				$A_{c1.3}$ & $\frac{i}{8}$  & $-\frac{i}{8}$ & $0$              & $-\frac{3i}{8}$ & $\frac{3i}{8}$  & $0$ &  $-\delta_{1}$      & $0$ \\
    				$A_{c1.4}$ & $\frac{1}{16}$  & $0$             & $0$             & $\frac{9}{16}$ & $0$            & $0$ & $-\delta_{1}$ & $-\delta_{2}$ \\
    				$A_{c1.5}$& $\frac{5}{16}$  & $0$             & $0$             & $-\frac{3}{16}$ & $0$            & $0$ & $-\delta_{1}$ & $-\delta_{2}$ \\
    			\bottomrule[1pt]
    	\end{tabular}}
    \end{table*}
\end{center}

\subsection{$D\bar{B}^{*}$ systems}

\begin{figure*}[htp]
   \centering
   \includegraphics[width=0.6\textwidth]{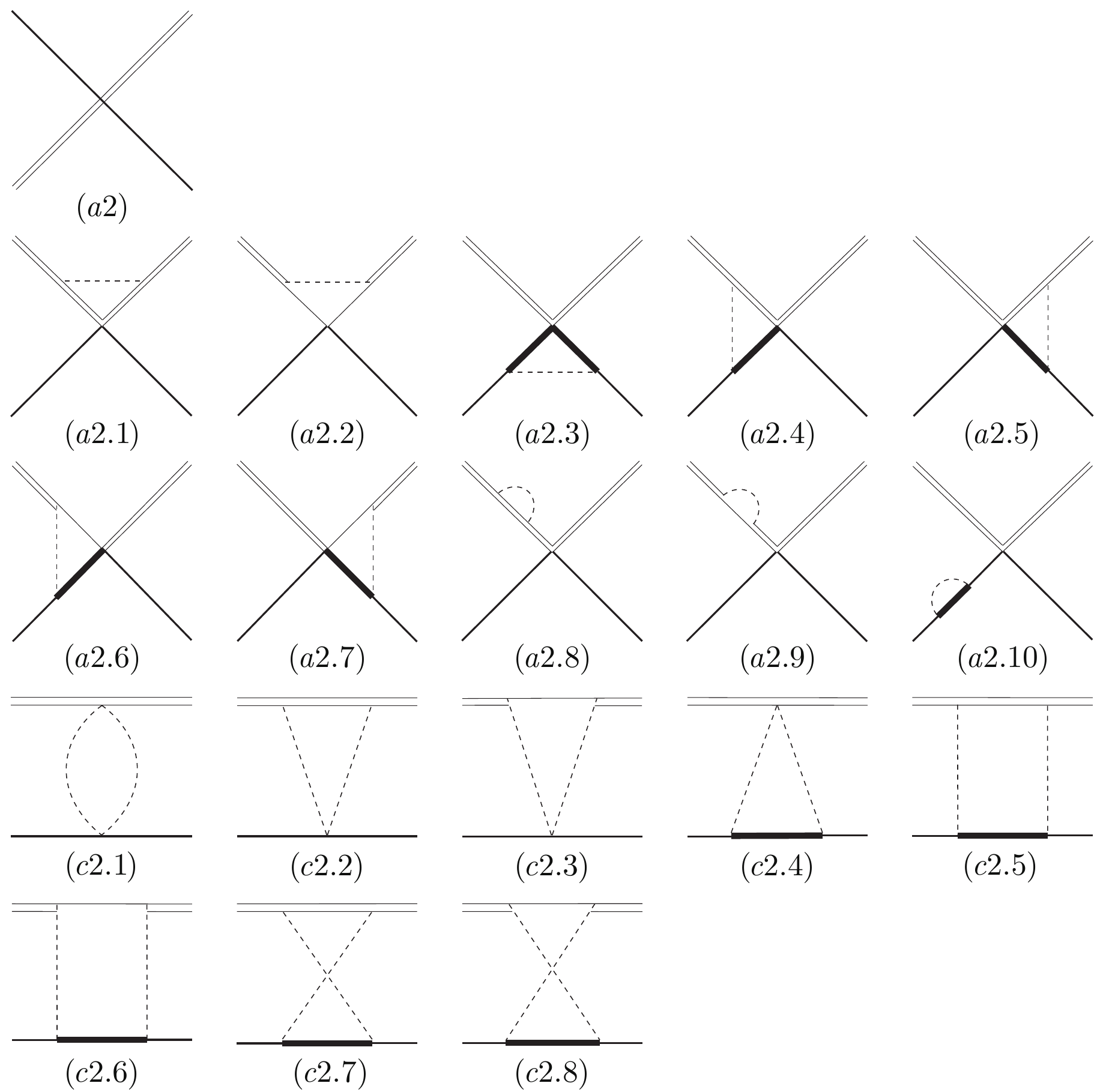}
   \captionsetup{justification=raggedright, singlelinecheck=false}
   \caption{LO contact ($a2$), NLO contact ($a2.1-a2.10$) and NLO TPE ($c2.1-c2.8$) diagrams of the process $D\bar{B}^{*} \to D\bar{B}^{*}$. The thin solid, double-thin solid, solid, thick solid, and dashed lines stand for the $\bar{B}$, $\bar{B}^{*}$, $D$, $D^{*}$, and a pion, respectively.}\label{BsD1}
\end{figure*}

We now focus on the $D\bar{B}^{*}$ system. As in the $D\bar{B}$ system, a contact term contributes to the $D\bar{B}^{*}$ scattering amplitudes at LO, as shown in Fig.~\ref{BsD1}. For the process $D(p_1)\bar{B}^{*}(p_2) \to D(p_3)\bar{B}^{*}(p_4)$, the scattering amplitude is given by
\begin{align}
     \mathcal{M}_{a2}^{(0)}& =-4(D_a+E_a)(\epsilon_{2} \cdot \epsilon^{*}_{4}),
\end{align}
with isospin $I=1$, and    
\begin{align}
     \mathcal{M}_{a2}^{(0)}& =-4(D_a-3E_a)(\epsilon_{2} \cdot \epsilon^{*}_{4}),
\end{align}
 with isospin $I=0$, where $\epsilon_{2}$ and $\epsilon^{*}_{4}$ are the polarization vectors of the initial $\bar{B}^{*}$ and final $\bar{B}^{*}$, respectively.

Next, in Fig.~\ref{BsD1}, we present the one-loop corrections to the contact diagram and the TPE diagrams of the $D\bar{B}^{*}$ system at NLO. We can calculate the amplitudes for these diagrams using the Feynman rules derived from the Lagrangian~\ref{L2}. The one-loop correction terms are: 
\begin{align}
    \mathcal{M}_{a2.1}^{(2)}=& -4\frac{g_{2}^{2}}{f^{2}}A J_{22}^{g} (\epsilon_{2} \cdot \epsilon^{*}_{4}), \label{a2.1} 
\end{align}

\begin{align}
    \mathcal{M}_{a2.2}^{(2)}=&~ 4\frac{g_{2}^{2}}{f^{2}}(d-3)(d-2)A J_{22}^{g}(\epsilon_{2} \cdot \epsilon^{*}_{4}), \label{a2.2} 
\end{align}

\begin{align}
    \mathcal{M}_{a2.3}^{(2)}=&~ 4\frac{g_{1}^{2}}{f^{2}}(d-1)A J_{22}^{g}(\epsilon_{2}\cdot \epsilon^{*}_{4}), \label{a2.3} 
\end{align}

\begin{align}
    \mathcal{M}_{a2.4}^{(2)}=& -4\frac{g_{1}g_{2}}{f^{2}}A J_{22}^{h} (\epsilon_{2} \cdot \epsilon^{*}_{4}), \label{a2.4}
\end{align}

\begin{align}
    \mathcal{M}_{a2.5}^{(2)}=& -4\frac{g_{1}g_{2}}{f^{2}}A J_{22}^{h} (\epsilon_{2} \cdot \epsilon^{*}_{4}), \label{a2.5}
\end{align}

\begin{align}
    \mathcal{M}_{a2.6}^{(2)}=& -4\frac{g_{1}g_{2}}{f^{2}}(d-3)(d-2)A J_{22}^{h} (\epsilon_{2} \cdot \epsilon^{*}_{4}), \label{a2.6}
\end{align}

\begin{align}
    \mathcal{M}_{a2.7}^{(2)}=& -4\frac{g_{1}g_{2}}{f^{2}}(d-3)(d-2)A J_{22}^{h} (\epsilon_{2} \cdot \epsilon^{*}_{4}), \label{a2.7} 
\end{align}

\begin{align}
    \mathcal{M}_{a2.(8+9)}^{(2)}=& ~\frac{3g_{2}^{2}}{2f^{2}}A [(d-2)\partial_{\omega} J_{22}^{b} (\omega_{1})+\partial_{\omega} J_{22}^{b}(\omega_{2})] \notag \\ 
    & \times (\epsilon_{2} \cdot \epsilon^{*}_{4}), \label{a2.8} 
\end{align}

\begin{align}
    \mathcal{M}_{a2.10}^{(2)}=&~ \frac{3g_{1}^{2}}{2f^{2}}(d-2) A \partial_{\omega} J_{22}^{b} (\epsilon_{2} \cdot \epsilon^{*}_{4}). \label{a2.10}
\end{align}
The TPE diagrams consist of football, triangle, planar, and crossed box diagrams. The corresponding amplitudes are given by:

\begin{align}
    \mathcal{M}_{c2.1}^{(2)}=&~ 4\frac{1}{f^{4}}[A_1(q_{0}^{2} J_{0}^{F}+J_{22}^F) \notag \\
    & +A_{15}(q_{0}^{2} J_{11}^{F}+q_{0}^{2} J_{21}^F+J_{22}^F) \notag \\
    & -A_{51}(q_{0}^{2} J_{11}^{F}+q_{0}^{2} J_{21}^F+J_{22}^F) \notag \\
    & +A_{5}(q_{0}^{2} J_{0}^{F}+2q_{0}^{2} J_{11}^{F} \notag \\
    & +q_{0}^{2} J_{21}^F+J_{22}^F)]\epsilon_{2} \cdot \epsilon^{*}_{4}, \label{c2.1} 
\end{align}

\begin{align}
    \mathcal{M}_{c2.2}^{(2)}=&~ i4\frac{g_{2}^{2}}{f^{4}} \{[A_{1}(q_{0} J_{22}^{S}+J_{24}^{S} +q_{0} J_{32}^{S}+J_{33}^{S}) \notag \\
    & -A_{5}(q_{0} J_{11}^{S} +2q_{0} J_{22}^{S}+J_{24}^{S} \notag \\
    & +q_{0} J_{32}^{S}+J_{33}^{S})](q \cdot \epsilon_{2})( q \cdot \epsilon^{*}_{4})  \notag \\
    & +[A_{1}(q_{0}J_{31}^{S}+J_{34}^{S}) \notag \\
    & -A_{5}(q_{0} J_{21}^{S}+q_{0} J_{31}^{S} +J_{34}^{S})](\epsilon_{2} \cdot \epsilon^{*}_{4})\}, \label{c2.2}
\end{align}

\begin{align}
    \mathcal{M}_{c2.3}^{(2)}=& -i4\frac{g_{2}^{2}}{f^{4}}(d-3)[A_{1}(q_{0} J_{22}^{S}+ J_{24}^{S}+q_{0} J_{32}^{S}+ J_{33}^{S})  \notag \\
    & -A_{5}(q_{0} J_{11}^{S}+2q_{0} J_{22}^{S}+J_{24}^{S} \notag \\
    & +q_{0} J_{32}^{S}+J_{33}^{S})](q \cdot \epsilon_{2})( q \cdot \epsilon^{*}_{4}) \notag \\
    & -i4\frac{g_{2}^{2}}{f^{4}}(d-3)\{A_{1}[q_{0}\vec{q}^{2} J_{22}^{S}+\vec{q}^{2} J_{24}^{S} \notag \\
    & +(2-d)q_{0} J_{31}^{S} +q_{0}\vec{q}^{2} J_{32}^{S}+\vec{q}^{2} J_{33}^{S} \notag \\
    & +(2-d) J_{34}^{S}] -A_{5}[q_{0}\vec{q}^{2} J_{11}^{S} \notag \\
    & +(2-d)q_{0} (J_{21}^{S}+J_{31}^{S})+2q_{0}\vec{q}^{2} J_{22}^{S} +\vec{q}^{2} J_{24}^{S}  \notag \\
    &  +q_{0}\vec{q}^{2} J_{32}^{S} + \vec{q}^{2} J_{33}^{S}+(2-d) J_{34}^{S}]\}(\epsilon_{2} \cdot \epsilon^{*}_{4}), \label{c2.3}
\end{align}

\begin{align}
    \mathcal{M}_{c2.4}^{(2)}=& -i4\frac{g_{1}^{2}}{f^{4}}\{A_{1}[q_{0}\vec{q}^{2} J_{22}^{T}+\vec{q}^{2}J_{24}^{T} +(1-d)q_{0}J_{31}^{T}  \notag \\
    & +q_{0}\vec{q}^{2} J_{32}^{T} +\vec{q}^{2}J_{33}^{T}+(1-d) J_{34}^{T}] -A_{5}[q_{0}\vec{q}^{2} J_{11}^{T} \notag \\
    & +(1-d)q_{0} (J_{21}^{T}+ J_{31}^{T})+2q_{0}\vec{q}^{2} J_{22}^{T} +\vec{q}^{2}J_{24}^{T}  \notag \\
    & +q_{0}\vec{q}^{2} J_{32}^{T}+\vec{q}^{2}J_{33}^{T}  +(1-d) J_{34}^{T} ]\}(\epsilon_{2} \cdot \epsilon^{*}_{4}), \label{c2.4} 
\end{align}

\begin{align}
    \mathcal{M}_{c2.5}^{(2)}=&~ 4\frac{g_{1}^{2}g_{2}^{2}}{f^{4}}A_{1}\{[J_{21}^{B}-\vec{q}^{2} J_{22}^{B} +(d+3)(J_{31}^{B}+J_{42}^{B})  \notag \\
    & -2\vec{q}^{2} J_{32}^{B} -\vec{q}^{2} J_{43}^{B}](q\cdot\epsilon_{2})(q \cdot \epsilon^{*}_{4})-[\vec{q}^{2} (J_{31}^{B} \notag \\
    & + J_{42}^{B})-(1+d) J_{41}^{B}](\epsilon_{2} \cdot \epsilon^{*}_{4})\},  \label{c2.5}
\end{align}

\begin{align}
    \mathcal{M}_{c2.6}^{(2)}=&~ 4\frac{g_{1}^{2}g_{2}^{2}}{f^{4}}(d-3)A_{1}\{[J_{21}^{B}-\vec{q}^{2} J_{22}^{B}+(d+3)J_{31}^{B} \notag \\
    &  -2\vec{q}^{2} J_{32}^{B}+(d+3)J_{42}^{B}-\vec{q}^{2} J_{43}^{B}](q\cdot\epsilon_{2})(q \cdot \epsilon^{*}_{4}) \notag \\
    & +[-\vec{q}^{2} J_{21}^{B}+\vec{q}^{4} J_{22}^{B} -(2d+1)\vec{q}^{2} (J_{31}^{B}+J_{42}^{B})  \notag \\
    & +2\vec{q}^{4} J_{32}^{B}+(d+1)(d-2) J_{41}^{B} +\vec{q}^{4} J_{43}^{B}](\epsilon_{2} \cdot \epsilon^{*}_{4})\}, \label{c2.6}
\end{align}

\begin{align}
    \mathcal{M}_{c2.7}^{(2)}=&~ 4\frac{g_{1}^{2}g_{2}^{2}}{f^{4}}A_{1}\{[J_{21}^{R}-\vec{q}^{2} J_{22}^{R}+(d+3)(J_{31}^{R}+ J_{42}^{R}) \notag \\
    & -2\vec{q}^{2} J_{32}^{R}-\vec{q}^{2} J_{43}^{R}](q\cdot\epsilon_{2})(q \cdot \epsilon^{*}_{4}) \notag \\
    & -[\vec{q}^{2} J_{31}^{R} -(1+d) J_{41}^{R} +\vec{q}^{2} J_{42}^{R}](\epsilon_{2} \cdot \epsilon^{*}_{4})\}, \label{c2.7} 
\end{align}

\begin{align}
    \mathcal{M}_{c2.8}^{(2)}=&~ 4\frac{g_{1}^{2}g_{2}^{2}}{f^{4}}(d-3)A_{1}\{[J_{21}^{R}-\vec{q}^{2} J_{22}^{R}+(d+3)J_{31}^{R} \notag \\
    & -2\vec{q}^{2} J_{32}^{B}+3J_{42}^{B}-\vec{q}^{2} J_{43}^{B}](q\cdot \epsilon_{2})(q \cdot \epsilon^{*}_{4})\notag \\
    & +[-\vec{q}^{2} J_{21}^{R} +\vec{q}^{4} J_{22}^{R} -(2d+1)\vec{q}^{2} (J_{31}^{R}+J_{42}^{R}) \notag \\
    & +2\vec{q}^{4} J_{32}^{R} +(d+1)(d-2) J_{41}^{R} +\vec{q}^{4} J_{43}^{R}](\epsilon_{2} \cdot \epsilon^{*}_{4})\}, \label{c2.8}
\end{align}

The coefficients in the above expressions are listed in Tables~\ref{3} and \ref{4}. In this work, we focus on the $S$-wave interactions, so we replace the terms $\epsilon _{2} \cdot \epsilon^{*}_{4}$ and $(q\cdot\epsilon_{2})(q \cdot \epsilon^{*}_{4})$ in the equations with \cite{Xu:2017tsr,Wang:2018atz,Xu:2021vsi} 
\begin{align}
     (\epsilon_{2}\cdot\epsilon_{4}^{*})\mapsto-1,\quad(q\cdot\epsilon_{2})(q\cdot\epsilon_{4}^{*})\mapsto\frac{1}{d-1}\vec{q}^{2}.
\end{align}

\renewcommand{\arraystretch}{1.5}
\begin{center}
    \begin{table}[hb]
          \centering
          \captionsetup{justification=raggedright, singlelinecheck=false}
          \caption{The coefficients in the NLO contact amplitudes in Eqs.~(\ref{a2.1})-(\ref{a2.10})}.\label{3}
    	\setlength{\tabcolsep}{2.5mm}{
    	\begin{tabular}{ccccrr}
    		\toprule[1pt]
    		& { $I=1$  } & { $I=0$ } &    & \\
    		&  $A$     &    $A$    &   $\omega_1$   & $\omega_2$ \\
    		\midrule[1pt]
    		$A_{a2.1}$ & $\frac{1}{4}(3D_{a}-E_{a})$  & $\frac{3}{4}(D_{a}+E_{a})$ & $\delta_{2}$ & $\delta_{2}$ \\
    		$A_{a2.2}$ & $\frac{1}{4}(3D_{a}-E_{a})$  & $\frac{3}{4}(D_{a}+E_{a})$ & $0$ & $0$  \\
    		$A_{a2.3}$ & $0$  & $0$  & $-\delta_{1}$ & $-\delta_{1}$ \\
    		$A_{a2.4}$ & $\frac{1}{4}(D_{b}+E_{b})$  & $\frac{-3}{4}(D_{b}-3E_{b})$  & $0$ & $-\delta_{1}$ \\
    		$A_{a2.5}$ & $\frac{1}{4}(D_{b}+E_{b})$  & $\frac{-3}{4}(D_{b}-3E_{b})$  & $-\delta_{1}$ & $0$ \\
    		$A_{a2.6}$ & $\frac{1}{4}(D_{b}+E_{b})$  & $\frac{-3}{4}(D_{b}-3E_{b})$  & $\delta_{2}$ & $-\delta_{1}$ \\
    		$A_{a2.7}$ & $\frac{1}{4}(D_{b}+E_{b})$  & $\frac{-3}{4}(D_{b}-3E_{b})$  & $-\delta_{1}$ & $\delta_{2}$ \\
    		$A_{a2.(8+9)}$ & $D_{a}+E_{a}$  & $D_{a}-3E_{a}$  & $0$ & $\delta_{2}$ \\
    		$A_{a2.10}$ & $D_{a}+E_{a}$  & $D_{a}-3E_{a}$  & $-\delta_{1}$ & $0$ \\
    		\bottomrule[1pt]
    	\end{tabular}}
    \end{table}
\end{center}

\renewcommand{\arraystretch}{1.4}
\begin{center}
    \begin{table*}[ht]
    \centering
    \captionsetup{justification=raggedright, singlelinecheck=false}
    \caption{The coefficients in the TPE amplitudes [Eqs.~(\ref{c2.1})-(\ref{c2.8})]. Note that we have $A_{51}=A_{15}$.}\label{4}
    	\setlength{\tabcolsep}{7.2mm}{
    	\begin{tabular}{ccrcrrccc}	
    		\toprule[1pt]
    		&  \multicolumn{3}{c}{ $I=1$ } &\multicolumn{3}{c}{ $I=0$ } &       & \\
    		\cline{2-4} \cline{5-7}
    		&  $A_1$   &  $A_5$   &  $A_{15}$   &   $A_1$   &  $A_5$   &  $A_{15}$   &  $\omega_1$   & $\omega_2$ \\
    		\midrule[1pt]
    		$A_{c2.1}$ & $\frac{1}{16} $  & $\frac{1}{16} $ & $-\frac{1}{16}$ & $-\frac{3}{16}$ & $-\frac{3}{16}$  & $\frac{3}{16}$ & $0$       & $0$ \\
    		$A_{c2.2}$ & $\frac{i}{8}$  & $\frac{-i}{8}$ & $0$              & $-\frac{3i}{8}$ & $\frac{3i}{8}$  & $0$ & $0$  &  $0$ \\
    		$A_{c2.3}$ & $\frac{i}{8}$  & $-\frac{i}{8}$ & $0$              & $-\frac{3i}{8}$ & $\frac{3i}{8}$  & $0$ &  $\delta_{2}$      & $0$ \\
    		$A_{c2.4}$ & $\frac{i}{8}$  & $-\frac{i}{8}$ & $0$              & $-\frac{3i}{8}$ & $\frac{3i}{8}$  & $0$ &  $-\delta_{1}$      & $0$ \\
    		$A_{c2.5}$ & $\frac{1}{16}$  & $0$             & $0$             & $\frac{9}{16}$ & $0$            & $0$ & $-\delta_{1}$ & $0$ \\
    		$A_{c2.6}$ & $\frac{1}{16}$  & $0$             & $0$             & $\frac{9}{16}$ & $0$            & $0$ & $-\delta_{1}$ & $\delta_{2}$ \\
    		$A_{c2.7}$& $\frac{5}{16}$  & $0$             & $0$             & $-\frac{3}{16}$ & $0$            & $0$ & $-\delta_{1}$ & $0$ \\
    		$A_{c2.8}$& $\frac{5}{16}$  & $0$             & $0$             & $\frac{-3}{16}$ & $0$            & $0$ & $-\delta_{1}$ & $\delta_{2}$ \\
    		\bottomrule[1pt]
    	\end{tabular}}
    \end{table*}
\end{center}

\subsection{$D^{*}\bar{B}^{*}$ systems}
   
\begin{figure*}[ht]
    \centering
    \includegraphics[width=0.9\textwidth]{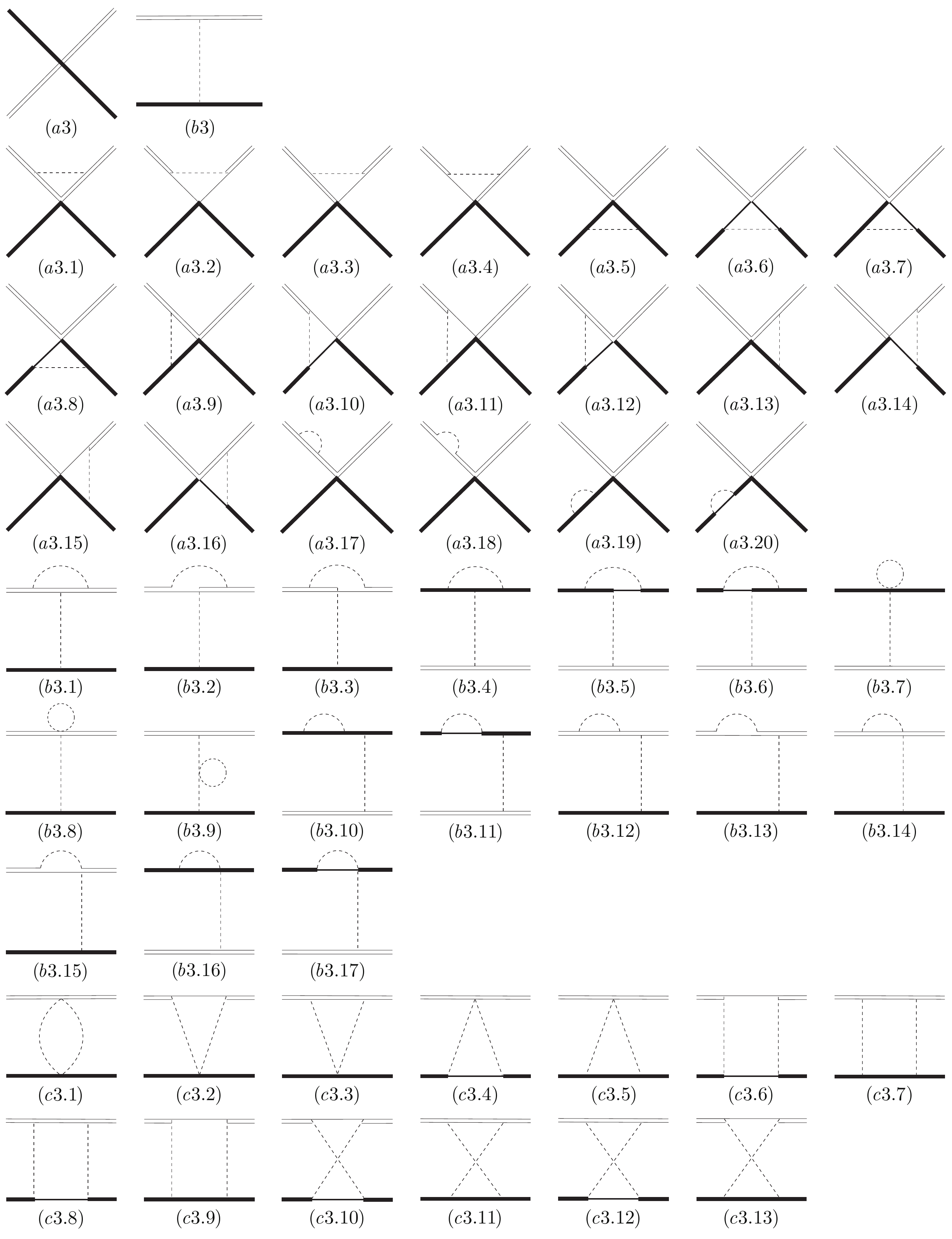}
    \captionsetup{justification=raggedright, singlelinecheck=false}
    \caption{LO contact ($a3$), LO OPE ($b3$), NLO contact ($a3.1-a3.2$0), NLO OPE ($b3.1-b3.17$), and NLO TPE ($c3.1-c3.13$) diagrams of the process $D^{*}\bar{B}^{*}$ to $D^{*}\bar{B}^{*}$. The thin solid, double-thin solid, solid, thick solid, and dashed lines stand for the $\bar{B}$, $\bar{B}^{*}$, $D$, $D^{*}$, and a pion, respectively.}\label{BsDs1}
\end{figure*}

Next, we consider the $D^{*}\bar{B}^{*}$ systems. At LO, both the contact and OPE diagrams contribute to the scattering amplitudes, as shown in Fig.~\ref{BsDs1}. At NLO, there are one-loop corrections to the LO diagrams and newly appeared TPE diagrams shown in Fig.~\ref{BsDs1}. Compared to the $D\bar{B}$ and $D\bar{B}^{*}$ systems, the $D^{*}\bar{B}^{*}$ systems include significantly more diagrams, resulting in more complex interactions. The corresponding scattering amplitudes for these diagrams are listed in Eqs.~(\ref{a31})-(\ref{b30}) and Eqs.~(\ref{a3.1})-(\ref{c3.13}).
    
At LO, utilizing the Lagrangian presented in Eqs.~(\ref{L1}) and (\ref{L2}), the contact and OPE contributions of the scattering process $D^{*}(p_1)\bar{B}^{*}(p_2) \to D^{*}(p_3)\bar{B}^{*}(p_4)$ depicted in Fig.~\ref{BsDs1} read  
\begin{align}
    \mathcal{M}_{a3}^{(0)}=&~4[(D_a+E_a)\mathcal{O}_1-(D_b+E_b)\mathcal{O}_2 \notag \\
    &+(D_b+E_b)\mathcal{O}_3],  \label{a31}
\end{align}

\begin{align}
    \mathcal{M}_{b3}^{(0)}=& -\frac{g_{1}g_{2}}{f^{2}} \frac{\mathcal{G}(q,\epsilon_{1},\epsilon_{2},\epsilon_{3}^{*},\epsilon_{4}^{*})} {q^{2}-m_{\pi}^{2}}, \label{b31} 
\end{align}
for isospin $I=1$, and we have
\begin{align}
    \mathcal{M}_{(a3)}^{(0)}=&~4[(D_a+E_a)\mathcal{O}_1-(D_b+E_b)\mathcal{O}_2 \notag \\
    &+(D_b+E_b)\mathcal{O}_3], \label{a30}
\end{align}

\begin{align}
    \mathcal{M}_{(b3)}^{(0)}=& -\frac{g_{1}g_{2}}{f^{2}}\frac{\mathcal{G}(q,\epsilon_{1},\epsilon_{2},\epsilon_{3}^{*},\epsilon_{4}^{*})}{q^{2}-m_{\pi}^{2}}, \label{b30}
\end{align}
for isospin $I=0$.

Here, $\epsilon_{1}(\epsilon_{2})$ and $\epsilon^{*}_{3}(\epsilon^{*}_{4})$ stand for the polarization vectors of the initial $D^{*}(\bar{B}^{*})$ and final $D^{*}(\bar{B}^{*})$ mesons, respectively. For convenience, we define 
\begin{align}
   \mathcal{O}_{1}&=(\epsilon_{1}\cdot\epsilon_{3}^{*})(\epsilon_{2}\cdot\epsilon_{4}^{*}), \quad \mathcal{O}_{2}=(\epsilon_{1}\cdot\epsilon_{4}^{*})(\epsilon_{2}\cdot\epsilon_{3}^{*}), \notag \\
   \mathcal{O}_{3}&=(\epsilon_{1}\cdot\epsilon_{2})(\epsilon_{3}^{*}\cdot\epsilon_{4}^{*}), \quad
   \mathcal{O}_{2}^{a}=(q\cdot\epsilon_{3}^{*})(q\cdot\epsilon_{2})(\epsilon_{1}\cdot\epsilon_{4}^{*}), \notag \\ 
   \mathcal{O}_{2}^{b}&=(q\cdot\epsilon_{1})(q\cdot\epsilon_{4}^{*})(\epsilon_{2}\cdot\epsilon_{3}^{*}), \notag \\
   \mathcal{O}_{3}^{a}&=(q\cdot\epsilon_{3}^{*})(q\cdot\epsilon_{4}^{*})(\epsilon_{1}\cdot\epsilon_{2}), \notag \\
   \mathcal{O}_{3}^{b}&=(q\cdot\epsilon_{1})(q\cdot\epsilon_{2})(\epsilon_{3}^{*}\cdot\epsilon_{4}^{*}), \label{tensor1}
\end{align}
and    
\begin{align}
   \mathcal{G}(q, \epsilon_1, \epsilon_2, \epsilon_3^*, \epsilon_4^*) = & ~\vec{q}^2 (\mathcal{O}_2-\mathcal{O}_3)+(\mathcal{O}_2^a-\mathcal{O}_3^a) \notag \\
   &+(\mathcal{O}_2^b-\mathcal{O}_3^b).
\end{align}

Similar to the definitions in (\ref{tensor1}), we also define
\begin{align}
    \mathcal{O}_{1}^{a}&=(q\cdot\epsilon_{1})(q\cdot\epsilon_{3}^{*})(\epsilon_{2}\cdot\epsilon_{4}^{*}), \notag \\
    \mathcal{O}_{1}^{b}&=(q\cdot\epsilon_{2})(q\cdot\epsilon_{4}^{*})(\epsilon_{1}\cdot\epsilon_{3}^{*}), \notag \\
    \mathcal{O}_{1}^{c}&=(q\cdot\epsilon_{1})(q\cdot\epsilon_{3}^{*})(q\cdot\epsilon_{2})(q\cdot\epsilon_{4}^{*}), \notag \\
    \mathcal{O}_{3}^{c}&=(q\cdot\epsilon_{1})(q\cdot\epsilon_{2})(q\cdot\epsilon_{3}^{*})(q\cdot\epsilon_{4}^{*}).
\end{align}    
    
The one-loop corrections to the contact amplitudes are listed as follows:
\begin{align}
    \mathcal{M}_{a_{3.1}}^{(2)}=&-4\frac{g_{2}^{2}}{f^{2}}[(d-3)(d-2)A_{1}\mathcal{O}_{1}-(d-3)A_{2}\mathcal{O}_{2}  \notag \\
    & +(d-3)A_{3}\mathcal{O}_{3}]J^{g}_{22}, \label{a3.1}
\end{align}

\begin{align}
    \mathcal{M}_{a_{3.2}}^{(2)}=&-4\frac{g_{2}^{2}}{f^{2}}A_{1}\mathcal{O}_{1}J^{g}_{22}, \label{a3.2}
\end{align}

\begin{align}
    \mathcal{M}_{a_{3.3}}^{(2)}=&-4\frac{g_{2}^{2}}{f^{2}}(d-3)[A_{1}\mathcal{O}_{2}-A_{2}\mathcal{O}_{3}]J^{g}_{22}, \label{a3.3}
\end{align}

\begin{align}
    \mathcal{M}_{a_{3.4}}^{(2)}=&-4\frac{g_{2}^{2}}{f^{2}}(d-3)[A_{1}\mathcal{O}_{2}-A_{2}\mathcal{O}_{3}]J^{g}_{22},  \label{a3.4}
\end{align}

\begin{align}
    \mathcal{M}_{a_{3.5}}^{(2)}=& -4\frac{g_{1}^{2}}{f^{2}}[(d-3)(d-2)A_{1}\mathcal{O}_{1}-(d-3)A_{2}\mathcal{O}_{2}  \notag \\
    & +(d-3)A_{3}\mathcal{O}_{3}]J^{g}_{22},  \label{a3.5} 
\end{align}

\begin{align}
    \mathcal{M}_{a_{3.6}}^{(2)}=&-4\frac{g_{1}^{2}}{f^{2}}A_{1}\mathcal{O}_{1}J^{g}_{22},  \label{a3.6}
\end{align}

\begin{align}
    \mathcal{M}_{a_{3.7}}^{(2)}=&-4\frac{g_{1}^{2}}{f^{2}}(d-3)[A_{1}\mathcal{O}_{2}-A_{2}\mathcal{O}_{3}]J^{g}_{22}, \label{a3.7}
\end{align}

\begin{align}
    \mathcal{M}_{a_{3.8}}^{(2)}=&-4\frac{g_{1}^{2}}{f^{2}}(d-3)[A_{1}\mathcal{O}_{2}-A_{2}\mathcal{O}_{3}]J^{g}_{22}, \label{a3.8} 
\end{align}

\begin{align}
    \mathcal{M}_{a_{3.9}}^{(2)}=&~4\frac{g_{1}g_{2}}{f^{2}}(d-3)[A_{1}\mathcal{O}_{1}-A_{2}\mathcal{O}_{2}+A_{3}\mathcal{O}_{3}]J^{h}_{22},  \label{a3.9}
\end{align}

\begin{align}
    \mathcal{M}_{a{3.10}}^{(2)}=&~4\frac{g_{1}^{2}}{f^{2}}A_{1}\mathcal{O}_{3}J^{h}_{22}, \label{a3.10}
\end{align}

\begin{align}
    \mathcal{M}_{a{3.11}}^{(2)}=&~4\frac{g_{1}^{2}}{f^{2}}(d-3)[A_{1}\mathcal{O}_{1}-A_{2}\mathcal{O}_{2}]J^{h}_{22}, \label{a3.11}
\end{align}

\begin{align}
    \mathcal{M}_{a{3.12}}^{(2)}=&~4\frac{g_{1}^{2}}{f^{2}}(d-3)[A_{1}\mathcal{O}_{1}-A_{2}\mathcal{O}_{2}]J^{h}_{22}, \label{a3.12}
\end{align}

\begin{align}
    \mathcal{M}_{a{3.13}}^{(2)}=&~4\frac{g_{1}g_{2}}{f^{2}}(d-3)[A_{1}\mathcal{O}_{1}-A_{2}\mathcal{O}_{2}+A_{3}\mathcal{O}_{3}]J^{h}_{22}, \label{a3.13}
\end{align}

\begin{align}
    \mathcal{M}_{a_{3.14}}^{(2)}=&~4\frac{g_{1}g_{2}}{f^{2}}A_{1} \mathcal{O}_{3}J^{h}_{22},  \label{a3.14}
\end{align}

\begin{align}
    \mathcal{M}_{a_{3.15}}^{(2)}=&~4\frac{g_{1}g_{2}}{f^{2}}(d-3)[A_{1}\mathcal{O}_{1}-A_{2}\mathcal{O}_{2}]J^{h}_{22}, \label{a3.15} 
\end{align}

\begin{align}
    \mathcal{M}_{a_{3.16}}^{(2)}=&~4\frac{g_{1}g_{2}}{f^{2}}(d-3)[A_{1}\mathcal{O}_{1}-A_{2}\mathcal{O}_{2}]J^{h}_{22}, \label{a3.16}
\end{align}

\begin{align}
    \mathcal{M}_{a_{3.(17+18)}}^{(2)}=&-\frac{3g_{2}^{2}}{2f^{2}}A_{1}[(d-2)\partial_{\omega}J^{b}_{22}(\omega_{1}) +\partial_{\omega}J^{b}_{22}(\omega_{2})], \label{a3.17}
\end{align}

\begin{align}
    \mathcal{M}_{a_{3.(19+20)}}^{(2)}=&-\frac{3g_{1}^{2}}{2f^{2}}A_{1}[(d-2)\partial_{\omega}J^{b}_{22}(\omega_{1})+\partial_{\omega}J^{b}_{22}(\omega_{2})],  \label{a3.19}
\end{align}    
where coefficients appearing in the above contact amplitudes are shown in Table~\ref{5}.

Next, we consider the NLO one-pion-exchange interactions in the $D^{*}\bar{B}^{*}$ systems. The corresponding diagrams are illustrated in Fig.~\ref{BsDs1}, and the amplitudes,
\begin{align}
    \mathcal{M}_{b{3.1}}^{(2)}=&~4\frac{g_{1}^{3}g_{2}}{f^{2}}A_{1}J^{g}_{22}\frac{\mathcal{G}(q,\epsilon_{1},\epsilon_{2},\epsilon_{3}^{*},\epsilon_{4}^{*})}{q^{2}-m_{\pi}^{2}}, \label{b3.1}\end{align}

\begin{align}
    \mathcal{M}_{b{3.2}}^{(2)}=&-4\frac{g_{1}^{3}g_{2}}{f^{2}}A_{1}J^{g}_{22}\frac{\mathcal{G}(q,\epsilon_{1},\epsilon_{2},\epsilon_{3}^{*},\epsilon_{4}^{*})}{q^{2}-m_{\pi}^{2}}, \label{b3.2} 
\end{align}

\begin{align}
    \mathcal{M}_{b{3.3}}^{(2)}=&-4\frac{g_{1}^{3}g_{2}}{f^{2}}A_{1}J^{g}_{22}\frac{\mathcal{G}(q,\epsilon_{1},\epsilon_{2},\epsilon_{3}^{*},\epsilon_{4}^{*})}{q^{2}-m_{\pi}^{2}}, \label{b3.3}
\end{align}

\begin{align}
    \mathcal{M}_{b{3.4}}^{(2)}=&~4\frac{g_{1}g_{2}^{3}}{f^{2}}A_{1}J^{g}_{22}\frac{\mathcal{G}(q,\epsilon_{1},\epsilon_{2},\epsilon_{3}^{*},\epsilon_{4}^{*})}{q^{2}-m_{\pi}^{2}}, \label{b3.4}\end{align}

\begin{align}
    \mathcal{M}_{b{3.5}}^{(2)}=&~4\frac{g_{1}g_{2}^{3}}{f^{2}}A_{1}J^{g}_{22}\frac{\mathcal{G}(q,\epsilon_{1},\epsilon_{2},\epsilon_{3}^{*},\epsilon_{4}^{*})}{q^{2}-m_{\pi}^{2}}, \label{b3.5}
\end{align}

\begin{align}
    \mathcal{M}_{b{3.6}}^{(2)}=&~4\frac{g_{1}g_{2}^{3}}{f^{2}}A_{1}J^{g}_{22}\frac{\mathcal{G}(q,\epsilon_{1},\epsilon_{2},\epsilon_{3}^{*},\epsilon_{4}^{*})}{q^{2}-m_{\pi}^{2}}, \label{b3.6} 
\end{align}

\begin{align}
    \mathcal{M}_{b{3.7}}^{(2)}=&-\frac{8g_{1}g_{2}}{3f^{2}}A_{1}J^{c}_{0}\frac{\mathcal{G}(q,\epsilon_{1},\epsilon_{2},\epsilon_{3}^{*},\epsilon_{4}^{*})}{q^{2}-m_{\pi}^{2}}, \label{b3.7}
\end{align}

\begin{align}
    \mathcal{M}_{b{3.8}}^{(2)}=&-\frac{8g_{1}g_{2}}{3f^{2}}A_{1}J^{c}_{0}\frac{\mathcal{G}(q,\epsilon_{1},\epsilon_{2},\epsilon_{3}^{*},\epsilon_{4}^{*})}{q^{2}-m_{\pi}^{2}}, \label{b3.8} 
\end{align}

\begin{align}
    \mathcal{M}_{b{3.9}}^{(2)}=&-\frac{8g_{1}g_{2}}{3f^{2}}A_{1}[2m^{2}L+\frac{m^{2}}{16\pi^{2}}{\rm{log}} (\frac{m^{2}}{\mu^{2}})] \notag \\
    &\times \frac{\mathcal{G}(q,\epsilon_{1}, \epsilon_{2}, \epsilon_{3}^{*}, \epsilon_{4}^{*})}{q^{2}-m_{\pi}^{2}}, \label{b3.9}\end{align}

\begin{align}
    \mathcal{M}_{b{3.(10+11)}}^{(2)}=&~\frac{3g_{1}g_{2}^{3}}{2f^{2}}A_{1}[(d-2)\partial_{\omega}J^{b}_{22}(\omega_{1})+\partial_{\omega}J^{b}_{22}(\omega_{2})]  \notag \\ 
    & \times \frac{\mathcal{G}(q,\epsilon_{1},\epsilon_{2},\epsilon_{3}^{*}, \epsilon_{4}^{*})}{q^{2}-m_{\pi}^{2}},  \label{b3.10}
\end{align}

\begin{align}
    \mathcal{M}_{b{3.(12+13)}}^{(2)}=&~\frac{3g_{1}g_{2}^{3}}{2f^{2}}A_{1}[(d-2)\partial_{\omega}J^{b}_{22}(\omega_{1})+\partial_{\omega}J^{b}_{22}(\omega_{2})] \notag \\
    &\times \frac{\mathcal{G}(q,\epsilon_{1},\epsilon_{2},\epsilon_{3}^{*},\epsilon_{4}^{*})}{q^{2}-m_{\pi}^{2}}, \label{b3.12} 
\end{align}

\begin{align}
    \mathcal{M}_{b3.14}^{(2)}=&~\mathcal{M}_{b3.15}^{(2)} =\mathcal{M}_{b3.16}^{(2)}=\mathcal{M}_{b3.17}^{(2)}=0. \label{b3.14}
\end{align}
    
The amplitudes for the TPE diagrams shown in Fig.~\ref{BsDs1} are calculated to be

\begin{align}
	\mathcal{M}_{c{3.1}}^{(2)}=& -4\frac{1}{f^{4}}[A_{1}(q^{2}_{0} J^{F}_{21} + J^{F}_{22})\notag \\
    &-(A_{15}+A_{51}+A_{5})(q^{2}_{0} J^{F}_{11} +q^{2}_{0} J^{F}_{21} + J^{F}_{22})\notag \\
    &+A_{5}(q^{2}_{0} J^{F}_{0} + q^{2}_{0} J^{F}_{11})]\mathcal{O}_{1}, \label{c3.1}
\end{align}

\begin{align}
	\mathcal{M}_{c{3.2}}^{(2)}=& -i4\frac{g_{2}^{2}}{f^{4}}\{[A_{1}(q_{0} J^{S}_{22} + J^{S}_{24}+ q_{0} J^{S}_{32} + J^{S}_{33})\notag \\
    &-A_{5}(q_{0} J^{S}_{11}+2q_{0} J^{S}_{22} + J^{S}_{24} + q_{0} J^{S}_{32} + J^{S}_{33})]\mathcal{O}^{a}_{1}\notag \\
    &+[A_{1}(q_{0} J^{S}_{31} + J^{S}_{34}) \notag \\
	& -A_{5}(q_{0} J^{S}_{21}+ q_{0} J^{S}_{31} + J^{S}_{34} )]\mathcal{O}_{1}\}, \label{c3.2}
\end{align}

\begin{align}
	\mathcal{M}^{(2)}_{c3.3}=&~ i4\frac{g^{2}_{2}}{f^{4}}(d-3)[A_{1}(q_{0}J^{S}_{22}+ J^{S}_{24}+q_{0}J^{S}_{32}+ J^{S}_{33})\notag \\
    & -{A}_{5}(q_{0} J^{S}_{11}+2q_{0} J^{S}_{22}+ J^{S}_{24}+q_{0} J^{S}_{32}+ J^{F}_{33})]\mathcal{O}^{a}_{1} \notag \\
    & +i4\frac{g^{2}_{2}}{f^{4}}(d-3)\{A_{1}[({q}_{0}\vec {q}^{2} J^{S}_{22}+\vec q^{2} J^{S}_{24}+(2-d)q_{0} J^{S}_{31} \notag \\
    & +q_{0}\vec q^{2} J^{S}_{32}+\vec {q}^{2} J^{S}_{33}+(2-d)q_{0} J^{S}_{34})] \notag \\
    & -{A}_{5}[q_{0}\vec q^{2} J^{S}_{11}+(2-d)q_{0} J^{S}_{21}+\vec q^{2} J^{S}_{24}+(2-d)q_{0} J^{S}_{31} \notag \\
    & +q_{0}\vec q^{2} J^{S}_{32}+\vec q^{2} J^{S}_{33}+(2-d)q_{0} J^{S}_{34}]\}\mathcal{O}_{1}, \label{c3.3}
\end{align}

\begin{align}
	\mathcal{M}_{c3.4}^{(2)}=&~ i4\frac{g_{1}^{2}}{f^{4}}\{[A_{1}(q_{0} J^{T}_{22} + J^{T}_{24}+ q_{0} J^{T}_{32} + J^{T}_{33}) \notag \\
    & -A_{5}(q_{0} J^{T}_{11}+2q_{0} J^{T}_{22} + J^{T}_{24} + q_{0} J^{T}_{32} + J^{T}_{33})]\mathcal{O}^{b}_{1}  \notag \\
    & +[A_{1}(q_{0} J^{T}_{31} + J^{T}_{34})\notag \\
    &-A_{5}(q_{0} J^{T}_{21}+ q_{0} J^{T}_{31} + J^{T}_{34} )]\mathcal{O}_{1}\}, \label{c3.4}
\end{align}

\begin{align}
	\mathcal{M}^{(2)}_{c3.5}=& -i4\frac{g_{1}^{2}}{f^{4}}(d-3)[A_{1}(q_{0} J^{T}_{22}+ J^{T}_{24}+q_{0} J^{T}_{32}+ J^{T}_{33}) \notag \\
    & -A_{5}(q_{0} J^{T}_{11}+2q_{0} J^{T}_{22}+J^{T}_{24}+{q}_{0} J^{T}_{32}+ J^{T}_{33})] \mathcal{O}^{a}_{1} \notag \\
	& -i4\frac{g^{2}_{1}}{f^{4}}(d-3)\{A_{1}[({q}_{0}\vec {q}^{2}J^{T}_{22}+\vec q^{2} J^{T}_{24} \notag \\
    & +(2-d)q_{0} J^{T}_{31}+q_{0}\vec q^{2} J^{T}_{32}+\vec {q}^{2} J^{T}_{33}+(2-d)q_{0} J^{T}_{34})] \notag \\
	& -A_{5}[q_{0}\vec q^{2} J^{T}_{11}+(2-d)q_{0} J^{T}_{21}+\vec q^{2} J^{T}_{24} \notag \\
    & +(2-d)q_{0} J^{T}_{31}+q_{0}\vec q^{2} J^{T}_{32}+\vec q^{2} J^{T}_{33} \notag \\
    & +(2-d)q_{0} J^{T}_{34}]\}\mathcal{O}_{1}, \label{c3.5}
\end{align}

\begin{align}
    \mathcal{M}^{(2)}_{c3.6}=& -4\frac{g_{1}^{2}g_{2}^{2}}{f^{4}}A_{1}[(\mathcal{O}_{1}+\mathcal{O}_{2}+\mathcal{O}_{3}) J^{B}_{41} \notag \\
    & +\mathcal{O}^{b}_{3}(J^{B}_{21} + 2J^{B}_{31} + J^{B}_{42}) \notag \\
    & +(\mathcal{O}^{a}_{1} + \mathcal{O}^{b}_{1}+\mathcal{O}^{a}_{2}+\mathcal{O}^{b}_{2})( J^{B}_{31} + J^{B}_{42})+\mathcal{O}^{a}_{3}J^{B}_{42} \notag \\
    & +\mathcal{O}^{c}_{3}(J^{B}_{22} + 2J^{B}_{32} + J^{B}_{43})], \label{c3.6}
\end{align}

\begin{align}
    \mathcal{M}^{(2)}_{c3.7}=& -4\frac{g_{1}^{2}g_{2}^{2}}{f^{4}}(d-3)^{2}A_{1}\{[(d-2)(d-1)\mathcal{O}_{1} \notag \\
    & +\mathcal{O}_{2}+\mathcal{O}_{3}] J^{B}_{41}+({\vec q}^{4}\mathcal{O}_{1}+{\vec q}^{2}\mathcal{O}^{a}_{1} \notag \\
    & +{\vec q}^{2}\mathcal{O}^{b}_{1}+\mathcal{O}^{c}_{1})(J^{B}_{22} + 2J^{B}_{32} + J^{B}_{43}) \notag \\
     & -({\vec q}^{2} \mathcal{O}_{1} +\mathcal{O}^{a}_{1} + \mathcal{O}^{b}_{1} -\mathcal{O}^{a}_{3}) J^{B}_{21} \notag \\
    &  -[2d{\vec q}^{2}\mathcal{O}_{1}+(d+2)\mathcal{O}^{a}_{1} +(d+2)\mathcal{O}^{b}_{1}-\mathcal{O}^{a}_{2} \notag \\
    & -\mathcal{O}^{b}_{2}-\mathcal{O}^{b}_{3}]( J^{B}_{31} + J^{B}_{42})  +\mathcal{O}^{b}_{3} J^{B}_{31} +\mathcal{O}^{a}_{3} J^{B}_{42}\}, \label{c3.7}
\end{align}

\begin{align}
     \mathcal{M}^{(2)}_{c3.8}=&~ 4\frac{g_{1}^{2}g_{2}^{2}}{f^{4}}(d-3)A_{1} \{(-d\mathcal{O}_{1} + \mathcal{O}_{2} + \mathcal{O}_{3}) J^{B}_{41} \notag \\
    & +({\vec q}^{2} \mathcal{O}^{b}_{1} + \mathcal{O}^{c}_{3}) (J^{B}_{22} + 2J^{B}_{32} + J^{B}_{43}) \notag \\
    & + (\mathcal{O}^{b}_{2} - \mathcal{O}^{b}_{1}) J^{B}_{21} +[{\vec q}^{2}\mathcal{O}_{1}+\mathcal{O}^{a}_{1}-(d+2)\mathcal{O}^{b}_{1}\notag \\
     & + \mathcal{O}^{b}_{2} + \mathcal{O}^{a}_{3} +\mathcal{O}^{b}_{3}]( J^{B}_{31} + J^{B}_{42}) +\mathcal{O}^{b}_{2} J^{B}_{31} + \mathcal{O}^{a}_{2} J^{B}_{42}\}, \label{c3.8}
\end{align}

\begin{align}
    \mathcal{M}^{(2)}_{c3.9}=&~ 4\frac{g_{1}^{2}g_{2}^{2}}{f^{4}}(d-3)A_{1} \{(-d\mathcal{O}_{1}+ \mathcal{O}_{2} + \mathcal{O}_{3}) J^{B}_{41}\notag \\
     &  +({\vec q}^{2} \mathcal{O}^{a}_{1} + \mathcal{O}^{c}_{3}) (J^{B}_{22} + 2J^{B}_{32} + J^{B}_{43})\notag \\
     &  + (\mathcal{O}^{a}_{2}-\mathcal{O}^{a}_{1}) J^{B}_{21} +[{\vec q}^{2} \mathcal{O}_{1}-(d+2) \mathcal{O}^{a}_{1} \notag \\
    &+\mathcal{O}^{b}_{1} +\mathcal{O}^{a}_{2} +\mathcal{O}^{a}_{3} + \mathcal{O}^{b}_{3}](J^{B}_{31} + J^{B}_{42})\notag \\
    & +\mathcal{O}^{a}_{2} J^{B}_{31} + \mathcal{O}^{b}_{2} J^{B}_{42}\}, \label{c3.9}
\end{align}

\begin{align}
    \mathcal{M}^{(2)}_{c3.10}=&~ 5\mathcal{M}^{(2)}_{c3.6}/.\{J^{B} \to J^{R}\}, \notag \\
    \mathcal{M}^{(2)}_{c3.11}=&~ 5\mathcal{M}^{(2)}_{c3.7}/.\{J^{B} \to J^{R}\}, \label{c3.10}
\end{align}

\begin{align}
    \mathcal{M}^{(2)}_{c2.12}=&~ 4\frac{g_{1}^{2}g_{2}^{2}}{f^{4}}(d-3)A_{1} \{(-d\mathcal{O}_{1} + \mathcal{O}_{2} + \mathcal{O}_{3}) J^{R}_{41} \notag \\
    & + ({\vec q}^{2} \mathcal{O}^{b}_{1} + \mathcal{O}^{c}_{3}) (J^{R}_{22} + 2J^{R}_{32} + J^{R}_{43})\notag \\
    & + ( \mathcal{O}^{b}_{2} - \mathcal{O}^{b}_{1}) J^{R}_{21}+[{\vec q}^{2}\mathcal{O}_{1}+\mathcal{O}^{a}_{1}-(d+2)\mathcal{O}^{b}_{1} \notag \\
    & +\mathcal{O}^{a}_{2} + \mathcal{O}^{b}_{2} +\mathcal{O}^{b}_{3}]( J^{R}_{31} + J^{R}_{42}) \notag \\
     &  +\mathcal{O}^{b}_{3} J^{R}_{31} + \mathcal{O}^{a}_{3} J^{R}_{42}\}, \label{c3.12}
\end{align}

\begin{align}
    \mathcal{M}^{(2)}_{c3.13}=&~ 4\frac{g_{1}^{2}g_{2}^{2}}{f^{4}}(d-3)A_{1} \{(-d\mathcal{O}_{1} + \mathcal{O}_{2} + \mathcal{O}_{3}) J^{R}_{41} \notag \\
     &  +({\vec q}^{2} \mathcal{O}^{b}_{1} + \mathcal{O}^{c}_{3}) (J^{R}_{22} + 2J^{R}_{32} + J^{R}_{43})  \notag \\
     & + (\mathcal{O}^{b}_{2} - \mathcal{O}^{b}_{1}) J^{R}_{21} +[{\vec q}^{2}\mathcal{O}_{1}+\mathcal{O}^{a}_{1}-(d+2)\mathcal{O}^{b}_{1}\notag \\
     & + \mathcal{O}^{a}_{2} + \mathcal{O}^{b}_{2} +\mathcal{O}^{a}_{3}]( J^{R}_{31} + J^{R}_{42})\notag \\
     &  +\mathcal{O}^{a}_{3} J^{R}_{31} + \mathcal{O}^{b}_{3} J^{R}_{42}\}. \label{c3.13}
\end{align}

The coefficients in the above NLO OPE and TPE amplitudes are listed in Tables \ref{6} and \ref{7}, respectively. To further evaluate the calculated amplitudes, we should deal with the terms in (\ref{tensor1}) according to the partial wave we considered. In the $S$-wave, for the terms like $(q\cdot\epsilon_{i})(q\cdot\epsilon_{j}^{*})(\epsilon_{k}\cdot\epsilon_{l}^{*})$, we can make the following substitutions:  
\begin{align*}
     (\epsilon_{i}\cdot\epsilon_{j})(q\cdot\epsilon_{k})(q\cdot\epsilon_{l})\mapsto-\frac{1}{d-1}\vec{q}^{2}(\epsilon_{i}\cdot\epsilon_{j})(\epsilon_{k}\cdot\epsilon_{l}), \notag \\
    (q\cdot\epsilon_{i})(q\cdot\epsilon_{j})(q\cdot\epsilon_{k})(q\cdot\epsilon_{l})\mapsto\frac{1}{(d-1)^{2}}\vec{q}^{4}(\epsilon_{i}\cdot\epsilon_{j})(\epsilon_{k}\cdot\epsilon_{l}),
\end{align*}
together with Table \ref{polar}.
\renewcommand{\arraystretch}{1.5}
\begin{center}
  \begin{table}[htbp]
  	\centering
        \captionsetup{justification=raggedright, singlelinecheck=false}
  	\caption{The values of the products of polarization vectors in the $S$-wave effective potentials with total angular momentum $J=0$, $1$, and $2$, respectively \cite{Wang:2018atz}.}\label{polar}
  	\setlength{\tabcolsep}{5.62mm}{
  	\begin{tabular}{ccccc}
  	  \toprule[1pt]
  		Terms & { $J=0$  } & { $J=1$ } &  { $J=2$ }   \\
  		\midrule[1pt]
  		$\mathcal{O}_{1}$ & $1$  & $1$   & $1$  \\
  		$\mathcal{O}_{2}$ & $1$  & $-1$  & $1$  \\
  		$\mathcal{O}_{3}$ & $3$  & $0$   & $0$  \\
  		\bottomrule[1pt]
  	\end{tabular}}
  \end{table}
\end{center}

\renewcommand{\arraystretch}{1.6}
\begin{center}
   	\begin{table*}[ht]
   	\centering
        \captionsetup{justification=raggedright, singlelinecheck=false}
   	\caption{The coefficients appearing in the contact term amplitudes at NLO [Eqs.~(\ref{a3.1})-(\ref{a3.19})].}\label{5}
   	\setlength{\tabcolsep}{0.8mm}{
   	\begin{tabular}{ccccccccc}
   		\toprule[1pt]
   		& \multicolumn{3}{c}{ $I=1$  } &\multicolumn{3}{c}{ $I=0$ } &        & \\
   		\cline{2-4} \cline{5-7}
   		&  $A_1$   &    $A_2$  &  $A_3$   &    $A_1$   &    $A_2$   &  $A_3$ &   $\omega_1$   & $\omega_2$ \\
   		\midrule[1pt]
   		$A_{a3.1}$ & $\frac{1}{4}(3D_{a}-E_{a})$ & $\frac{1}{4}(3D_{b}-E_{b})$ & $\frac{1}{4}(3D_{b}-E_{b}$) & $\frac{3}{4}(D_{a}+E_{a})$ & $\frac{3}{4}(D_{b}+E_{b})$ & $\frac{3}{4}(D_{b}+E_{b})$ & $0$ & $0$  \\
   		$A_{a3.2}$ & $\frac{1}{4}(3D_{a}-E_{a})$ & $0$ & $0$ & $\frac{3}{4}(D_{a}+E_{a})$ & $0$ & $0$ & $\delta_{2}$ & $\delta_{2}$ \\
   		$A_{a3.3(4)}$ & $\frac{1}{4}(3D_{b}-E_{b})$ & $\frac{1}{4}(3D_{b}-E_{b})$ & $0$ & $\frac{3}{4}(D_{b}+E_{b})$ & $\frac{3}{4}(D_{b}+E_{b})$ & $0$ & $0(\delta_{2})$ & $\delta_{2}(0)$ \\
   		$A_{a3.5}$ & $0$ & $0$ & $0$ & $0$ & $0$ & $0$ &  $0$ & $0$ \\
   		$A_{a3.6}$ & $0$ & $0$ & $0$ & $0$ & $0$ & $0$ & $\delta_{1}$ & $\delta_{1}$ \\
   		$A_{a3.7(8)}$ & $0$ & $0$ & $0$ & $0$ & $0$ & $0$ & $0(\delta_{1})$ & $\delta_{1}(0)$ \\
   		$A_{a3.9}$ & $\frac{1}{4}(D_{b}+E_{b})$ & $\frac{1}{4}(D_{a}+E_{a})$ & \makecell[c]{$\frac{1}{4}[(d-3)(D_{b}+E_{b})$\\$+(D_{a}+E_{a})]$}  & $\frac{-3}{4}(D_{b}-3E_{b})$ & $\frac{-3}{4}(D_{a}-3E_{a})$ & \makecell[c]{$\frac{-3}{4}[(d-3)(D_{b}-3E_{b})$\\$+(D_{a}-3E_{a})]$} & $0$ & $0$ \\
   		$A_{a3.10}$ & $\frac{1}{4}(D_{b}+E_{b})$ & $0$ & $0$ & $\frac{-3}{4}(D_{b}-3E_{b})$ & $0$ & $0$ & $\delta_{1}$ & $\delta_{2}$ \\
   		$A_{a3.11(12)}$ & $\frac{1}{4}(D_{b}+E_{b})$ & $\frac{1}{4}(D_{b}+E_{b})$ & $0$ & $\frac{-3}{4}(D_{b}-3E_{b})$ & $\frac{-3}{4}(D_{b}-3E_{b})$ & $0$ &  $0(\delta_{1})$ & $\delta_{2}(0)$ \\
   		$A_{a3.13}$ & $\frac{1}{4}(D_{b}+E_{b})$ & $\frac{1}{4}(D_{a}+E_{a})$ & \makecell[c]{$\frac{1}{4}[(d-3)(D_{b}+E_{b})$\\$+(D_{a}+E_{a})]$}  & $\frac{-3}{4}(D_{b}-3E_{b})$ & $\frac{-3}{4}(D_{a}-3E_{a})$ & \makecell[c]{$\frac{-3}{4}[(d-3)(D_{b}-3E_{b})$\\$+(D_{a}-3E_{a})]$} & $0$ & $0$ \\
   		$A_{a3.14}$ & $\frac{1}{4}(D_{b}+E_{b})$ & $0$ & $0$ & $\frac{-3}{4}(D_{b}-3E_{b})$ & $0$ & $0$ & $\delta_{1}$ & $\delta_{2}$ \\
   		$A_{a3.15(16)}$ & $\frac{1}{4}(D_{b}+E_{b})$ & $\frac{1}{4}(D_{b}+E_{b})$ & $0$ & $\frac{-3}{4}(D_{b}-3E_{b})$ & $\frac{-3}{4}(D_{b}-3E_{b})$ & $0$ & $0(\delta_{1})$ & $\delta_{2}(0)$ \\
   		$A_{a3.(17+18)}$ & $D_{a}+E_{a}$ & $D_{b}+E_{b}$ & $D_{b}+E_{b}$ & $D_{a}-3E_{a}$ & $D_{b}-3E_{b}$ & $D_{b}-3E_{b}$ & $0$ & $\delta_{2}$ \\
   		$A_{a3.(19+20)}$ & $D_{a}+E_{a}$ & $D_{b}+E_{b}$ & $D_{b}+E_{b}$ & $D_{a}-3E_{a}$ & $D_{b}-3E_{b}$ & $D_{b}-3E_{b}$ & $0$ & $\delta_{1}$ \\
   		\bottomrule[1pt]
   	\end{tabular}}
   \end{table*}
\end{center}  

\begin{center} 
   \renewcommand{\arraystretch}{1.5}
   \begin{table*}[htbp]
   	\centering
        \captionsetup{justification=raggedright, singlelinecheck=false}
   	\caption{The constants $A$ (as well as $\omega_{1,2}$) appearing in the OPE amplitudes [Eqs.~(\ref{b3.1})-(\ref{b3.14})].}\label{6}
   	\setlength{\tabcolsep}{2.4mm}{
   	\begin{tabular}{ccccccccccccc}	
   	\toprule[1pt]
   	 & $A_{b3.1}$ & $A_{b3.2(3)}$ & $A_{b3.4}$ &$A_{b3.5(6)}$ & $A_{b3.7}$ & $A_{b3.8}$ & $A_{b3.9}$ & $A_{b3.(10+11)}$ & $A_{b3.(12+13)}$ & $A_{b3.14(15)}$ & $A_{b3.16(17)}$  \\	
   	\midrule[1pt]
   	 $I=1$ & $-\frac{1}{16}$  & $-\frac{1}{16}$  & $-\frac{1}{16}$  & $-\frac{1}{16}$  & $-\frac{1}{12}$ & $-\frac{1}{12}$ & 
         $\frac{1}{4}$ & $\frac{1}{4}$ & $\frac{1}{4}$ & $0$ & $0$  \\
   	 $I=0$ & $\frac{3}{16}$ & $\frac{3}{16}$ & $\frac{3}{16}$ & 
         $\frac{3}{16}$ & $\frac{1}{4}$ & $\frac{1}{4}$ & $-\frac{3}{4}$ & $-\frac{3}{4}$ & $-\frac{3}{4}$ & $0$ & $0$ \\
   	 $\omega_1$ & $0$ & $\delta_{2}(0)$ & $0$ & $0(\delta_{1})$ & 
         $0$ & $0$ & $0$ & $0$ & $0$ & $0$ & $0$   \\
   	 $\omega_2$ & $0$ & $0(\delta_{2})$ & $0$ & $\delta_{1}(0)$ & 
         $0$ & $0$ & $0$ & $\delta_{1}$ & $\delta_{2}$ & $0$ & $0$   \\
   	\bottomrule[1pt]
   	\end{tabular}}
   	\end{table*}    
\end{center} 
   
\renewcommand{\arraystretch}{1.4}
\begin{table*}[hbtp]
	\centering
	\captionsetup{justification=raggedright, singlelinecheck=false}
	\caption{The constants appearing in the TPE amplitudes [Eqs.~(\ref{c3.1})-(\ref{c3.13})]. Note that we have $A_{51}=A_{15}$.}\label{7}
	\setlength{\tabcolsep}{7.4mm}{
	\begin{tabular}{ccrcrrccc}	
		\toprule[1pt]
		&  \multicolumn{3}{c}{ $I=1$ } &\multicolumn{3}{c}{ $I=0$ } &       & \\
		\cline{2-4} \cline{5-7}
	  &  $A_1$   &  $A_5$   &  $A_{15}$   &   $A_1$   &  $A_5$   &  $A_{15}$   &  $\omega_1$   & $\omega_2$ \\
		\midrule[1pt]
    $A_{c3.1}$ & $\frac{1}{16} $  & $\frac{1}{16} $ & $-\frac{1}{16}$ & $-\frac{3}{16}$ & $-\frac{3}{16}$  & $\frac{3}{16}$ & $0$       & $0$ \\
    $A_{c3.2}$ & $\frac{i}{8}$  & $-\frac{i}{8}$ & $0$              & $-\frac{3i}{8}$ & $\frac{3i}{8}$  & $0$ & $\delta_2$ &     $0$ \\
    $A_{c3.3}$ & $\frac{i}{8}$  & $-\frac{i}{8}$ & $0$              & $-\frac{3i}{8}$ & $\frac{3i}{8}$  & $0$ & $0$    & $0$ \\
    $A_{c3.4}$ & $\frac{i}{8}$  & $-\frac{i}{8}$ & $0$         & $-\frac{3i}{8}$ & $\frac{3i}{8}$  & $0$ & $\delta_1$ & $0$ \\
    $A_{c3.5}$ & $\frac{i}{8}$  & $-\frac{i}{8}$ & $0$              & $-\frac{3i}{8}$ & $\frac{3i}{8}$  & $0$ & $0$  &  $0$ \\
    $A_{c3.6}$ & $\frac{1}{16}$   & $0$        & $0$            & $\frac{9}{16}$   & $0$    & $0$ &  $\delta_1$      & $\delta_2$ \\
    $A_{c3.7}$ & $\frac{1}{16}$  & $0$             & $0$             & $\frac{9}{16}$ & $0$            & $0$ & $0$ & $0$ \\
    $A_{c3.8}$ & $\frac{1}{16}$   & $0$             & $0$            & $\frac{9}{16}$ & $0$            & $0$ & $\delta_{1}$ & $0$   \\
    $A_{c3.9}$ & $\frac{1}{16}$  & $0$             & $0$             & $\frac{9}{16}$ & $0$            & $0$ & $0$          & $\delta_{2}$ \\
    $A_{c3.10}$ & $\frac{5}{16}$  & $0$             & $0$             & $-\frac{3}{16}$ & $0$            & $0$ & $\delta_1$          & $\delta_{2}$ \\
    $A_{c3.11}$& $\frac{5}{16}$  & $0$             & $0$             & $-\frac{3}{16}$ & $0$            & $0$ & $0$          & $0$ \\
    $A_{c3.12}$& $\frac{5}{16}$  & $0$             & $0$             & $-\frac{3}{16}$ & $0$            & $0$ & $\delta_{1}$ & $0$ \\
    $A_{c3.13}$ & $\frac{5}{16}$  & $0$             & $0$             & $-\frac{3}{16}$ & $0$            & $0$ & $0$          & $\delta_2$ \\
	    \bottomrule[1pt]
         \end{tabular}
    }
\end{table*}

\section{Numerical results and discussions}
\label{Sec: Results}	
   
\subsection{Potentials in coordinate space and possible bound states}
   
After obtaining the scattering amplitudes, we now evaluate the effective potentials and analyze their behavior in coordinate space. To obtain the numerical results, we use the following LECs, determined by the resonance saturation model~\cite{Xu:2017tsr, Wang:2018atz, Xu:2021vsi, Du:2016tgp, Ecker:1988te, Ecker:1989yg, Donoghue:1988ed}:
\begin{align}
   	D_{a}=-13.23~\rm GeV^{-2}, \it{E}_{a}=\rm{-11.49~ GeV^{-2}}.
\end{align}   
Other parameters include: $m_{\pi} = 0.139~\rm{GeV}$, the pion decay constant $f_{\pi} = 0.086~\rm{GeV}$, the renormalization scale $\mu = 4\pi f_{\pi}$, $\delta_{1} = 0.142~\rm{GeV}$, $\delta_{2} = 0.045~\rm{GeV}$, and the coupling constants $g_{1} = 0.65$ and $g_{2} = 0.52$ \cite{Xu:2017tsr, Wang:2018atz}.

Then we will substitute the potentials into the Schr$\ddot{\rm o}$dinger equation and search for the bound states. In this work, we regularize the effective potentials using the Gauss regulator $\mathcal{F}(\mathbf{q}) = \exp(-\mathbf{q}^{2n}/ \Lambda^{2n})$ to prevent the divergence at high-momentum transfer. Usually, the value of the cutoff parameter $\Lambda$ in chiral effective field theory is around $0.5~\rm{GeV}$ \cite{Machleidt:2011zz, Meng:2019ilv, Epelbaum:2014efa}.

\subsubsection{$D\bar{B}$ systems}
   
\begin{figure*}[ht]
    \centering
    \includegraphics[width=1.0\textwidth]{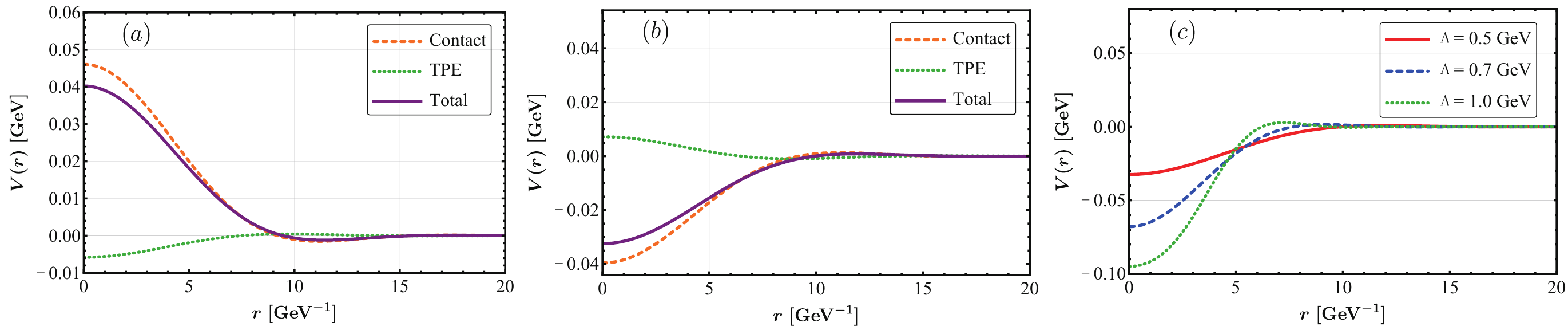}
    \captionsetup{justification=raggedright, singlelinecheck=false}
    \caption{$S$-wave potential of the $D\bar{B}$ systems and the dependence of the total potential on the cutoff parameter, namely, (a) the potential of the $I(J^{P})=1(0^+)$ $D\bar{B}$ system at $\Lambda=0.5~\rm{GeV}$, (b) the potential of the $I(J^{P})=0(0^+)$ $D\bar{B}$ system at $\Lambda=0.5~\rm{GeV}$, and (c) the total potential of the $D\bar{B}[1(0^+)]$ system with different cutoffs. The green dashed, red dashed, and purple solid lines represent the contact, TPE, and full potential, respectively.}\label{BD}
\end{figure*}

Firstly, we analyze the behaviors of the effective potentials of the $D\bar{B}$ systems, which receive contributions from the contact and the TPE terms. There are two channels: $I(J^{P}) = 1(0^{+})$ and $I(J^{P}) = 0(0^{+})$, and their effective potentials in coordinate space are shown in Figs.~\ref{BD}(a) and \ref{BD}(b), respectively. Here, we adopt the cutoff parameter $\Lambda$ with $0.5~\rm{GeV}$. 

For the $I(J^{P}) = 1(0^{+})$ state in Fig.~\ref{BD}(a), we observe that although attractive, the TPE interaction is much weaker than the repulsive contact interaction, resulting in an overall repulsive potential. Accordingly, no bound states are found in this system.
 
For the $I(J^{P}) = 0(0^{+})$ system, the effective potentials are shown in Fig.~\ref{BD}(b). Compared to the $I(J^{P}) = 1(0^{+})$ system, we can find that the TPE potential is weakly repulsive, while the contact potential is attractive. However, the summed attraction is too weak, so no bound state is found in this system.

It is worth noting that the effective potentials of the $I(J^{P}) = 1(0^{+})$ $D\bar{B}$ system have similar behaviors with the potentials of the $I(J^{P}) = 1(0^{+})$ $\bar{B}\bar{B}$ system in Ref.~\cite{Wang:2018atz}, which is a consequence of heavy flavor symmetry between the charm and bottom quarks. As shown in Fig.~\ref{BD}(c), the total potentials become more and more attractive as the cutoff parameter $\Lambda$ increases. The bound states emerge at $\Lambda = 0.7~\rm{GeV}$ and $\Lambda = 1.0~\rm{GeV}$, lying $0.8~\rm{MeV}$ and $3.6~\rm{MeV}$ below the $D\bar{B}$ mass threshold, respectively. However, if we change the cutoff to $0.54~\rm{GeV}$, the very shallow bound state can emerge in the $D \bar{B}[0(0^{+})]$ channel.

\subsubsection{$D\bar{B}^{*}$ systems}
   
\begin{figure*}[ht]
    \centering
    \includegraphics[width=1.0\textwidth]{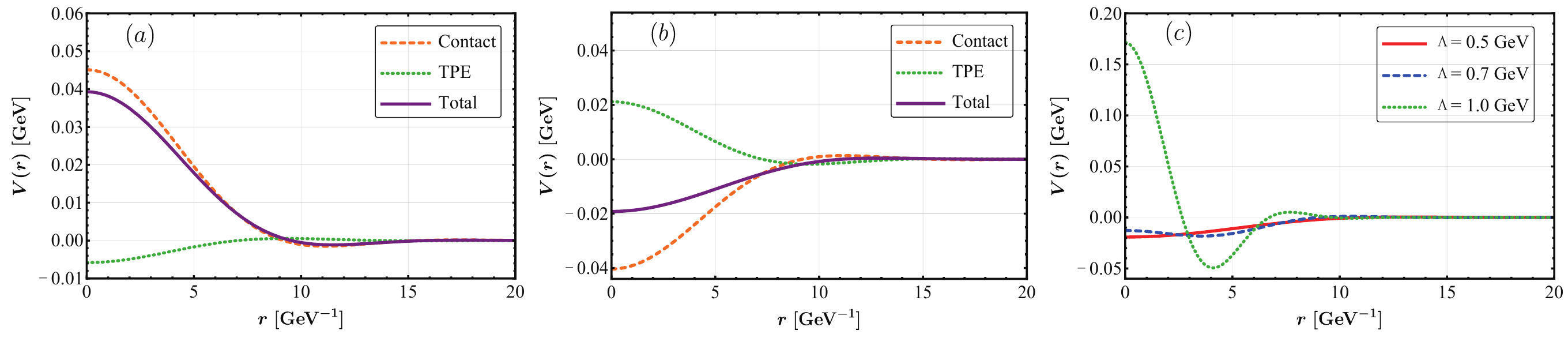}
    \captionsetup{justification=raggedright, singlelinecheck=false}
    \caption{$S$-wave potential of the $D\bar{B}^{*}$ systems and the dependence of the total potential on the cutoff parameter, namely, (a) the potential of the $I(J^{P})=1(1^+)$ $D\bar{B}^{*}$ system at $\Lambda=0.5~\rm{GeV}$, (b) the potential of the $I(J^{P})=0(1^+)$ $D\bar{B}^{*}$ system at $\Lambda=0.5~\rm{GeV}$, and (c) the total potential of the $D\bar{B}^{*}[0(1^{+})]$ system with different cutoffs. The green dashed and red dashed lines represent the contact and TPE contributions, respectively. The purple solid lines stand for the full potential.}\label{BsD}
\end{figure*}

In this section, we investigate the interactions of the $D\bar{B}^{*}$ systems with $I(J^{P}) = 1(1^{+})$ and $I(J^{P}) = 0(1^{+})$. As shown in Fig.~\ref{BsD1} before, only the contact diagram contributes to the amplitudes at LO. The one-loop corrections to the contact diagrams and TPE diagrams appear at NLO. Their results are shown in Figs.~\ref{BsD}(a) and \ref{BsD}(b), where we adopt the cutoff parameter $\Lambda$ with $0.5~\rm{GeV}$.

Comparing Fig.~\ref{BsD}(a) with Fig.~\ref{BD}(a), we observe that the behavior of the $D\bar{B}^{*}[1(1^{+})]$ effective potential is similar to the $D\bar{B}[1(0^{+})]$ effective potential. For the $D\bar{B}^{*}[1(1^{+})]$ system, the repulsive contact interaction is much stronger than the attractive TPE interaction, which leads to a net repulsive potential. As a result, no bound state exists in this system.

In the case of the $D\bar{B}^{*}[0(1^{+})]$ system, the effective potentials are shown in Fig.~\ref{BsD}(b). Compared to the $1(1^{+})$ system, the behaviors of the $0(1^{+})$ contact and TPE potential are both reversed, which results in an overall attractive interaction. However, this potential is weaker even than the one in the previous $D\bar{B}[0(0^{+})]$ system. As shown in Fig.~\ref{BsD}(c), the total potentials at $\Lambda = 0.7~\rm{GeV}$ and at $\Lambda = 0.5~\rm{GeV}$ are of similar strength. However, when $\Lambda = 1.0~\rm{GeV}$, the effective potential is repulsive near the origin but becomes attractive at intermediate distances. This behavior appears because the TPE contribution at large momentum is more sensitive to cutoff $\Lambda$ variations, which leads to a repulsive potential at short distances. In our calculation, no bound state is found in the $D\bar{B}^{*} [0(1^{+})]$ system.

Therefore, we conclude that the interaction in the $D \bar{B}^{*}[0(1^{+})]$ system is weaker than that in the $D \bar{B}[0(0^{+})]$ system. This is consistent with the predictions of the one-boson-exchange potential model \cite{Li:2012ss}, but contrasts with the conclusions drawn 
from lattice QCD calculations \cite{Alexandrou:2023cqg}.

\subsubsection{$D^{*}\bar{B}^{*}$ systems}

In this section, we explore the possibility of bound states in the $D^{*}\bar{B}^{*}$ systems, which includes six different isospin-spin states: $1(0^{+})$, $0(0^{+})$, $1(1^{+})$, $0(1^{+})$, $1(2^{+})$, and $0(2^{+})$. At LO, both contact and OPE contribute to the effective potential. At NLO, the one-loop corrections to the contact and OPE diagrams shown in Fig.~\ref{BsDs1} as well as TPE diagrams all contribute.

\begin{figure*}[ht]
	 \centering
    \includegraphics[width=1.0\textwidth]{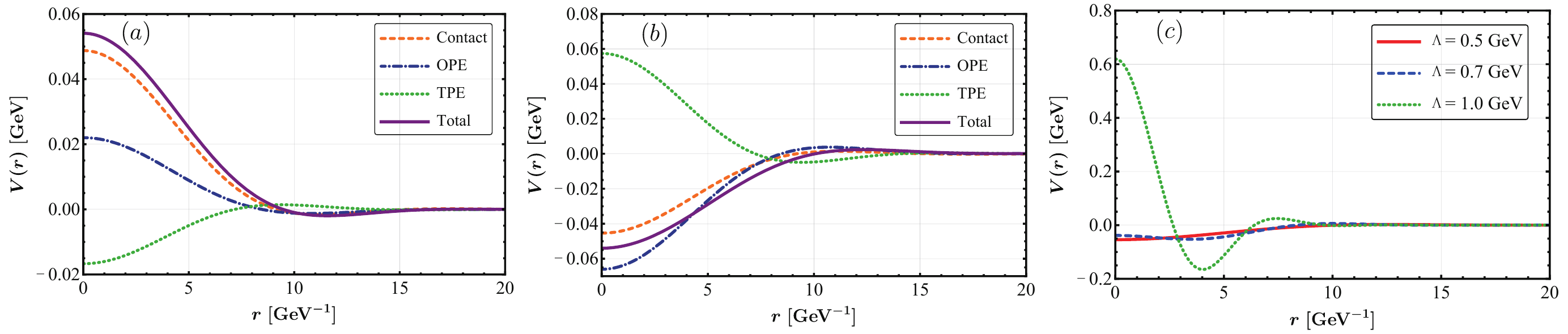}
    \captionsetup{justification=raggedright, singlelinecheck=false}
	 \caption{$S$-wave potential of $J=0$ $D^{*}\bar{B}^{*}$ systems and the dependence of the total potential on the cutoff parameter, namely, (a) the potential of the $I(J^{P})=1(0^+)$ $D^{*}\bar{B}^{*}$ system at $\Lambda=0.5~\rm{GeV}$, (b) the potential of the $I(J^{P})=0(0^+)$ $D^{*}\bar{B}^{*}$ system at $\Lambda=0.5~\rm{GeV}$, and (c) the total potential of the $D^{*}\bar{B}^{*}[0(0^{+})]$ system with different cutoffs. The green dashed, blue dashed, and red dashed lines represent the contact, OPE, and TPE contributions, respectively. The purple solid lines stand for the full potential.}\label{BsDs 0}
\end{figure*}

First, we focus on the $J^{P}=0^{+}$ $D^{*}\bar{B}^{*}$ state. The effective potentials are shown in Figs.~\ref{BsDs 0}(a) and \ref{BsDs 0}(b), where we adopt the cutoff parameter $\Lambda$ with $0.5~\rm{GeV}$. In Fig.~\ref {BsDs 0}(a), we observe that both the contact and OPE interactions are repulsive, while the TPE interaction is attractive. However, the total potential is repulsive, meaning no bound state can exist in the $1(0^{+})$ $D^{*}\bar{B}^{*}$ system. In contrast, for the $0(0^{+})$ system in Fig.~\ref{BsDs 0}(b), the contact and OPE interactions provide an attractive force, while the TPE interaction is repulsive. The repulsive TPE contribution partially counteracts the attractive contact interaction, but the OPE interaction remains strong enough to form a bound state. The resulting binding energy is $\Delta E = 6.0~\rm{MeV}$. As shown in Fig.~\ref{BsDs 0}, the OPE terms can contribute to the total interaction of $D^{*}\bar{B}^{*}$ systems, which is a departure from the $D\bar{B}^{(*)}$ systems. 

The $\Lambda$ dependence of the $I(J^{P})=0(0^{+})$ total potential is shown in Fig.~\ref{BsDs 0}(c). We find that the total interaction is sensitive to the cutoff, particularly near the origin, which stems from the TPE contribution. As the cutoff increases to $1.0~\rm{GeV}$, the binding energy becomes $35.9~\rm{MeV}$.

\begin{figure*}[htbp]
     \centering
    \includegraphics[width=1.0\textwidth]{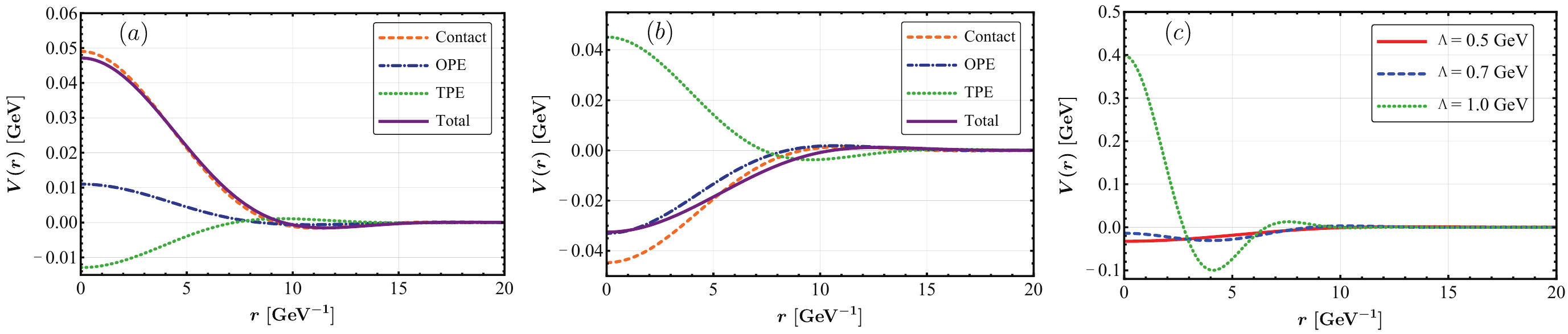}
    \captionsetup{justification=raggedright, singlelinecheck=false}
     \captionsetup{justification=raggedright, singlelinecheck=false}
     \caption{$S$-wave potential of $J=1$ $D^{*}\bar{B}^{*}$ systems and the dependence of the total potential on the cutoff parameter, namely, (a) the potential of the $I(J^{P})=1(1^+)$ $D^{*}\bar{B}^{*}$ system at $\Lambda=0.5~\rm{GeV}$, (b) the potential of the $I(J^{P})=0(1^+)$ $D^{*}\bar{B}^{*}$ system at $\Lambda=0.5~\rm{GeV}$, and (c) the total potential of the $D^{*}\bar{B}^{*}[0(1^{+})]$ system on the different cutoffs. The green dashed, blue dashed, and red dashed lines represent the contact, OPE, and TPE contributions, respectively. The purple solid lines stand for the full potential.}\label{BsDs1.1}
\end{figure*}

Next, we analyze the interactions in the $D^{*}\bar{B}^{*}[1^{+}]$ system. For the $1(1^{+})$ state, as seen in Fig.~\ref{BsDs1.1}(a), the OPE contribution is largely offset by the TPE contribution, with the repulsive contact potential remaining. It results in a fully repulsive total potential, meaning no bound state exists in the $D^{*}\bar{B}^{*}[1(1^{+})]$ system. For the $D^{*}\bar{B}^{*}[0(1^{+})]$ system, the effective potentials, shown in Fig.~\ref{BsDs1.1}(b), exhibit similar behaviors to those of the $D^{*}\bar{B}^{*}[0(0^{+})]$ system. Although the attraction is not as strong as in the $D^{*}\bar{B}^{*}[0(0^{+})]$ system, it is still sufficient to form a shallow bound state. By solving the Schr\"odinger equation, we obtain the $D^{*}\bar{B}^{*}[0(1^{+})]$ binding energy $\Delta E = 0.6~\rm{MeV}$ at $\Lambda = 0.5~\rm{GeV}$.

\begin{figure*}[ht]
     \centering
    \includegraphics[width=1.0\textwidth]{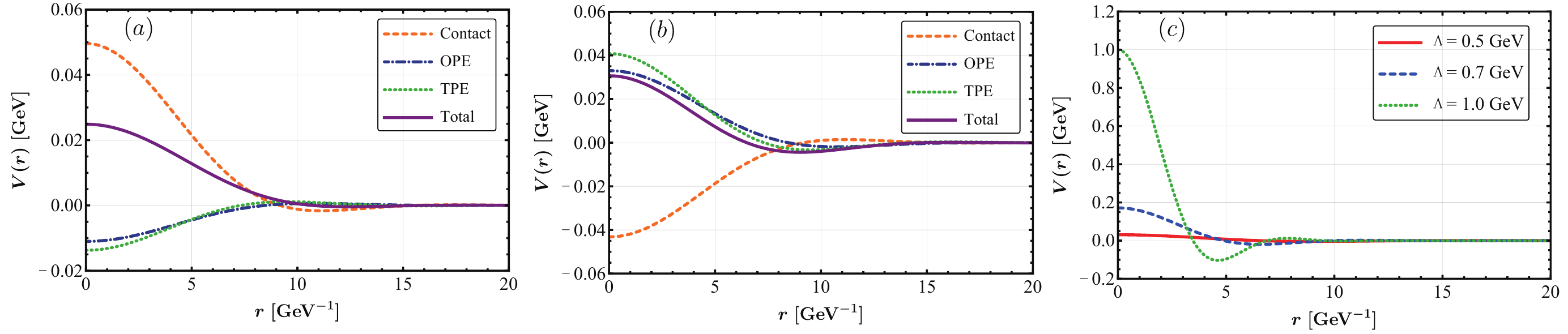}
    \captionsetup{justification=raggedright, singlelinecheck=false}
     \caption{$S$-wave potential of $J=2$ $D^{*}\bar{B}^{*}$ systems and the dependence of the total potential on the cutoff parameter, namely, (a) the potential of the $I(J^{P})=1(2^+)$ $D^{*}\bar{B}^{*}$ system at $\Lambda=0.5~\rm{GeV}$, (b) the potential of the $I(J^{P})=0(2^+)$ $D^{*}\bar{B}^{*}$ system at $\Lambda=0.5~\rm{GeV}$, and (c) the total potential of the $D^{*}\bar{B}^{*}[0(2^{+})]$ system with the different cutoffs. The green dashed, blue dashed, and red dashed lines represent the contact, OPE, and TPE contributions, respectively. The purple solid lines stand for the full potential.}\label{BsDs1.2}
\end{figure*}   

Finally, we examine the possibility of bound state formations in the $J^{P} = 2^{+}$ $D^{*}\bar{B}^{*}$ system. 
In Fig.~\ref{BsDs1.2}(a), for the $1(2^{+})$ $D^{*}\bar{B}^{*}$ system, both the OPE and TPE potentials are mildly attractive, while the dominant contact potential remains repulsive, resulting in a repulsive total potential.

We also calculate the effective potentials for the $D^{*}\bar{B}^{*}[0(2^{+})]$ system, with the potential profiles shown in Fig.~\ref{BsDs1.2}(b). In this situation, both the OPE and TPE act as repulsive forces that counter the attractive force from the contact term, leading to an overall repulsive potential. This behavior contrasts with the attractive effective potential of the other $D^{(*)}\bar{B}^{(*)}$ systems with $I = 0$. Consequently, there is no evident bound state in the $D^{*}\bar{B}^{*}[0(2^{+})]$ system.

Similarly, the effective potentials of $D^{*}\bar{B}^{*}[0(2^{+})]$ at different cutoff parameters are shown in Fig.~\ref{BsDs1.2}(c). Although the total interaction is repulsive at small distances, as seen in other channels, the potential becomes attractive in the intermediate range at $\Lambda = 1.0~\rm{GeV}$. This attraction allows the $D^{*}$-$\bar{B}^{*}$ mesons to form a bound state, which lies $4.3~\rm{MeV}$ below the threshold.

Based on the above calculations and analyses, we have obtained the coordinate potentials and determined the possible bound states by solving the Schr$\ddot{\rm o}$dinger equation. The total interactions are repulsive in the $I = 1$ channels and attractive in the $I = 0$ channels. Then, we focus on the $I = 0$ $D^{(*)}\bar{B}^{(*)}$ systems to search for bound states. For the $D \bar{B}[0(0^{+})]$, $D \bar{B}^{*}[0(1^{+})]$, and $D^{*}\bar{B}^{*}[0(2^{+})]$ systems, the attractive potentials are too weak to bind the mesons, and no definite bound states are found in these channels at a cutoff of $\Lambda = 0.5~\rm{GeV}$. However, if we change the cutoff to $0.54~\rm{GeV}$, the very shallow bound state can emerge in the $D \bar{B}[0(0^{+})]$ channel. As for the $D \bar{B}^{*}[0(1^{+})]$, the bound state appears at $\Lambda = 1.02~\rm{GeV}$, and the binding energy will increase as the $\Lambda$ gets larger. 

Besides, in the $D^{*}\bar{B}^{*}[0(0^{+})]$ and $D^{*}\bar{B}^{*}[0(1^{+})]$ channels, two shallow bound states are found with binding energies of $6.0~\rm{MeV}$ and $0.6~\rm{MeV}$, respectively. Therefore, we suggest that future theoretical and experimental studies should focus on these two systems to find new heavy tetraquark states. Compared to the total potentials of the five $I=0$ channels, we can also conclude that the interaction in the $D^{*}\bar{B}^{*}[0(0^{+})]$ channel is more attractive than the other four channels. The mass and the binding energy of each $D^{(*)}\bar{B}^{(*)}[I=0]$ system under different cutoffs are listed in Table~\ref{mass}.

\renewcommand{\arraystretch}{1.5}
\begin{center}
  \begin{table*}[ht]
      \centering
      \captionsetup{justification=raggedright, singlelinecheck=false}
      \caption{The bound states in the five $D^{(*)}\bar{B}^{(*)}$ systems with $I=0$. The cutoff parameter $\Lambda$, state mass $M$, and binding energy $E$ are in units of $\rm{GeV}$, $\rm{MeV}$, and $\rm{MeV}$ respectively. ``$\cdots$" means that there is no bound state.}\label{mass}
  	\setlength{\tabcolsep}{5.0mm}{
  	\begin{tabular}{cccccccccccc}
  	  \toprule[1pt]
		&  \multicolumn{2}{c}{$D\bar{B}$} & \multicolumn{2}{c}{$D\bar{B}^{*}$} &  \multicolumn{2}{c}{$D^{*}\bar{B}^{*}(J=0)$}  & \multicolumn{2}{c}{$D^{*}\bar{B}^{*}(J=1)$}  & \multicolumn{2}{c}{$D^{*}\bar{B}^{*}(J=2)$} \\
		\cline{2-3}  \cline{4-5} \cline{6-7} \cline{8-9} \cline{10-11}
  		$\Lambda$ & {$M$} & {$E$} & {$M$} & {$E$} & {$M$} & {$E$} & {$M$} & {$E$} & {$M$} & {$E$} \\
  		\midrule[1pt]
  		$0.5$ & $\cdots$ & $\cdots$ &  $\cdots$  & $\cdots$ & $7235.7$ & $6.0$  & $7241.1$ & $0.6$ & $\cdots$ & $\cdots$  \\
  		$0.7$ & $7143.6$ & $0.8$   & $\cdots$  & $\cdots$ & $7230.6$ & $11.1$  & $7240.0$ & $1.7$ & $\cdots$ & $\cdots$  \\
  		$1.0$ & $7140.8$ & $3.6$   & $\cdots$  & $\cdots$ & $7205.8$ & $35.9$ & $7232.3$ & $9.4$ & $7237.4$ & $4.3$  \\
  		\bottomrule[1pt]
  	\end{tabular}}
  \end{table*}
\end{center}

\subsection{Coupled-channel effects}
\label{cce}

We have considered interactions of the $D^{(*)}\bar{B}^{(*)}$ systems in the single-channel case. In this section, we will investigate how much their coupled-channel effects affect the above results. We categorize the $I=0$ coupled-channel $D^{(*)}\bar{B}^{(*)}$ systems into two groups according to their $0(J^{P})$ quantum numbers: \{$D\bar{B}(^{1}S_{0})$, $D^{*}\bar{B}^{*}(^{1}S_{0})$\}$[0(0^+)]$ and \{$D\bar{B}^{*}(^{3}S_{1})$, $D^{*}\bar{B}(^{3}S_{1})$, $D^{*}\bar{B}^{*}(^{3}S_{1})$\}$[0(1^+)]$.

We first consider the $D^{(*)}\bar{B}^{(*)}[0(0^+)]$ system composed of \{$D\bar{B}(^{1}S_{0})$, $D^{*}\bar{B}^{*}(^{1}S_{0})$\}. In this case, the effective potentials of the inelastic scattering process $D\bar{B} \to D^{*}\bar{B}^{*}$ and $D^{*}\bar{B}^{*} \to D\bar{B}$ are needed. We depict the $D\bar{B} \to D^{*}\bar{B}^{*}$ diagrams in Fig.~\ref{BD-BsDs}. At LO, the contact and OPE diagrams contribute to the scattering amplitudes. At NLO, there are one-loop corrections to these LO diagrams and TPE diagrams. The calculated scattering amplitudes for these diagrams are listed in Appendix \ref{A1}. The $D^{(*)}\bar{B}^{(*)}[0(0^+)]$ coupled-channel potential in the coordinate space is 
\begin{align*}
        \begin{pmatrix}
	V_{D\bar{B}-D\bar{B}}   & V_{D\bar{B}-D^{*}\bar{B}^{*}}  \\
	V_{D^{*}\bar{B}^{*}-D\bar{B}}   &  V_{D^{*}\bar{B}^{*}-D^{*}\bar{B}^{*}}
	\end{pmatrix} ,
\end{align*}
where the inelastic scattering $D\bar{B} \to D^{*}\bar{B}^{*}$ and $D^{*}\bar{B}^{*} \to D\bar{B}$ appear as nondiagonal elements in the $2\times2$ potential matrix. Then we solve the coupled-channel Schr\"odinger equation and obtain the numerical results, including the binding energy $E$, and the probability of the individual channel $P_{i}$, which are shown in Table~\ref{cch1}.

Here, we consider the variation on the cutoff $\Lambda$ just as in the single-channel case. Compared with the single-channel $D\bar{B}[0(0^+]$ results in Table \ref{mass}, we find coupled channels somewhat strengthen the binding energy: from 0.8 GeV in the single channel to 1.1 GeV in the coupled channels at $\Lambda=0.7$ GeV, for example. Also, because of the large mass gap (about $190~{\rm{MeV}}$) between the $D\bar{B}$ and $D^{*}\bar{B}^{*}$ thresholds, the contribution of $D^{*}\bar{B}^{*}(^{1}S_{0})$ is $0.1\%$ at $\Lambda = 0.5~\rm{GeV}$. If we raise the cutoff to $0.7~\rm{GeV}$ and $1.0~\rm{GeV}$, the probability of $D^{*}\bar{B}^{*}(^{1}S_{0})$ increases to $0.3\%$ and $0.9\%$, respectively. 

For the $D^{(*)}\bar{B}^{(*)}[0(1^+)]$ state, all the coupled-channel diagrams and corresponding scattering amplitudes are shown in Appendix \ref{A1}-\ref{A5}. Using these amplitudes, we can obtain a $3\times3$ coupled-channel potential matrix, 
\begin{align*}
       \begin{pmatrix}
	V_{D\bar{B}^{*}-D\bar{B}^{*}}    & V_{D\bar{B}^{*}-D^{*}\bar{B}}      &     V_{D\bar{B}^{*}-D^{*}\bar{B}^{*}} \\
	V_{D^{*}\bar{B}-D\bar{B}^{*}}    & V_{D^{*}\bar{B}-D^{*}\bar{B}}      &   V_{D^{*}\bar{B}-D^{*}\bar{B}^{*}} \\
       V_{D^{*}\bar{B}^{*}-D\bar{B}^{*}} & V_{D^{*}\bar{B}^{*}-D^{*}\bar{B}}  & V_{D^{*}\bar{B}^{*}-D^{*}\bar{B}^{*}}
	\end{pmatrix}.
\end{align*}
We still take the values of the cutoff parameter at $0.5~\rm{GeV}$, $0.7~\rm{GeV}$, and $1.0~\rm{GeV}$ to evaluate binding energy and probability $P_{i}$, respectively. The numerical results are listed in Table~\ref{cch2}. We find no bound state in the $D^{(*)}\bar{B}^{(*)}[0(1^+)]$ system, which is consistent with the single-channel $D\bar{B}^{*}[0(1^+)]$ calculation. The dominant channel is $D\bar{B}^{*}(^{3}S_{1})$, with a probability of $99.9\%-99.5\%$.

According to the above analyses, we can find that the coupled-channel effects do not change our main conclusions obtained in the single-channel case.

\renewcommand{\arraystretch}{1.5}
\begin{center}
  \begin{table}[ht]
  	\centering
        \captionsetup{justification=raggedright, singlelinecheck=false}
  	\caption{The numerical results for $D^{(*)}\bar{B}^{(*)}[0(0^+)]$ systems in the $S$-wave. The cutoff parameter $\Lambda$ and binding energy $E$ relative to the threshold of $D\bar{B}$ are in units of $\rm{GeV}$ and $\rm{MeV}$, respectively. ``$\cdots$" means that there is no bound solution.}\label{cch1}
  	\setlength{\tabcolsep}{7.0mm}{
  	\begin{tabular}{ccccc}
  	  \toprule[1pt]
  		$\Lambda$ & { $0.5$ } & { $0.7$ } &  { $1.0$ }   \\
  		\midrule[1pt]
  		$ E $ & $\cdots$  & $1.1$   & $7.3$  \\
  		$P_{1}$ & $99.9\%$  & $99.7\%$  & $99.1\%$  \\
  		$P_{2}$ & $0.1\%$   & $0.3\%$   & $0.9\%$  \\
  		\bottomrule[1pt]
  	\end{tabular}}
  \end{table}
\end{center}

\renewcommand{\arraystretch}{1.5}
\begin{center}
  \begin{table}[ht]
  	\centering
        \captionsetup{justification=raggedright, singlelinecheck=false}
  	\caption{The numerical results for $D^{(*)}\bar{B}^{(*)}[0(1^+)]$ systems in the $S$-wave. The cutoff parameter $\Lambda$ and binding energy $E$ relative to the threshold of $D\bar{B}^{*}$ are in units of $\rm{GeV}$ and $\rm{MeV}$, respectively. ``$\cdots$" means that there is no bound solution.}\label{cch2}
  	\setlength{\tabcolsep}{7.0mm}{
  	\begin{tabular}{ccccc}
  	  \toprule[1pt]
  		$\Lambda$ & { $0.5$ } & { $0.7$ } &  { $1.0$ }   \\
  		\midrule[1pt]
  		$E$ & $\cdots$  & $\cdots$   & $\cdots$  \\
  		$P_{1}$ & $99.9\%$  & $99.7\%$  & $99.5\%$  \\
  		$P_{2}$ & $0.1\%$   & $0.3\%$   & $0.5\%$  \\
            $P_{3}$ & $0.0\%$   & $0.0\%$   & $0.0\%$  \\
  		\bottomrule[1pt]
  	\end{tabular}}
  \end{table}
\end{center}

\subsection{Two-body scattering of $D^{(*)}\bar{B}^{(*)}$ systems}

\begin{figure*}[ht]
    \centering
    \includegraphics[width=0.66\textwidth]{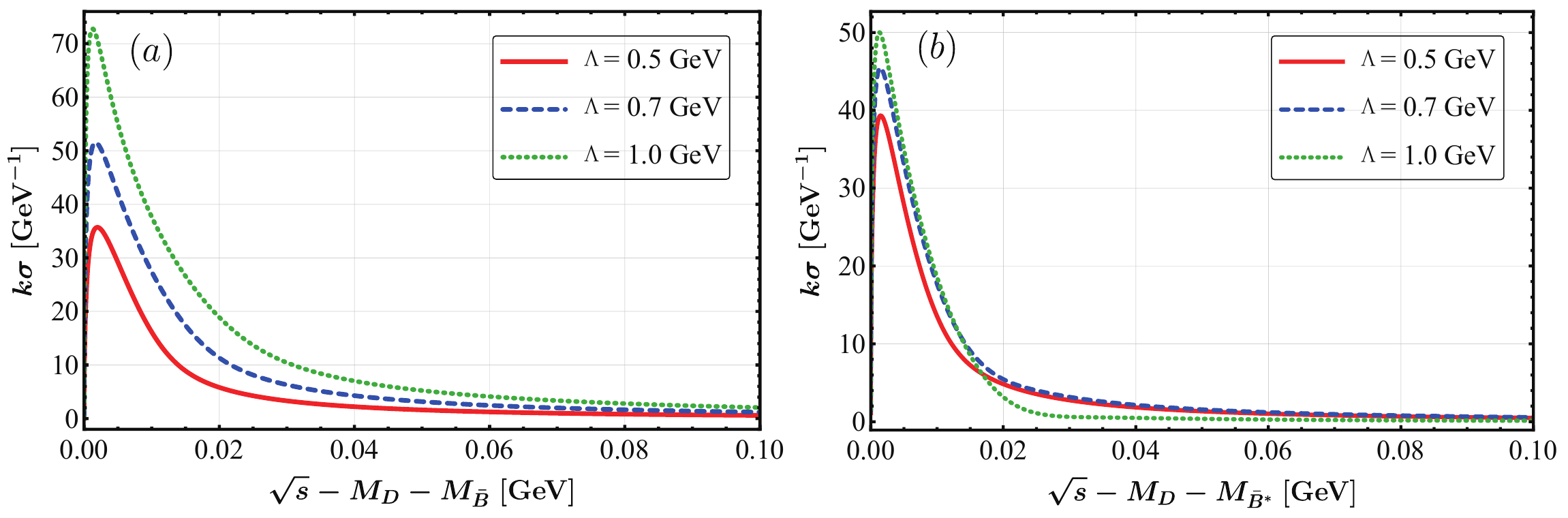}
    \captionsetup{justification=raggedright, singlelinecheck=false}
    \caption{The dependencies of the $S$-wave scattering rates for the $D \bar{B} [0(0^{+})]$ system and $D\bar{B}^{*}[0(1^{+})]$ system on the cutoff $\Lambda$.}\label{ps1}
\end{figure*}

\begin{figure*}[ht]
    \centering
    \includegraphics[width=1.0\textwidth]{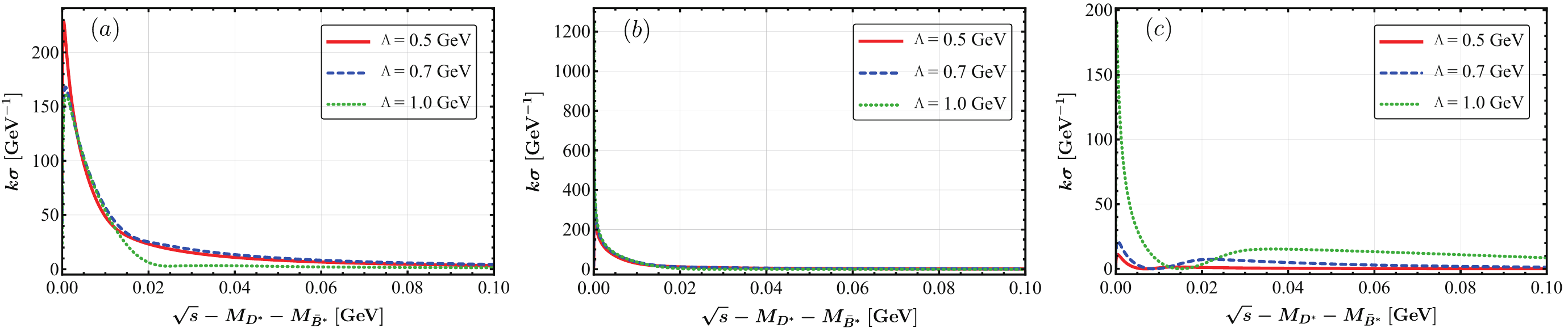}
    \captionsetup{justification=raggedright, singlelinecheck=false}
    \caption{The dependence of the $S$-wave scattering rate $k\sigma$ for the $D^{*}\bar{B}^{*}[0(0^{+})]$, $D^{*}\bar{B}^{*}[0(1^{+})]$, and $D^{*}\bar{B}^{*}[0(2^{+})]$ systems on the cutoff $\Lambda$.}\label{BsDsps}
\end{figure*}

In addition to the binding energies, we further derive information about $D^{(*)}\bar{B}^{(*)}$ elastic scattering by solving the Lippmann-Schwinger equation. In quantum theory, the scattering rate represents the probability per unit time for a scattering event, and it is proportional to the cross section, the number of targets, and the flux \cite{Thomson:2013zua}. Using Eqs.~(\ref{LSE})-(\ref{sigma}), we now calculate the scattering rates between $D^{(*)}$ and $\bar{B}^{(*)}$. The variations of the scattering rates $k\sigma(k)$ with cutoffs are shown in Figs.~\ref{ps1} and \ref{BsDsps}, as functions of the center-of-mass energy. 

The scattering length and effective range can be related to the interaction between particles through the scattering amplitude or phase shift \cite{Taylor:1972pty}. Based on Eq.~(\ref{ERE}), we also evaluate the scattering lengths and effective ranges at different cutoff values $\Lambda$, with numerical results listed in Table~\ref{length}. 

Let us focus on the scattering rates $k\sigma(k)$ for the $D\bar{B}[0(0^{+})]$ and $D\bar{B}^{*}[0(1^{+})]$ systems in Figs.~\ref{ps1}(a) and \ref{ps1}(b). The enhancement of $k\sigma(k)$ near the threshold is gradual, which is insufficient to confirm the presence of a bound or virtual state. Looking at the scattering lengths and effective ranges in Table~\ref{length}, we find that the scattering lengths $a$ for different cutoffs are negative, which indicates that the phase shifts satisfy $0 \le \delta < \pi/2$ \cite{Zhai:2021uap}. As a result, there are no evident bound states in the two systems, which aligns with the outcomes found in the previous section by solving the Schr\"odinger equation.

We now move to the scattering of the two vector meson systems. The scattering rates for the $D^{*}\bar{B}^{*}[0(0^{+})]$, $D^{*}\bar{B}^{*}[0(1^{+})]$, and $D^{*}\bar{B}^{*}[0(2^{+})]$ systems are shown in Fig.~\ref{BsDsps}. Near the thresholds, there are significant enhancements, which may indicate the presence of possible bound or virtual states. 

In Table~\ref{length}, we find that, in the $D^{*}\bar{B}^{*}[0(0^{+})]$ system, the scattering lengths are positive for different cutoff values, indicating the presence of unambiguous bound states, which supports the calculations in the above section.

For the $D^{*}\bar{B}^{*}[0(1^{+})]$ system, the scattering length is negative at $\Lambda = 0.5~\rm{GeV}$ and $\Lambda = 0.7~\rm{GeV}$, but it takes on a large positive value at $\Lambda = 1.0~\rm{GeV}$, signaling the existence of a shallow bound state.

Additionally, we evaluate the corresponding phase shifts for these $S$-wave bound systems according to Eq.~(\ref{phase}). We find that 
\begin{align}
    \lim_{k\to 0} \delta_{0} = \pi,
\end{align}
which is consistent with Levinson’s theorem \cite{Taylor:1972pty}, describing the physical connection between phase shifts and the existence of bound states.

For the $D^{*}\bar{B}^{*}[0(2^{+})]$ system at different cutoffs, the effective interaction is attractive at small momentum (large distance) but repulsive at large momentum (small distance). It initially causes a positive phase shift at low momentum and then a negative phase shift at high momentum, so that the scattering rate drops to zero, and then gradually increases against the center-of-momentum energy. The negative scattering length indicates that the attractive effect is not strong enough to bind $D^{*}$ and $\bar{B}^{*}$ together in the $0(2^{+})$ channel.

\renewcommand{\arraystretch}{1.5}
\begin{center}
  \begin{table*}[ht]
      \centering
      \captionsetup{justification=raggedright, singlelinecheck=false}
      \caption{The scattering length $a$ and effective range $r$ in the five $D^{(*)}\bar{B}^{(*)}[I=0]$ scattering channels. The cutoff parameter $\Lambda$, $a$, and $r$ are in units of $\rm{GeV}$, $\rm{fm}$, and $\rm{fm}$, respectively.}\label{length}
  	\setlength{\tabcolsep}{4.8mm}{
  	\begin{tabular}{ccccccccccc}
  	  \toprule[1pt]
  		&  \multicolumn{2}{c}{$D\bar{B}$} & \multicolumn{2}{c}{$D\bar{B}^{*}$} &  \multicolumn{2}{c}{$D^{*}\bar{B}^{*}(J=0)$}  & \multicolumn{2}{c}{$D^{*}\bar{B}^{*}(J=1)$}  & \multicolumn{2}{c}{$D^{*}\bar{B}^{*}(J=2)$} \\
		\cline{2-3}  \cline{4-5} \cline{6-7} \cline{8-9} \cline{10-11}
  		$\Lambda$ & $a$ & $r$ & $a$ & $r$ & $a$ & $r$ & $a$ & $r$ & $a$ & $r$ \\
  		\midrule[1pt]
  		$0.5$ & $-1.64$  & $2.60$ & $-1.86$   & $2.05$ & $6.37$ & $1.91$ & $-9.87$ & $0.54$ & $-1.25$  & $11.05$ \\
  		$0.7$ & $-2.12$  & $2.21$ & $-2.03$  & $1.97$ & $4.48$ & $1.97$ & $-24.06$ &  $0.71$ & $-1.76$ & $7.97$  \\
  		$1.0$ & $-2.74$  & $1.85$ & $-2.22$  & $2.08$ & $4.38$  & $0.92$  & $140.94$ & $1.19$ & $-7.31$ &  $3.67$ \\
  		\bottomrule[1pt]
  	\end{tabular}}
  \end{table*}
\end{center}

\subsection{Estimation of the contact term LECs}

For the $D\bar{B}[0(0^{+})]$ and $D\bar{B}^{*}[0(1^{+})]$ systems, the above calculations show that the total potentials are not strong enough to form bound states for the heavy mesons. However, this does not imply that no bound state can exist in these systems. Note that the contact terms are determined using the resonance saturation model, therefore, we now explore alternative methods to redetermine the LECs of the contact terms in this work.

Recently, a lattice QCD study conducted by Alexandrou {\it et al.} \cite{Alexandrou:2023cqg} found shallow bound states for both $J=0$ and $J=1$ $\bar{b}\bar{c}ud$ tetraquarks, with binding energies of $0.5_{-1.5}^{+0.4}\rm{MeV}$ and $2.4^{+2.0}_{-0.7}~\rm{MeV}$, respectively. Padmanath {\it et al.} reported a lattice QCD study indicating that an $I(J^{P})=0(1^{+})$ $\bar{b}\bar{c}ud$ bound state with a binding energy of $43(^{+7}_{-6})(^{+24}_{-14})~\rm{MeV}$ can exist below the $B^{*}\bar{D}$ threshold \cite{Padmanath:2023rdu}. And, they presented another study searching for tetraquark candidates with exotic quark content $bc\bar{u}\bar{d}$ in the $I=0$ and $J^{P}=0^{+}$ channels, and found a subthreshold pole in the $S$-wave $D\bar{B}$ scattering amplitude, which corresponds to the binding energy of $39(^{+4}_{-6})(^{+8}_{-18})\rm{MeV}$ below the $D\bar{B}$ threshold \cite{Radhakrishnan:2024ihu}. The above results provide valuable references for redetermining the contact terms' LECs.

\begin{figure}[htbp]
   \centering
   \includegraphics[width=0.45\textwidth]{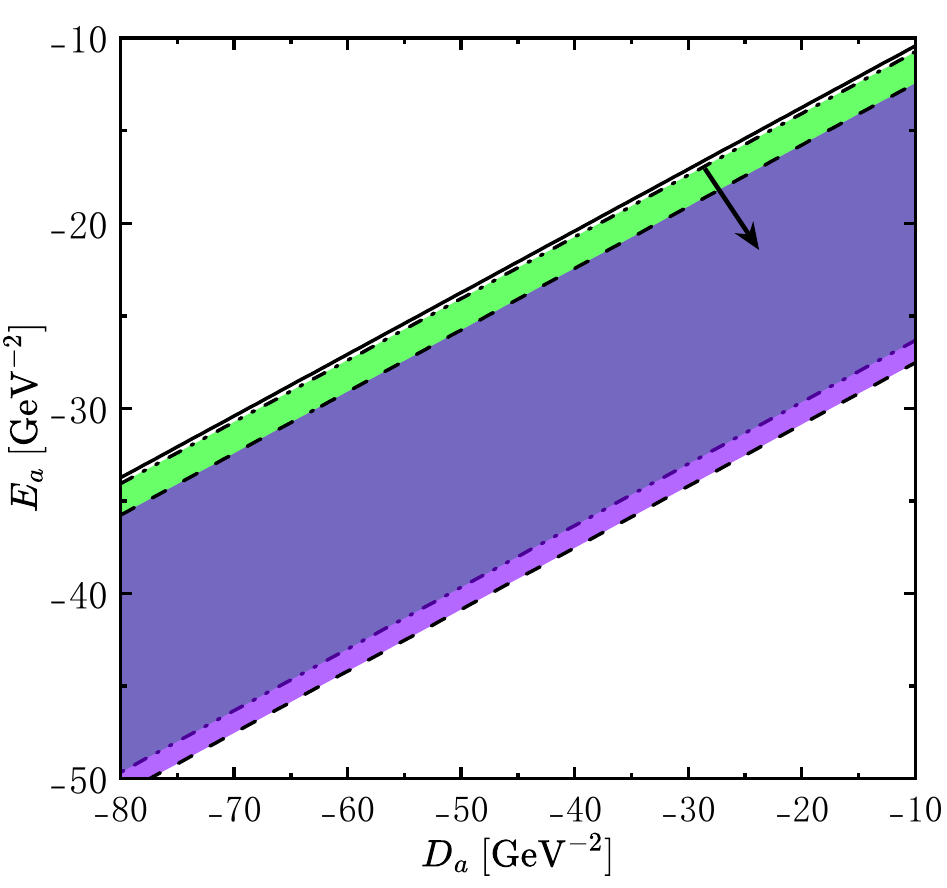}
   \captionsetup{justification=raggedright, singlelinecheck=false}
   \caption{The binding energies of the $D\bar{B}[0(0^{+})]$ and $D\bar{B}^{*}[0(1^{+})]$ states vary with the redetermined contact LECs $D_{a}$ and $E_{a}$ with cutoff $\Lambda=0.5~\rm{GeV}$. The purple and green bands correspond to the $D\bar{B}$ and $D\bar{B}^{*}$ systems, respectively. The black solid line represents the contact LECs determined by the resonance saturation model. The parallel dashed (dotted) lines at the boundaries of the purple (green) band stand for the regions of parameters within the binding energies $0~\rm{MeV}$ and $30~\rm{MeV}$. The arrow stands for the direction that binding energy increases in for both $D\bar{B}[0(0^{+})]$ and $D\bar{B}^{*}[0(1^{+})]$ systems. } \label{contact}
\end{figure}

To ascertain the possible parameter region which allows bound states to exist in the $D\bar{B}[0(0^{+})]$ and $D\bar{B}^{*}[0(1^{+})]$ systems at $\Lambda = 0.5~\rm{GeV}$, we can vary the LECs $D_a$ and $E_a$ within the ranges $[-80, 10]~\rm{GeV^{-2}}$ and $[-50, -10]~\rm{GeV^{-2}}$, respectively. We find that when the LECs satisfy the relation
\begin{align}
    D_{a}-3E_{a}=22.2~\rm{GeV^{-2}},
\end{align}
a bound state forms with almost zero binding energy. When the LECs satisfy 
\begin{align}
    D_{a}-3E_{a}=69.0~\rm{GeV^{-2}},
\end{align}
a bound state with a binding energy of 30 $\rm MeV$ below the $D\bar{B}$ threshold can be obtained. The green band in Fig.~\ref{contact} represents the corresponding parameter region. Similarly, we can also determine the contact term LECs for the $D\bar{B}^{*}[0(1^{+})]$ system. For binding energy of $0~\rm{MeV}$, we have
\begin{align}
    D_{a}-3E_{a}=27.3~\rm{GeV^{-2}},
\end{align}
and for a binding energy of $30~\rm{MeV}$,
\begin{align}
    D_{a}-3E_{a}=72.6~\rm{GeV^{-2}}.
\end{align}

The corresponding parameter variations are shown as the purple band in Fig.~\ref{contact}. The arrow indicates the direction in which the binding energy increases for both the $D\bar{B}[0(0^{+})]$ and $D\bar{B}^{*}[0(1^{+})]$ systems.

Notably, in this analysis, the attraction between $D$ and $\bar{B}$ is stronger than that between $D$ and $\bar{B}^{*}$. This contrasts the results reported in Ref.~\cite{Alexandrou:2023cqg} but is consistent with the results obtained using the one-boson-exchange model \cite{Li:2012ss}.

\section{summary}
\label{Sec: Summary}

In this work, we systematically investigated the interactions of the $S$-wave $D^{(*)}\bar{B}^{(*)}$ systems within the framework of ChEFT using the heavy hadron formalism. We calculated the effective potentials, including contributions from the contact, OPE, and TPE terms, up to NLO at the one-loop level, adopting Weinberg's formalism. By performing a Fourier transformation, we analyzed the behaviors of the effective potentials in coordinate space in detail and employed the Gauss form factor to regularize the divergence in the integral at large momentum. We then inserted the coordinate space potentials into the Schrödinger equation to search for potential $D^{(*)}\bar{B}^{(*)}$ bound states. 

The results showed that all total potentials in the $I=1$ channels are repulsive, whereas those in the $I=0$ channels are attractive. Further calculations examined the variation of interactions with the cutoff parameter $\Lambda$ across five $I=0$ $D^{(*)}\bar{B}^{(*)}$ channels. We found that bound states are more likely to exist in the $D\bar{B}[0(0^{+})]$, $D^{*}\bar{B}^{*}[0(0^{+})]$, and $D^{*}\bar{B}^{*}[0(1^{+})]$ channels. Notably, the TPE contributions are more sensitive to the cutoff in the $I=0$ $D^{*}\bar{B}^{*}$ channels and play a dominant role in these channels, unlike the case in $D\bar{B}[0(0^{+})]$. We list the mass and binding energies in Table~\ref{mass}.

We also discussed the $S$-wave coupled-channel effects of $D^{(*)}\bar{B}^{(*)}[0(0^{+})]$ and $D^{(*)}\bar{B}^{(*)}[0(1^{+})]$ systems. The numerical results listed in Table~\ref{cch1} indicate that, for the $D^{(*)}\bar{B}^{(*)}[0(0^{+})]$ system, the $D\bar{B}[0(0^{+})]$ channel is dominant,  and the coupled channels can somewhat strengthen the binding of the single channel. For $D^{(*)}\bar{B}^{(*)}[0(1^{+})]$ system, $D\bar{B}^{*}[0(1^+)]$ is dominant, and we obtained no binding solution, which is consistent with the single-channel calculation. In a word, the inclusion of the coupled channels does not change our main conclusions given in the single-channel case.

By substituting the momentum space potentials into the Lippmann-Schwinger equation, we calculated the scattering $T$-matrices and the associated scattering phase shifts. To gain further insight into the $S$-wave $D^{(*)}\bar{B}^{(*)}$ interactions, we calculated the scattering rate, scattering length, and effective range, and provide their numerical results in Table~\ref{length} based on the $T$-matrix and scattering phase. Accordingly, we studied the dependence of these physical quantities on the cutoff parameter $\Lambda$. In our investigation, the shallow bound state is more likely to exist in the $D\bar{B}[I(J^{P})=0(0^{+})]$ system than in the $D\bar{B}^{*}[I(J^{P})=0(1^{+})]$ system. 

Based on the above calculations, we found that the interactions in the $D^{*}\bar{B}^{*}[I=0]$ system are more attractive than those in the $D\bar{B}^{(*)}[I=0]$ systems, and $D^{*}\bar{B}^{*}[I(J^{P})=0(0^{+})]$ and $D^{*}\bar{B}^{*}[I(J^{P})=0(1^{+})]$ systems possess large binding energies and positive scattering lengths, which suggests strong bound state formations in these channels. Therefore, we strongly recommend the experiment to find the $D^{*}\bar{B}^{*}[I=0]$ tetraquark systems. 

Considering other theoretical and lattice QCD studies, we redetermined the contact term LECs and calculated the effective potentials for the $0(0^{+})$ $D\bar{B}$ system and the $0(1^{+})$ $D\bar{B}^{*}$ system. In Fig.~\ref{contact}, we present the dependencies of the binding energies of these two charm-bottom systems on the redetermined contact LECs $D_{a}$ and $E_{a}$.

Building on our calculations and analysis, we can comprehensively compare previous theoretical and lattice QCD studies. Our results provide valuable insights to inform future experimental research.

\section*{Acknowledgments}

Zhe Liu expresses sincere gratitude to Prof. Zhan-Wei Liu for his insightful discussions and extends thanks to Zi-Le Zhang and Ri-Qing Qian for their valuable contributions to the discussions. This work is supported by the National Natural Science Foundation of China under Grant No. 12335001, No. 12247101, No. 12465016, and No. 12005168, the ``111 Center" under Grant No. B20063, the Natural Science Foundation of Gansu Province (No. 22JR5RA389 and No. 22JR5RA171), the fundamental Research Funds for the Central Universities (Grant No. lzujbky-2023-stlt01), and the project for top-notch innovative talents of Gansu province.

\section{ DATA AVAILABILITY}

The data that support the findings of this article are openly available \cite{Zhe:2025dat}.

\appendix

\section{Coupled-channel calculations} \label{SecAppA}

In this appendix, we show the coupled-channel calculations of $D^{(*)}\bar{B}^{(*)}$ systems. The effective potentials derived from Lagrangians in Eq.~(\ref{L1}) and Eq.~(\ref{L3}) for $D\bar{B} \to D^{*}\bar{B}^{*}$ and $D^{(*)}\bar{B}^{(*)} \to D^{*}\bar{B}^{*}$ scattering are presented in this appendix.  

\subsection{$D\bar{B} \to D^{*}\bar{B}^{*}$}
\label{A1}

\begin{figure*}[ht]
     \centering
     \includegraphics[width=1.0\textwidth]{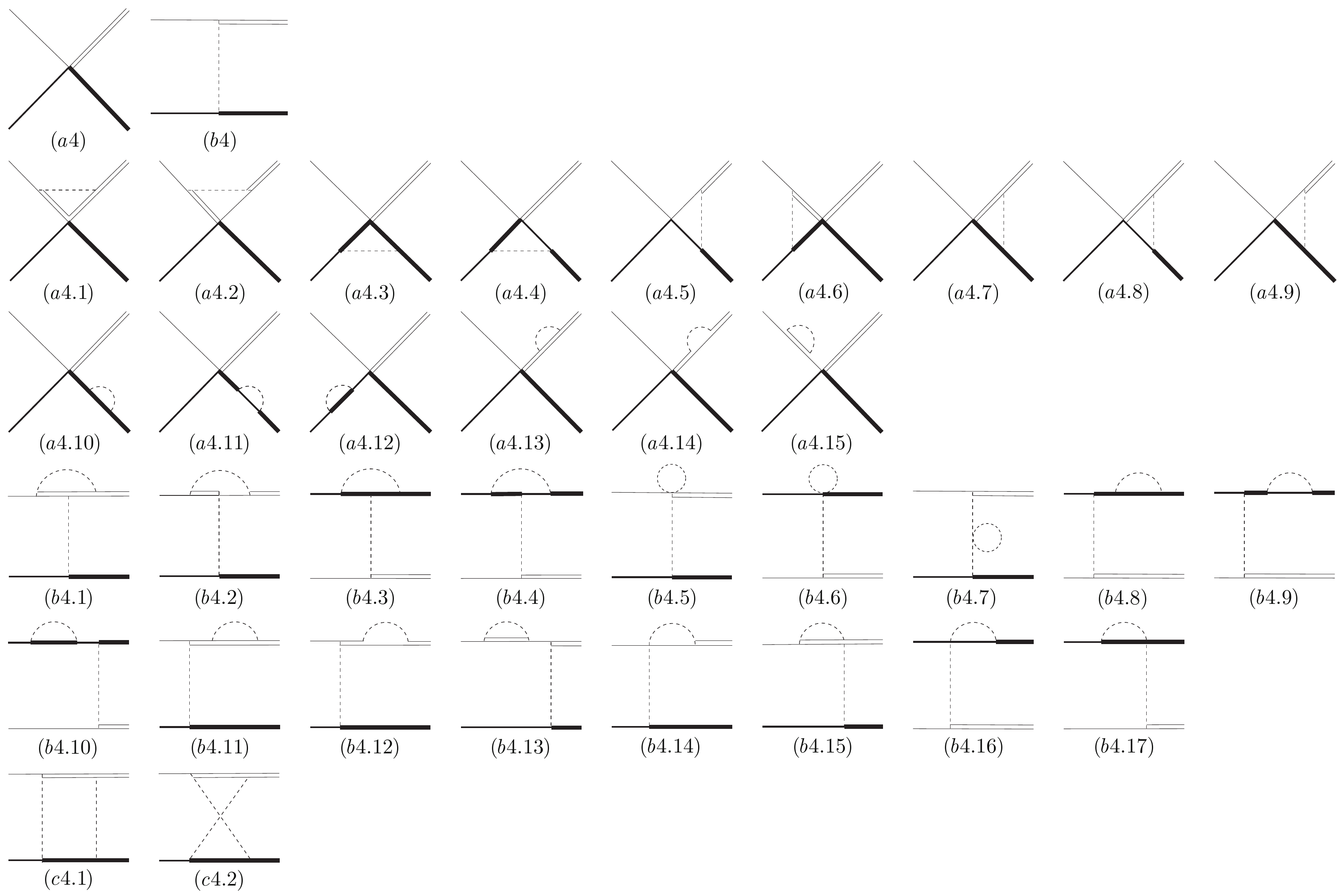}
     \captionsetup{justification=raggedright, singlelinecheck=false}
     \caption{LO contact ($a4$), LO OPE ($b4$), NLO contact ($a4.1-a4.15$), NLO OPE ($b4.1-b4.17$), and NLO TPE ($c4.1-c4.2$) diagrams of the process $D\bar{B} \to D^{*}\bar{B}^{*}$.}\label{BD-BsDs}
\end{figure*} 

For the $D(p_1)\bar{B}(p_2) \to D^{*}(p_3)\bar{B}^{*}(p_4)$ process, the corresponding diagrams are illustrated in Fig.~\ref{BD-BsDs}. The scattering amplitudes of LO diagrams are given by

\begin{align}
     \mathcal{M}_{a4}^{(0)}=&-4(D_a+E_a)(\epsilon_{3}^{*} \cdot \epsilon^{*}_{4}),
\end{align}

\begin{align}
     \mathcal{M}_{b4}^{(0)}=&~\frac{g_{1}g_{2}}{f^{2}}\frac{q_{\mu}q_{\nu}}{q^{2}-m^{2}}\epsilon_{3}^{*\mu}(p_{3}) \epsilon^{*\nu}_{4}(p_{4}),
\end{align}
with isospin $I=1$, and    
\begin{align}
     \mathcal{M}_{a4}^{(0)}=&-4(D_a-3E_a)(\epsilon_{3}^{*} \cdot \epsilon^{*}_{4}),
\end{align}

\begin{align}
     \mathcal{M}_{b4}^{(0)}=&-3\frac{g_{1}g_{1}}{f^{2}}\frac{q_{\mu}q_{\nu}}{q^{2}-m^{2}}\epsilon_{3}^{*\mu}(p_{3}) \epsilon^{*\nu}_{4}(p_{4}),
\end{align}
with isospin $I=0$, where $\epsilon_{3}^{*}$ and $\epsilon^{*}_{4}$ are the polarization vectors of the final $D^{*}$ and $\bar{B}^{*}$, respectively. 

The one-loop corrections to the contact amplitudes are listed as

\begin{align}
     \mathcal{M}_{a4.1}^{(2)}=& -4\frac{g_{2}^{2}}{f^{2}}(d-3)(d-2)AJ_{22}^{g}(\epsilon_{3}^{*} \cdot \epsilon^{*}_{4}),
\end{align}

\begin{align}
     \mathcal{M}_{a4.2}^{(2)}=&~ 4\frac{g_{2}^{2}}{f^{2}}AJ_{22}^{g}(\epsilon_{3}^{*} \cdot \epsilon^{*}_{4}),
\end{align}

\begin{align}
     \mathcal{M}_{a4.3}^{(2)}=& -4\frac{g_{1}^{2}}{f^{2}}(d-3)(d-2)AJ_{22}^{g}(\epsilon_{3}^{*} \cdot \epsilon^{*}_{4}),
\end{align}

\begin{align}
     \mathcal{M}_{a4.4}^{(2)}=&~ 4\frac{g_{1}^{2}}{f^{2}}AJ_{22}^{g}(\epsilon_{3}^{*} \cdot \epsilon^{*}_{4}),
\end{align}

\begin{align}
     \mathcal{M}_{a4.5}^{(2)}=& -4\frac{g_{1}g_{2}}{f^{2}}AJ_{22}^{h}(\epsilon_{3}^{*} \cdot \epsilon^{*}_{4}),
\end{align}

\begin{align}
     \mathcal{M}_{a4.6}^{(2)}=& -4\frac{g_{1}g_{2}}{f^{2}}AJ_{22}^{h}(\epsilon_{3}^{*} \cdot \epsilon^{*}_{4}),
\end{align}

\begin{align}
     \mathcal{M}_{a4.7}^{(2)}=& -4\frac{g_{1}g_{2}}{f^{2}}(d-3)(d-2)AJ_{22}^{h}(\epsilon_{3}^{*} \cdot \epsilon^{*}_{4}),
\end{align}

\begin{align}
     \mathcal{M}_{a4.8}^{(2)}=&~ \mathcal{M}_{a4.9}^{(2)}=0,
\end{align}

\begin{align}
     \mathcal{M}_{a4.(10+11)}^{(2)}=& ~\frac{3g_{1}^{2}}{2f^{2}}A [(d-2)\partial_{\omega} J_{22}^{b} (\omega_{1})+\partial_{\omega} J_{22}^{b}(\omega_{2})] \notag \\  & \times (\epsilon^{*}_{3} \cdot \epsilon^{*}_{4}),
\end{align}

\begin{align}
     \mathcal{M}_{a4.12}^{(2)}=& ~\frac{3g_{1}^{2}}{2f^{2}}(d-1)A \partial_{\omega} J_{22}^{b} (\omega_{1}) (\epsilon^{*}_{3} \cdot \epsilon^{*}_{4}),
\end{align}

\begin{align}
     \mathcal{M}_{a4.(13+14)}^{(2)}=& ~\frac{3g_{2}^{2}}{2f^{2}}A [(d-2)\partial_{\omega} J_{22}^{b} (\omega_{1})+\partial_{\omega} J_{22}^{b}(\omega_{2})] \notag \\  & \times (\epsilon^{*}_{3} \cdot \epsilon^{*}_{4}),
\end{align}

\begin{align}
     \mathcal{M}_{a4.15}^{(2)}=&~\frac{3g_{2}^{2}}{2f^{2}}(d-1)A \partial_{\omega} J_{22}^{b} (\omega_{1}) (\epsilon^{*}_{3} \cdot \epsilon^{*}_{4}),
\end{align}
where coefficients appearing in the above contact amplitudes are shown in Table~\ref{13}.
The NLO OPE diagrams are illustrated in Fig.~\ref{BD-BsDs}, and the corresponding amplitudes
\begin{align}
     \mathcal{M}_{b4.1}^{(2)}=& -4\frac{g_{1}g_{2}^{3}}{f^{2}}A  J_{22}^{g}\frac{q_{\mu}q_{\nu}}{q^{2}-m^{2}}\epsilon^{*\mu}_{3}\epsilon^{*\nu}_{4},
\end{align}

\begin{align}
     \mathcal{M}_{b4.2}^{(2)}=&~ 4\frac{g_{1}g_{2}^{3}}{f^{2}}(d-3)(d-2)A J_{22}^{g}\frac{q_{\mu}q_{\nu}}{q^{2}-m^{2}}\epsilon^{*\mu}_{3}\epsilon^{*\nu}_{4},
\end{align}

\begin{align}
     \mathcal{M}_{b4.3}^{(2)}=& -4\frac{g_{1}^{3}g_{2}}{f^{2}}A J_{22}^{g}\frac{q_{\mu}q_{\nu}}{q^{2}-m^{2}}\epsilon^{*\mu}_{3}\epsilon^{*\nu}_{4}
\end{align}

\begin{align}
     \mathcal{M}_{b4.4}^{(2)}=&~ 4\frac{g_{1}^{3}g_{2}}{f^{2}}(d-3)(d-2)A J_{22}^{g}\frac{q_{\mu}q_{\nu}}{q^{2}-m^{2}}\epsilon^{*\mu}_{3}\epsilon^{*\nu}_{4},
\end{align}

\begin{align}
     \mathcal{M}_{b4.5}^{(2)}=&~ 4\frac{g_{1}g_{2}}{f^{2}}A J_{0}^{c}\frac{q_{\mu}q_{\nu}}{q^{2}-m^{2}}\epsilon^{*\mu}_{3}\epsilon^{*\nu}_{4},
\end{align}

\begin{align}
     \mathcal{M}_{b4.6}^{(2)}=&~ 4\frac{g_{1}g_{2}}{f^{2}}A J_{0}^{c}\frac{q_{\mu}q_{\nu}}{q^{2}-m^{2}}\epsilon^{*\mu}_{3}\epsilon^{*\nu}_{4},
\end{align}

\begin{align}
     \mathcal{M}_{b4.7}^{(2)}=&~ \frac{8g_{1}g_{2}}{3f^{2}}A [2m^{2}L+\frac{m^{2}}{8\pi^{2}}{\rm{log}}(\frac{m}{\mu})]\frac{q_{\mu}q_{\nu}}{q^{2}-m^{2}} \notag \\ 
     &\times\epsilon^{*\mu}_{3}\epsilon^{*\nu}_{4},
\end{align}

\begin{align}
     \mathcal{M}_{b4.(8+9)}^{(2)}=& -4\frac{g_{1}^{3}g_{2}}{f^{2}}A [(d-2)\partial_{\omega} J_{22}^{b}(\omega_{1})+\partial_{\omega} J_{22}^{b}(\omega_{2})] \notag \\ 
     &\times \frac{q_{\mu}q_{\nu}}{q^{2}-m^{2}} \epsilon^{*\mu}_{3}\epsilon^{*\nu}_{4},
\end{align}
    
\begin{align}
     \mathcal{M}_{b4.10}^{(2)}=& -\frac{3g_{1}^{3}g_{2}}{2f^{2}}(d-1)A \partial_{\omega} J_{22}^{b}(\omega_{1}) \frac{q_{\mu}q_{\nu}}{q^{2}-m^{2}} \notag \\ 
     &\times \epsilon^{*\mu}_{3}\epsilon^{*\nu}_{4},
\end{align}

\begin{align}
     \mathcal{M}_{b4.(11+12)}^{(2)}=& -4\frac{g_{1}g_{2}^{3}}{f^{2}}A [(d-2)\partial_{\omega} J_{22}^{b}(\omega_{1})+\partial_{\omega} J_{22}^{b}(\omega_{2})] \notag \\ 
     &\times \frac{q_{\mu}q_{\nu}}{q^{2}-m^{2}} \epsilon^{*\mu}_{3}\epsilon^{*\nu}_{4},
\end{align}

\begin{align}
     \mathcal{M}_{b4.13}^{(2)}=&-\frac{3g_{1}g_{2}^{3}}{2f^{2}} (d-1)A \partial_{\omega} J_{22}^{b}(\omega_{1}) \frac{q_{\mu}q_{\nu}}{q^{2}-m^{2}} \notag \\ 
     &\times\epsilon^{*\mu}_{3}\epsilon^{*\nu}_{4},
\end{align}

\begin{align}
     \mathcal{M}_{b4.14}^{(2)}=&~\frac{8g_{1}g_{2}}{f^{2}}A   [2m^{2}L+\frac{m^{2}}{8\pi^{2}}{\rm{log}}(\frac{m}{\mu})]\frac{q_{\mu}q_{\nu}}{q^{2}-m^{2}}\notag \\ 
     &\times\epsilon^{*\mu}_{3}\epsilon^{*\nu}_{4},
\end{align}

\begin{align}
     \mathcal{M}_{b4.15}^{(2)}=&~\frac{8g_{1}g_{2}}{f^{2}}A [2m^{2}L+\frac{m^{2}}{8\pi^{2}}{\rm{log}}(\frac{m}{\mu})]\frac{q_{\mu}q_{\nu}}{q^{2}-m^{2}} \notag \\ 
     &\times\epsilon^{*\mu}_{3}\epsilon^{*\nu}_{4},
\end{align}

\begin{align}
     \mathcal{M}_{b4.16}^{(2)}=&~\frac{8g_{1}g_{2}}{f^{2}}A [2m^{2}L+\frac{m^{2}}{8\pi^{2}}{\rm{log}}(\frac{m}{\mu})]\frac{q_{\mu}q_{\nu}}{q^{2}-m^{2}} \notag \\ 
     &\times\epsilon^{*\mu}_{3}\epsilon^{*\nu}_{4},
\end{align}
    
\begin{align}
     \mathcal{M}_{b4.17}^{(2)}=&~\frac{8g_{1}g_{2}}{f^{2}}A [2m^{2}L+\frac{m^{2}}{8\pi^{2}}{\rm{log}} (\frac{m}{\mu})]\frac{q_{\mu}q_{\nu}}{q^{2}-m^{2}} \notag \\ 
     &\times\epsilon^{*\mu}_{3}\epsilon^{*\nu}_{4}.
\end{align}

The TPE amplitudes at NLO are given by
\begin{align}
   \mathcal{M}_{c4.1}^{(2)}=&-4\frac{g_{1}^2g_{2}^2}{f^{4}}(d-3)A [(q \cdot \epsilon^{*}_{3})(q \cdot \epsilon^{*}_{4}) \notag \\ 
   &+q^{2}(\epsilon^{*}_{3} \cdot \epsilon^{*}_{4})]J^{B}_{21}(-\delta_{1},-\delta_{2}),
\end{align}

\begin{align}
   \mathcal{M}_{c4.2}^{(2)}=&~4\frac{g_{1}^2g_{2}^2}{f^{4}}(d-3)A [(q \cdot \epsilon^{*}_{3})(q \cdot \epsilon^{*}_{4}) \notag \\ 
   &+q^{2}(\epsilon^{*}_{3} \cdot \epsilon^{*}_{4})]J^{R}_{21}(-\delta_{1},-\delta_{2}).
\end{align}

\renewcommand{\arraystretch}{1.5}
\begin{center}
    \begin{table}[ht]
          \centering
          \captionsetup{justification=raggedright, singlelinecheck=false}
          \caption{The coefficients in the NLO contact amplitudes for the $D\bar{B} \to D^{*}\bar{B}^{*}$ process.}\label{13}
    	\setlength{\tabcolsep}{2.2mm}{
    	\begin{tabular}{ccccrr}
     		\toprule[1pt]
    		& { $I=1$  } & { $I=0$ } &    & \\
    		&  $A$     &    $A$    &   $\omega_1$   & $\omega_2$ \\
    		\midrule[1pt]
    		$A_{a4.1(2)}$ & $0$  & $0$  & $-\delta_{2}$ & $0(\delta_{2})$ \\
    		$A_{a4.3(4)}$ & $0$  & $0$  & $-\delta_{1}$ & $0(\delta_{1})$ \\
    		$A_{a4.5}$ & $\frac{1}{4}(D_{b}+E_{b})$  & $\frac{-3}{4}(D_{b}-3E_{b})$  & $\delta_{1}$ & $\delta_{2}$ \\
    		$A_{a4.6}$ & \makecell[c]{$\frac{1}{4}(D_{a}+E_{a}$\\$+2D_{b}+2E_{b})$}  & \makecell[c]{$\frac{-3}{4}(D_{a}-3E_{a}$\\$+2D_{b}-6E_{b})$}  & $-\delta_{1}$ & $-\delta_{2}$ \\
    		$A_{a4.7}$ & $0$  & $0$  & $0$ & $0$ \\
    		$A_{a4.(10+11)}$ & $D_{b}+E_{b}$  & $D_{b}-3E_{b}$  & $0$ & $\delta_{1}$ \\
                $A_{a4.12}$ & $D_{b}+E_{b}$  & $D_{b}-3E_{b}$  & $-\delta_{1}$ & $0$ \\
                $A_{a4.(13+14)}$ & $D_{b}+E_{b}$  & $D_{b}-3E_{b}$  & $\delta_{2}$ & $0$  \\
                $A_{a4.15}$ & $D_{b}+E_{b}$  & $D_{b}-3E_{b}$  & $0$ & $-\delta_{2}$  \\
    		\bottomrule[1pt]
    	\end{tabular}}
    \end{table}
\end{center}

\renewcommand{\arraystretch}{1.5}
\begin{center}
    \begin{table}[b]
          \centering
          \captionsetup{justification=raggedright, singlelinecheck=false}
          \caption{The coefficients in the NLO OPE amplitudes for the $D\bar{B} \to D^{*}\bar{B}^{*}$ process .} \label{14}
    	\setlength{\tabcolsep}{4mm}{
    	\begin{tabular}{ccccrr}
    		\toprule[1pt]
    		& { $I=1$  } & { $I=0$ } &    & \\
    		&  $A$     &    $A$    &   $\omega_1$   & $\omega_2$ \\
    		\midrule[1pt]
    		$A_{b4.1(2)}$ & $-\frac{1}{16}$  & $\frac{3}{16}$ & $-\delta_{2}$  & $0(\delta_{2})$  \\
    		$A_{b4.3(4)}$ & $-\frac{1}{16}$  & $\frac{3}{16}$  & $-\delta_{1}$  & $0(\delta_{1})$ \\
    		$A_{b4.5(6)}$ & $-\frac{1}{12}$  & $\frac{1}{4}$  & $0$ & $0$ \\
    		$A_{b4.7}$ & $\frac{1}{4}$  & $-\frac{3}{4}$  & $0$ & $0$ \\
    		$A_{b4.(8+9)}$ & $\frac{1}{4}$  & $-\frac{3}{4}$  & $0$ & $\delta_{1}$ \\
    		$A_{b4.10}$ & $\frac{1}{4}$  & $-\frac{3}{4}$  & $-\delta_{1}$ & $0$ \\
                $A_{b4.(11+12)}$ & $\frac{1}{4}$  & $-\frac{3}{4}$  & $0$ & $\delta_{2}$ \\
    	    $A_{b4.13}$ & $\frac{1}{4}$  & $-\frac{3}{4}$  & $-\delta_{2}$ & $0$ \\
                $A_{b4.14(15)}$ & $\frac{1}{4}$  & $-\frac{3}{4}$  & $0$ & $0$ \\
                $A_{b4.16(17)}$ & $\frac{1}{4}$  & $-\frac{3}{4}$  & $0$ & $0$ \\
    		\bottomrule[1pt]
    	\end{tabular}}
    \end{table}
\end{center}
In Table~\ref{14}, we list the coefficients of the OPE amplitudes in the $D\bar{B} \to D^{*}\bar{B}^{*}$ process. For the TPE amplitudes, the coefficients are
\begin{align}
    A_{c4.1}=1/16, \quad A_{c4.2}=5/16,
\end{align}
with $I=1$, and
\begin{align}
    A_{c4.1}=9/16, \quad A_{c4.2}=-3/16,
\end{align}
with $I=0$.

\subsection{$D^{*}\bar{B} \to D^{*}\bar{B}$}
\label{A2}

\begin{figure*}[ht]
     \centering
     \includegraphics[width=0.6\textwidth]{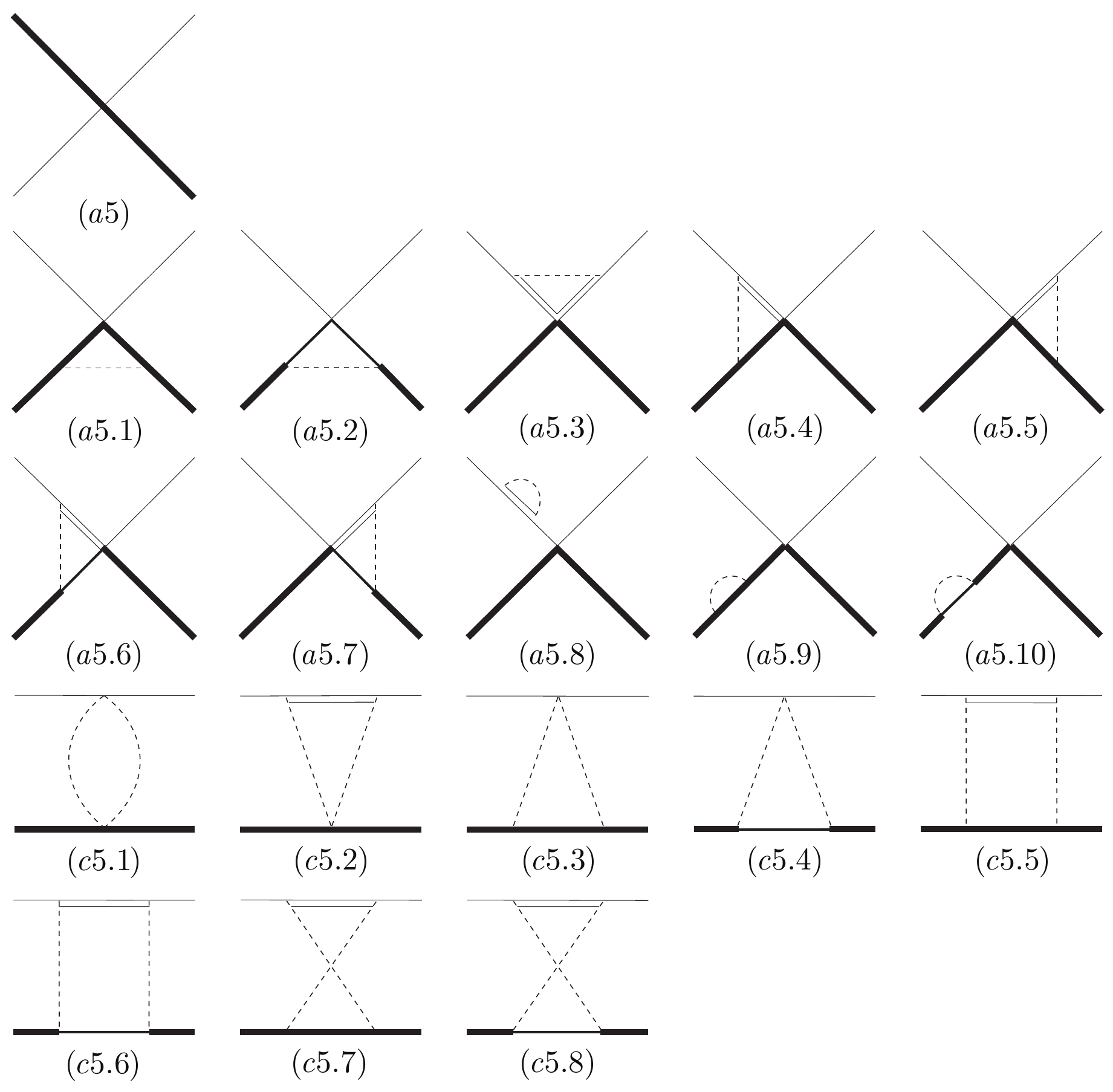}
     \captionsetup{justification=raggedright, singlelinecheck=false}
     \caption{LO contact ($a5$), NLO contact ($a5.1-a5.10$) and NLO TPE ($c5.1-c5.8$) diagrams of the process $D^{*}\bar{B} \to D^{*}\bar{B}$.}\label{BDs}
\end{figure*} 

For the $D^{*}(p_1)\bar{B}(p_2) \to D^{*}(p_3) \bar{B}(p_4) $ process, the scattering amplitude of the LO contact diagram is
\begin{align}
    \mathcal{M}_{a5}^{(0)}= -4(D_{a}+E_{a})(\epsilon_{1} \cdot \epsilon^{*}_{3}),
\end{align}
with isospin $I=1$, and
\begin{align}
    \mathcal{M}_{a5}^{(0)}= -4(D_{a}-3E_{a})(\epsilon_{1} \cdot \epsilon^{*}_{3}),
\end{align}
with isospin $I=0$, where $\epsilon_{1}$ and $\epsilon^{*}_{4}$ are the polarization vectors of the initial $D$ and final $\bar{B}^{*}$, respectively.
Next, we consider the NLO interactions in the $D^{*}(p_1)\bar{B}(p_2) \to D^{*}(p_3) \bar{B}^(p_4) $ process. The corresponding diagrams are illustrated in Fig.~\ref{BDs}. The one-loop corrections to the contact term are
\begin{align}
    \mathcal{M}_{a5.1}^{(2)}=&~ 4\frac{g_{1}^{2}}{f^{2}}(d-3)(d-2)A J_{22}^{g}(\epsilon_{1} \cdot \epsilon^{*}_{3}), 
\end{align}

\begin{align}
    \mathcal{M}_{a5.2}^{(2)}=& -4\frac{g_{1}^{2}}{f^{2}}A J_{22}^{g} (\epsilon_{1} \cdot \epsilon^{*}_{3}),
\end{align}

\begin{align}
    \mathcal{M}_{a5.3}^{(2)}=&~ 4\frac{g_{2}^{2}}{f^{2}}(d-1)A J_{22}^{g}(\epsilon_{1} \cdot \epsilon^{*}_{3}), 
\end{align}

\begin{align}
    \mathcal{M}_{a5.4}^{(2)}=& -4\frac{g_{1}g_{2}}{f^{2}}(d-3)(d-2)A J_{22}^{h} (\epsilon_{1} \cdot \epsilon^{*}_{3}), 
\end{align}

\begin{align}
    \mathcal{M}_{a5.5}^{(2)}=& -4\frac{g_{1}g_{2}}{f^{2}}(d-3)(d-2)A J_{22}^{h} (\epsilon_{1} \cdot \epsilon^{*}_{3}), 
\end{align}

\begin{align}
    \mathcal{M}_{a5.6}^{(2)}=& -4\frac{g_{1}g_{2}}{f^{2}}A J_{22}^{h} (\epsilon_{1} \cdot \epsilon^{*}_{3}), 
\end{align}

\begin{align}
    \mathcal{M}_{a5.7}^{(2)}=& -4\frac{g_{1}g_{2}}{f^{2}}A J_{22}^{h} (\epsilon_{1} \cdot \epsilon^{*}_{3}), 
\end{align}

\begin{align}
    \mathcal{M}_{a5.8}^{(2)}=&~ \frac{3g_{2}^{2}}{2f^{2}}(d-1) A \partial_{\omega} J_{22}^{b} (\epsilon_{1} \cdot \epsilon^{*}_{3}), 
\end{align}

\begin{align}
    \mathcal{M}_{a5.(9+10)}^{(2)}=& ~\frac{3g_{1}^{2}}{2f^{2}}A [(d-2)\partial_{\omega} J_{22}^{b} (\omega_{1})+\partial_{\omega} J_{22}^{b}(\omega_{2})] \notag \\ 
    & \times (\epsilon_{1} \cdot \epsilon^{*}_{3}), 
\end{align}
where coefficients appearing in the above contact amplitudes are shown in Table~\ref{15}. The TPE amplitudes at NLO are given by

\begin{align}
    \mathcal{M}_{c5.1}^{(2)}=&~ 4\frac{1}{f^{4}}[A_1(q_{0}^{2} J_{0}^{F}+J_{22}^F) +A_{15}(q_{0}^{2} J_{11}^{F}+q_{0}^{2} J_{21}^F+J_{22}^F) \notag \\
    & -A_{51}(q_{0}^{2} J_{11}^{F}+q_{0}^{2} J_{21}^F+J_{22}^F)+A_{5}(q_{0}^{2} J_{0}^{F}+2q_{0}^{2}J_{11}^{F}  \notag \\
    &   + q_{0}^{2} J_{21}^F+J_{22}^F)]\epsilon_{1} \cdot \epsilon^{*}_{3},  
\end{align}

\begin{align}
    \mathcal{M}_{c5.2}^{(2)}=& -i4\frac{g_{1}^{2}}{f^{4}}\{A_{1}[q_{0}\vec{q}^{2} J_{22}^{T}+\vec{q}^{2}J_{24}^{T} +(1-d)q_{0}J_{31}^{T}  \notag \\
    & +q_{0}\vec{q}^{2} J_{32}^{T} +\vec{q}^{2}J_{33}^{T}+(1-d) J_{34}^{T}]  \notag \\
    & -A_{5}[q_{0}\vec{q}^{2} J_{11}^{T} +(1-d)q_{0} (J_{21}^{T}+ J_{31}^{T})\notag \\
    & +2q_{0}\vec{q}^{2} J_{22}^{T} +\vec{q}^{2}J_{24}^{T} +q_{0}\vec{q}^{2} J_{32}^{T}+\vec{q}^{2}J_{33}^{T} \notag \\
    &  +(1-d) J_{34}^{T} ]\}(\epsilon_{1} \cdot \epsilon^{*}_{3}), 
\end{align}

\begin{align}
    \mathcal{M}_{c5.3}^{(2)}=& -i4\frac{g_{1}^{2}}{f^{4}}(d-3)[A_{1}(q_{0} J_{22}^{S}+ J_{24}^{S}+q_{0} J_{32}^{S}+ J_{33}^{S}) \notag \\
    & -A_{5}(q_{0} J_{11}^{S}+2q_{0} J_{22}^{S}+J_{24}^{S} \notag \\
    &  +q_{0} J_{32}^{S}+J_{33}^{S})](q \cdot \epsilon_{1})( q \cdot \epsilon^{*}_{3}) \notag \\
    & -i4\frac{g_{1}^{2}}{f^{4}}(d-3)\{A_{1}[q_{0}\vec{q}^{2} J_{22}^{S}+\vec{q}^{2} J_{24}^{S} \notag \\
    & +(2-d)q_{0} J_{31}^{S} +q_{0}\vec{q}^{2} J_{32}^{S}+\vec{q}^{2} J_{33}^{S}+(2-d) J_{34}^{S}] \notag \\
    & -A_{5}[q_{0}\vec{q}^{2} J_{11}^{S}+(2-d)q_{0} (J_{21}^{S}+J_{31}^{S}) \notag \\
    & +2q_{0}\vec{q}^{2} J_{22}^{S} +\vec{q}^{2} J_{24}^{S}  +q_{0}\vec{q}^{2} J_{32}^{S} + \vec{q}^{2} J_{33}^{S} \notag \\
    & +(2-d) J_{34}^{S}]\}(\epsilon_{1} \cdot \epsilon^{*}_{3}),
\end{align}

\begin{align}
    \mathcal{M}_{c5.4}^{(2)}=&~ i4\frac{g_{1}^{2}}{f^{4}} \{[A_{1}(q_{0} J_{22}^{S}+J_{24}^{S} +q_{0} J_{32}^{S}+J_{33}^{S}) \notag \\
    & -A_{5}(q_{0} J_{11}^{S}  +2q_{0} J_{22}^{S}+J_{24}^{S}+q_{0} J_{32}^{S} \notag \\
    & +J_{33}^{S})](q \cdot \epsilon_{1})( q \cdot \epsilon^{*}_{3}) +[A_{1}(q_{0}J_{31}^{S}+J_{34}^{S}) \notag \\
    & -A_{5}(q_{0} J_{21}^{S}+q_{0} J_{31}^{S} +J_{34}^{S})](\epsilon_{1} \cdot \epsilon^{*}_{3})\}, 
\end{align}

\begin{align}
    \mathcal{M}_{c5.5}^{(2)}=&~ 4\frac{g_{1}^{2}g_{2}^{2}}{f^{4}}(d-3) A_{1} \{[J_{21}^{B}-\vec{q}^{2} J_{22}^{B}+(d+3)J_{31}^{B} \notag \\
    & -2\vec{q}^{2} J_{32}^{B}+(d+3)J_{42}^{B}-\vec{q}^{2} J_{43}^{B}](q\cdot\epsilon_{1})(q \cdot \epsilon^{*}_{3}) \notag \\
    & +[-\vec{q}^{2} J_{21}^{B}+\vec{q}^{4} J_{22}^{B} -(2d+1)\vec{q}^{2} (J_{31}^{B}+J_{42}^{B}) \notag \\
    &  +2\vec{q}^{4} J_{32}^{B}+(d+1)(d-2) J_{41}^{B} +\vec{q}^{4} J_{43}^{B}](\epsilon_{1} \cdot \epsilon^{*}_{3})\},  
\end{align}
    
\begin{align}
    \mathcal{M}_{c5.6}^{(2)}=&~ 4\frac{g_{1}^{2}g_{2}^{2}}{f^{4}}A_{1}\{[J_{21}^{B}-\vec{q}^{2} J_{22}^{B} +(d+3)(J_{31}^{B}+J_{42}^{B}) \notag \\
    &  -2\vec{q}^{2} J_{32}^{B} -\vec{q}^{2} J_{43}^{B}](q \cdot \epsilon_{1})(q \cdot \epsilon^{*}_{3}) \notag \\
    & -[\vec{q}^{2} (J_{31}^{B}+ J_{42}^{B})-(1+d) J_{41}^{B}](\epsilon_{1} \cdot \epsilon^{*}_{3})\}, 
\end{align}

\begin{align}
    \mathcal{M}_{c5.7}^{(2)}=&~ 4\frac{g_{1}^{2}g_{2}^{2}}{f^{4}}(d-3)A_{1}\{[J_{21}^{R}-\vec{q}^{2} J_{22}^{R}+(d+3)J_{31}^{R} \notag \\
    &  -2\vec{q}^{2} J_{32}^{B}+3J_{42}^{B}-\vec{q}^{2} J_{43}^{B}](q\cdot \epsilon_{1})(q \cdot \epsilon^{*}_{3})\notag \\
    & +[-\vec{q}^{2} J_{21}^{R} +\vec{q}^{4} J_{22}^{R} -(2d+1)\vec{q}^{2} (J_{31}^{R}+J_{42}^{R})\notag \\
    & +2\vec{q}^{4} J_{32}^{R} +(d+1)(d-2) J_{41}^{R} +\vec{q}^{4} J_{43}^{R}](\epsilon_{1} \cdot \epsilon^{*}_{3})\}, 
\end{align}

\begin{align}
    \mathcal{M}_{c5.8}^{(2)}=&~ 4\frac{g_{1}^{2}g_{2}^{2}}{f^{4}}A_{1}\{[J_{21}^{R}-\vec{q}^{2} J_{22}^{R}+(d+3)(J_{31}^{R}+ J_{42}^{R})  \notag \\
    &-2\vec{q}^{2} J_{32}^{R}-\vec{q}^{2} J_{43}^{R}](q\cdot\epsilon_{1})(q \cdot \epsilon^{*}_{3}) \notag \\
    & -[\vec{q}^{2} J_{31}^{R} -(1+d) J_{41}^{R} +\vec{q}^{2} J_{42}^{R}](\epsilon_{1} \cdot \epsilon^{*}_{3})\},
\end{align}

where coefficients appearing in the above contact amplitudes are shown in Table~\ref{16}.

\renewcommand{\arraystretch}{1.5}
\begin{center}
    \begin{table}[hb]
          \centering
          \captionsetup{justification=raggedright, singlelinecheck=false}
          \caption{The coefficients in the NLO contact amplitudes for the $D^{*}\bar{B} \to D^{*} \bar{B} $ process.}\label{15}
    	\setlength{\tabcolsep}{2.5mm}{
    	\begin{tabular}{ccccrr}
    		\toprule[1pt]
    		& { $I=1$  } & { $I=0$ } &    & \\
    		&  $A$     &    $A$    &   $\omega_1$   & $\omega_2$ \\
    		\midrule[1pt]
    		$A_{a5.1}$ & $0$  & $0$  & $0$ & $0$ \\
    		$A_{a5.2}$ & $0$  & $0$  & $\delta_{1}$ & $\delta_{1}$ \\
    		$A_{a5.3}$ & $0$  & $0$  & $-\delta_{2}$ & $-\delta_{2}$ \\
    		$A_{a5.4}$ & $\frac{1}{4}(D_{b}+E_{b})$  & $\frac{-3}{4}(D_{b}-3E_{b})$  & $-\delta_{2}$ & $0$ \\
    		$A_{a5.5}$ & $\frac{1}{4}(D_{b}+E_{b})$  & $\frac{-3}{4}(D_{b}-3E_{b})$  & $0$ & $-\delta_{2}$ \\
    		$A_{a5.6}$ & $\frac{1}{4}(D_{b}+E_{b})$  & $\frac{-3}{4}(D_{b}-3E_{b})$  & $-\delta_{2}$ & $\delta_{1}$ \\
    		$A_{a5.7}$ & $\frac{1}{4}(D_{b}+E_{b})$  & $\frac{-3}{4}(D_{b}-3E_{b})$  & $\delta_{1}$ & $-\delta_{2}$ \\
    		$A_{a5.8}$ & $D_{a}+E_{a}$  & $D_{a}-3E_{a}$  & $0$ & $-\delta_{2}$ \\
    		$A_{a5.(9+10)}$ & $D_{a}+E_{a}$  & $D_{a}-3E_{a}$  & $ \delta_{1}$ & $0$ \\
    		\bottomrule[1pt]
    	\end{tabular}}
    \end{table}
\end{center}

\renewcommand{\arraystretch}{1.5}
\begin{center}
    \begin{table*}[ht]
    \centering
    \captionsetup{justification=raggedright, singlelinecheck=false}
    \caption{The coefficients in the TPE amplitudes for the $D^{*}(p_1)\bar{B}(p_2) \to D^{*}(p_3) \bar{B}^(p_4) $ process. Note that we have $A_{51}=A_{15}$.}\label{16}
    	\setlength{\tabcolsep}{7.0mm}{
    	\begin{tabular}{ccrcrrccc}	
    		\toprule[1pt]
    		&  \multicolumn{3}{c}{ $I=1$ } &\multicolumn{3}{c}{ $I=0$ } &       & \\
    		\cline{2-4} \cline{5-7}
    		&  $A_1$   &  $A_5$   &  $A_{15}$   &   $A_1$   &  $A_5$   &  $A_{15}$   &  $\omega_1$   & $\omega_2$ \\
    		\midrule[1pt]
    		$A_{c5.1}$ & $\frac{1}{16} $  & $\frac{1}{16} $ & $-\frac{1}{16}$ & $-\frac{3}{16}$ & $-\frac{3}{16}$  & $\frac{3}{16}$ & $0$       & $0$ \\
    		$A_{c5.2}$ & $\frac{i}{8}$  & $\frac{-i}{8}$ & $0$              & $-\frac{3i}{8}$ & $\frac{3i}{8}$  & $0$ & $-\delta_{2}$  &  $0$ \\
    		$A_{c5.3}$ & $\frac{i}{8}$  & $-\frac{i}{8}$ & $0$              & $-\frac{3i}{8}$ & $\frac{3i}{8}$  & $0$ &  $0$      & $0$ \\
    		$A_{c5.4}$ & $\frac{i}{8}$  & $-\frac{i}{8}$ & $0$              & $-\frac{3i}{8}$ & $\frac{3i}{8}$  & $0$ &  $\delta_{1}$      & $0$ \\
    		$A_{c5.5}$ & $\frac{1}{16}$  & $0$             & $0$             & $\frac{9}{16}$ & $0$            & $0$ & $0$ & $-\delta_{2}$ \\
    		$A_{c5.6}$ & $\frac{1}{16}$  & $0$             & $0$             & $\frac{9}{16}$ & $0$            & $0$  &  $\delta_{1}$ & $-\delta_{2}$ \\
    		$A_{c5.7}$& $\frac{5}{16}$  & $0$             & $0$             & $-\frac{3}{16}$ & $0$            & $0$ & $0$ & $-\delta_{2}$ \\
    		$A_{c5.8}$& $\frac{5}{16}$  & $0$             & $0$             & $-\frac{3}{16}$ & $0$            & $0$ & $\delta_{1}$ & $-\delta_{2}$ \\
    		\bottomrule[1pt]
    	\end{tabular}}
    \end{table*}
\end{center}

\subsection{$D\bar{B}^{*} \to D^{*}\bar{B}$}
\label{A3}
 
\begin{figure*}[ht]
     \centering
     \includegraphics[width=1.0\textwidth]{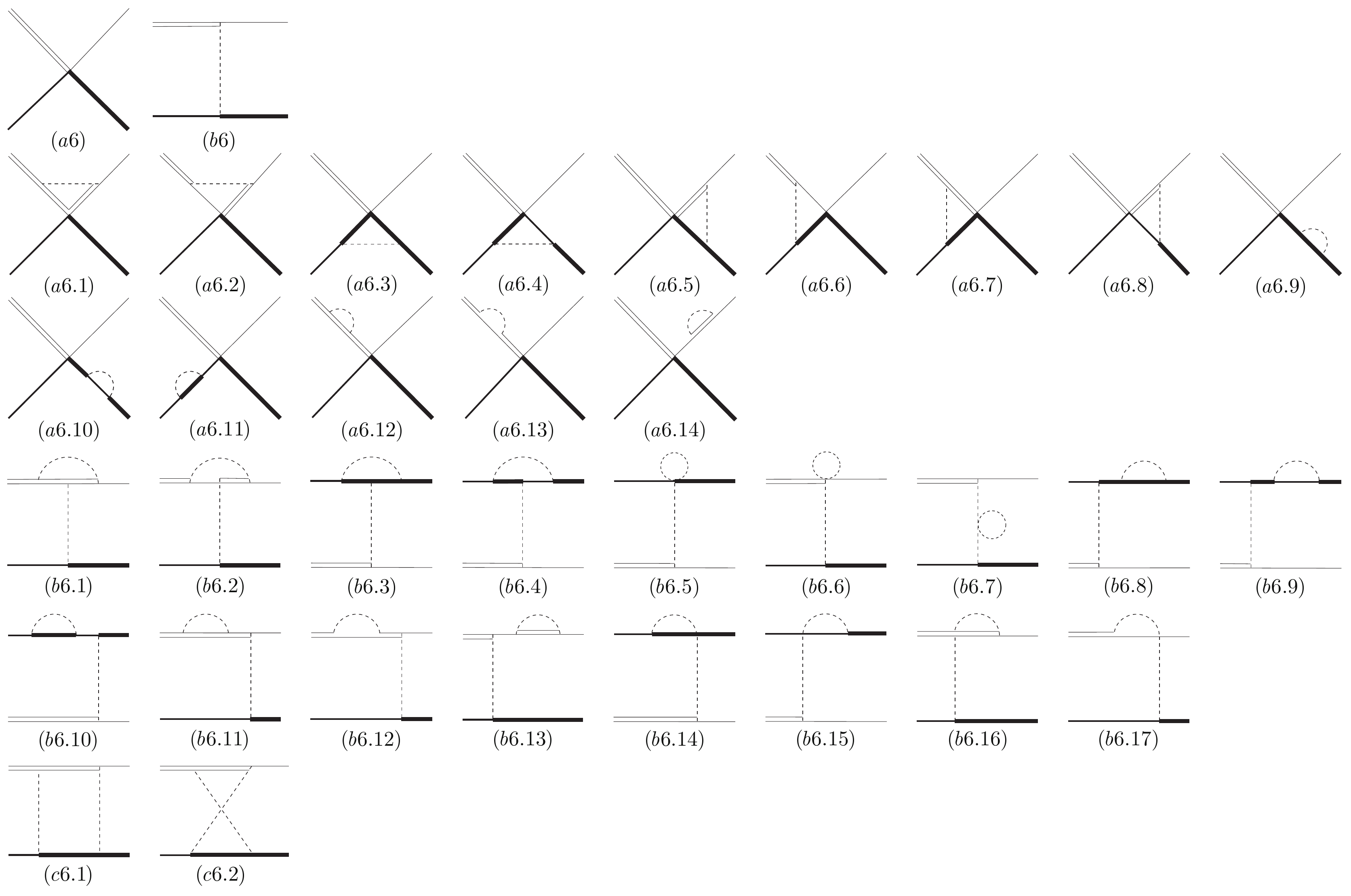}
     \captionsetup{justification=raggedright, singlelinecheck=false}
     \caption{LO contact ($a6$), LO OPE ($b6$), NLO contact ($a6.1-a6.14$), NLO OPE ($b6.1-b6.17$), and NLO TPE ($c6.1-c6.2$) diagrams of the process $D\bar{B}^{*} \to D^{*}\bar{B}$.}\label{BsD-BDs}
\end{figure*}

Now we focus on the process $D(p_1)\bar{B}^{*}(p_2) \to D^{*}(p_3) \bar{B}(p_4) $, and the corresponding diagrams are illustrated in Fig.~\ref{BsD-BDs}. The scattering amplitudes of LO diagrams are given by
\begin{align}
    \mathcal{M}_{a6}^{(0)}=&~ 4(D_{b}+E_{b})(\epsilon_{2} \cdot \epsilon^{*}_{3}),
\end{align}

\begin{align}
    \mathcal{M}_{b6}^{(0)}=& -\frac{g_{1}g_{2}}{f^2}\frac{q_{\mu}q_{\nu}}{q^{2}-m^{2}} \epsilon_{2}^{\mu} \epsilon^{*\nu}_{3},
\end{align}
with isospin $I=1$, and
\begin{align}
    \mathcal{M}_{a6}^{(0)}=&~ 4(D_{b}-3E_{b})(\epsilon_{2} \cdot \epsilon^{*}_{3}),
\end{align}

\begin{align}
    \mathcal{M}_{b6}^{(0)}=&~ \frac{3g_{1}g_{2}}{f^2}\frac{q_{\mu}q_{\nu}}{q^{2}-m^{2}} \epsilon_{2}^{\mu} \epsilon^{*\nu}_{3},
\end{align}
with isospin $I=0$.

The one-loop corrections to the contact amplitudes are
\begin{align}
    \mathcal{M}_{a6.1}^{(2)}=&~ 4\frac{g_{2}^{2}}{f^{2}}(d-3)(d-2)A J_{22}^{g}(\epsilon_{2} \cdot \epsilon^{*}_{3}),
\end{align}

\begin{align}
    \mathcal{M}_{a6.2}^{(2)}=& -4\frac{g_{2}^{2}}{f^{2}}A J_{22}^{g} (\epsilon_{2} \cdot \epsilon^{*}_{3}), 
\end{align}

\begin{align}
    \mathcal{M}_{a6.3}^{(2)}=&~ 4\frac{g_{1}^{2}}{f^{2}}(d-3)(d-2)A J_{22}^{g}(\epsilon_{2} \cdot \epsilon^{*}_{3}), 
\end{align}

\begin{align}
    \mathcal{M}_{a6.4}^{(2)}=& -4\frac{g_{1}^2}{f^{2}}A J_{22}^{g} (\epsilon_{2} \cdot \epsilon^{*}_{3}), 
\end{align}

\begin{align}
    \mathcal{M}_{a6.5}^{(2)}=&~ 4\frac{g_{1}g_{2}}{f^{2}}(d-3)(d-2)A J_{22}^{h} (\epsilon_{2} \cdot \epsilon^{*}_{3}), 
\end{align}

\begin{align}
    \mathcal{M}_{a6.6}^{(2)}=&~ 4\frac{g_{1}g_{2}}{f^{2}}A J_{22}^{h} (\epsilon_{2} \cdot \epsilon^{*}_{3}), 
\end{align}

\begin{align}
    \mathcal{M}_{a6.7}^{(2)}=&~ 4\frac{g_{1}g_{2}}{f^{2}}(d-3)(d-2)A J_{22}^{h} (\epsilon_{2} \cdot \epsilon^{*}_{3}), 
\end{align}

\begin{align}
    \mathcal{M}_{a6.8}^{(2)}=&~ 4\frac{g_{1}g_{2}}{f^{2}}A J_{22}^{h} (\epsilon_{2} \cdot \epsilon^{*}_{3}), 
\end{align}

\begin{align}
    \mathcal{M}_{a6.(9+10)}^{(2)}=& -\frac{3g_{1}^{2}}{2f^{2}}A [(d-2)\partial_{\omega} J_{22}^{b} (\omega_{1})+\partial_{\omega} J_{22}^{b}(\omega_{2})] \notag \\ 
    & \times (\epsilon_{2} \cdot \epsilon^{*}_{3}), 
\end{align}

\begin{align}
    \mathcal{M}_{a6.11}^{(2)}=& -\frac{3g_{1}^{2}}{2f^{2}}(d-1) A \partial_{\omega} J_{22}^{b} (\epsilon_{2} \cdot \epsilon^{*}_{3}), 
\end{align}

\begin{align}
    \mathcal{M}_{a6.(12+13)}^{(2)}=& -\frac{3g_{2}^{2}}{2f^{2}}A [(d-2)\partial_{\omega} J_{22}^{b} (\omega_{1})+\partial_{\omega} J_{22}^{b}(\omega_{2})] \notag \\ 
    & \times (\epsilon_{2} \cdot \epsilon^{*}_{3}), 
\end{align}

\begin{align}
    \mathcal{M}_{a6.14}^{(2)}=& -\frac{3g_{2}^{2}}{2f^{2}}(d-1)A \partial_{\omega} J_{22}^{b} (\epsilon_{2} \cdot \epsilon^{*}_{3}).
\end{align}

From Fig.~\ref{BsD-BDs}, we can obtain the scattering amplitudes of NLO OPE diagrams,
\begin{align}
    \mathcal{M}_{b6.1}^{(2)}=& -4\frac{g_{1}g_{2}^{3}}{f^{2}}(d-3)(d-2)A J_{22}^{g}\frac{q_{\mu}q_{\nu}}{q^{2}-m^{2}} \epsilon_{2}^{\mu} \epsilon^{*\nu}_{3}, 
\end{align}

\begin{align}
    \mathcal{M}_{b6.2}^{(2)}=&~ 4\frac{g_{1}g_{2}^{3}}{f^{2}}A J_{22}^{g} \frac{q_{\mu}q_{\nu}}{q^{2}-m^{2}} \epsilon_{2}^{\mu} \epsilon^{*\nu}_{3}, 
\end{align}

\begin{align}
    \mathcal{M}_{b6.3}^{(2)}=& -4\frac{g_{1}^{3}g_{2}}{f^{2}}(d-3)(d-2)A J_{22}^{g}\frac{q_{\mu}q_{\nu}}{q^{2}-m^{2}} \epsilon_{2}^{\mu} \epsilon^{*\nu}_{3},
\end{align}

\begin{align}
    \mathcal{M}_{b6.4}^{(2)}=&~ 4\frac{g_{1}^{3}g_{2}}{f^{2}}A J_{22}^{g} \frac{q_{\mu}q_{\nu}}{q^{2}-m^{2}} \epsilon_{2}^{\mu} \epsilon^{*\nu}_{3}, 
\end{align}

\begin{align}
    \mathcal{M}_{b6.5}^{(2)}=& -4\frac{g_{1}g_{2}}{f^{2}}A J_{0}^{c}\frac{q_{\mu}q_{\nu}}{q^{2}-m^{2}} \epsilon_{2}^{\mu} \epsilon^{*\nu}_{3}, 
\end{align}

\begin{align}
    \mathcal{M}_{b6.6}^{(2)}=& -4\frac{g_{1}g_{2}}{f^{2}}A J_{0}^{c}\frac{q_{\mu}q_{\nu}}{q^{2}-m^{2}} \epsilon_{2}^{\mu} \epsilon^{*\nu}_{3}, 
\end{align}

\begin{align}
    \mathcal{M}_{b6.7}^{(2)}=& -\frac{8g_{1}g_{2}}{f^{2}}A [2m^{2}L+\frac{m^{2}}{8\pi^{2}}{\rm{log}} (\frac{m}{\mu})]\frac{q_{\mu}q_{\nu}}{q^{2}-m^{2}} \notag \\ 
     & \times \epsilon_{2}^{\mu} \epsilon^{*\nu}_{3}, 
\end{align}

\begin{align}
    \mathcal{M}_{b6.(8+9)}^{(2)}=& -\frac{3g_{1}g_{2}}{2f^{2}}A [(d-2)\partial_{\omega} J_{22}^{b} (\omega_{1})+\partial_{\omega} J_{22}^{b}(\omega_{2})] \notag \\ 
     & \times \frac{q_{\mu}q_{\nu}}{q^{2}-m^{2}} \epsilon_{2}^{\mu} \epsilon^{*\nu}_{3}, 
\end{align}

\begin{align}
    \mathcal{M}_{b6.10}^{(2)}=&~ \frac{3g_{1}g_{2}}{2f^{2}}(d-1)A \partial_{\omega} J_{22}^{b} \frac{q_{\mu}q_{\nu}}{q^{2}-m^{2}} \epsilon_{2}^{\mu} \epsilon^{*\nu}_{3}, 
\end{align}

\begin{align}
    \mathcal{M}_{b6.(11+12)}^{(2)}=& -\frac{3g_{1}g_{2}}{2f^{2}}A [(d-2)\partial_{\omega} J_{22}^{b} (\omega_{1})+\partial_{\omega} J_{22}^{b}(\omega_{2})] \notag \\ 
     & \times \frac{q_{\mu}q_{\nu}}{q^{2}-m^{2}} \epsilon_{2}^{\mu} \epsilon^{*\nu}_{3},
\end{align}
    
\begin{align}
    \mathcal{M}_{b6.13}^{(2)}=&~ \frac{3g_{1}g_{2}}{2f^{2}}(d-1)A \partial_{\omega} J_{22}^{b} \frac{q_{\mu}q_{\nu}}{q^{2}-m^{2}} \epsilon_{2}^{\mu} \epsilon^{*\nu}_{3}, 
\end{align}

\begin{align}
    \mathcal{M}_{b6.14}^{(2)}=& -\frac{8g_{1}g_{2}}{f^{2}}A [2m^{2}L+\frac{m^{2}}{8\pi^{2}}{\rm{log}} (\frac{m}{\mu})]\frac{q_{\mu}q_{\nu}}{q^{2}-m^{2}} \notag \\ 
     & \times \epsilon_{2}^{\mu} \epsilon^{*\nu}_{3}, 
\end{align}

\begin{align}
     \mathcal{M}_{b6.15}^{(2)}=& -\frac{8g_{1}g_{2}}{f^{2}}A [2m^{2}L+\frac{m^{2}}{8\pi^{2}}{\rm{log}} (\frac{m}{\mu})]\frac{q_{\mu}q_{\nu}}{q^{2}-m^{2}} \notag \\ 
     & \times \epsilon_{2}^{\mu} \epsilon^{*\nu}_{3}, 
\end{align}

\begin{align}
     \mathcal{M}_{b6.16}^{(2)}=& -\frac{8g_{1}g_{2}}{f^{2}}A [2m^{2}L+\frac{m^{2}}{8\pi^{2}}{\rm{log}} (\frac{m}{\mu})]\frac{q_{\mu}q_{\nu}}{q^{2}-m^{2}} \notag \\ 
     & \times \epsilon_{2}^{\mu} \epsilon^{*\nu}_{3}, 
\end{align}

\begin{align}
     \mathcal{M}_{b6.17}^{(2)}=& -\frac{8g_{1}g_{2}}{f^{2}}A [2m^{2}L+\frac{m^{2}}{8\pi^{2}}{\rm{log}} (\frac{m}{\mu})]\frac{q_{\mu}q_{\nu}}{q^{2}-m^{2}} \notag \\ 
     & \times \epsilon_{2}^{\mu} \epsilon^{*\nu}_{3}. 
\end{align}

The TPE amplitudes are given by

\begin{align}
   \mathcal{M}_{c6.1}^{(2)}=&-4\frac{g_{1}^2g_{2}^2}{f^{4}}(d-3)A [(q \cdot \epsilon_{2})(q \cdot \epsilon^{*}_{3}) \notag \\ 
     & +q^{2}(\epsilon_{2} \cdot \epsilon^{*}_{3})]J^{B}_{21}(-\delta_{1},-\delta_{2}),
\end{align}

\begin{align}
   \mathcal{M}_{c6.2}^{(2)}=&~4\frac{g_{1}^2g_{2}^2}{f^{4}}(d-3)A [(q \cdot \epsilon_{2})(q \cdot \epsilon^{*}_{3}) \notag \\ 
     & +q^{2}(\epsilon_{2} \cdot \epsilon^{*}_{3})]J^{R}_{21}(-\delta_{1},-\delta_{2}).
\end{align}

In Tables~\ref{17} and \ref{18}, we list the coefficients of the contact amplitudes and OPE amplitudes at NLO, respectively. For the TPE terms, the coefficients are

\begin{align}
    A_{c6.1}=1/16, \quad A_{c6.2}=5/16,
\end{align}
with $I=1$, and
\begin{align}
    A_{c6.1}=9/16, \quad A_{c6.2}=-3/16,
\end{align}
with $I=0$.

\renewcommand{\arraystretch}{1.5}
\begin{center}
    \begin{table}[ht]
          \centering
          \captionsetup{justification=raggedright, singlelinecheck=false}
          \caption{The coefficients in the NLO contact amplitudes for the $D\bar{B}^{*} \to D^{*}\bar{B}$ process.}\label{17}
    	\setlength{\tabcolsep}{2.0mm}{
    	\begin{tabular}{ccccrr}
     		\toprule[1pt]
    		& { $I=1$  } & { $I=0$ } &    & \\
    		&  $A$     &    $A$    &   $\omega_1$   & $\omega_2$ \\
    		\midrule[1pt]
    		$A_{a6.1(2)}$ & $0$  & $0$  & $0(\delta_{2})$ & $-\delta_{2}$ \\
    		$A_{a6.3(4)}$ & $0$  & $0$  & $-\delta_{1}$ & $0(\delta_{1})$ \\
    		$A_{a6.5}$ & $\frac{1}{4}(D_{b}+E_{b})$  & $\frac{-3}{4}(D_{b}-3E_{b})$  & $0$ & $-\delta_{2}$ \\
    		$A_{a6.6}$ & $\frac{1}{4}(D_{a}+E_{a})$  & $\frac{-3}{4}(D_{a}-3E_{a})$  & $\delta_{2}$ & $-\delta_{1}$ \\
    		$A_{a6.7}$ & $\frac{1}{4}(D_{b}+E_{b})$  & $\frac{-3}{4}(D_{b}-3E_{b})$  & $0$ & $-\delta_{1}$ \\
    		$A_{a6.8}$ & $\frac{1}{4}(D_{a}+E_{a})$  & $\frac{-3}{4}(D_{a}-3E_{a})$  & $\delta_{1}$ & $-\delta_{2}$ \\
    		$A_{a6.(9+10)}$ & $D_{b}+E_{b}$  & $D_{b}-3E_{b}$  & $0$ & $\delta_{1}$ \\
                $A_{a6.11}$ & $D_{b}+E_{b}$  & $D_{b}-3E_{b}$  & $-\delta_{1}$ & $0$ \\
    	    $A_{a6.(12+13)}$ & $D_{b}+E_{b}$  & $D_{b}-3E_{b}$  & $0$ & $\delta_{2}$ \\
                $A_{a6.14}$ & $D_{b}+E_{b}$  & $D_{b}-3E_{b}$  & $-\delta_{2}$ & $0$  \\
    		\bottomrule[1pt]
    	\end{tabular}}
    \end{table}
\end{center}

\renewcommand{\arraystretch}{1.5}
\begin{center}
    \begin{table}[!ht]
          \centering
          \captionsetup{justification=raggedright, singlelinecheck=false}
          \caption{The coefficients in the NLO OPE amplitudes for the $D\bar{B}^{*} \to D^{*}\bar{B}$ process.}\label{18}
    	\setlength{\tabcolsep}{3.8mm}{
    	\begin{tabular}{ccccrr}
    		\toprule[1pt]
    		& { $I=1$  } & { $I=0$ } &    & \\
    		&  $A$     &    $A$    &   $\omega_1$   & $\omega_2$ \\
    		\midrule[1pt]
    		$A_{b6.1(2)}$ & $-\frac{1}{16}$  & $\frac{3}{16}$  & $0(\delta_{2})$ & $-\delta_{2}$ \\
    		$A_{b6.3(4)}$ & $-\frac{1}{16}$  & $\frac{3}{16}$  & $-\delta_{1}$ & $0(\delta_{1})$ \\
    		$A_{b6.5(6)}$ & $-\frac{1}{12}$  & $\frac{1}{4}$  & $0$ & $0$ \\
    		$A_{b6.7}$ & $\frac{1}{4}$  & $-\frac{3}{4}$  & $0$ & $0$ \\
    		$A_{b6.(8+9)}$ & $\frac{1}{4}$  & $-\frac{3}{4}$  & $0$ & $\delta_{1}$ \\
    		$A_{b6.10}$ & $\frac{1}{4}$  & $-\frac{3}{4}$  & $-\delta_{1}$ & $0$ \\
                $A_{b6.(11+12)}$ & $\frac{1}{4}$  & $-\frac{3}{4}$  & $0$ & $\delta_{2}$ \\
    	    $A_{b6.13}$ & $\frac{1}{4}$  & $-\frac{3}{4}$  & $-\delta_{2}$ & $0$ \\
                $A_{b6.14(15)}$ & $\frac{1}{4}$  & $-\frac{3}{4}$  & $0$ & $0$ \\
                $A_{b6.16(17)}$ & $\frac{1}{4}$  & $-\frac{3}{4}$  & $0$ & $0$ \\
    		\bottomrule[1pt]
    	\end{tabular}}
    \end{table}
\end{center}

\subsection{$D\bar{B}^{*} \to D^{*}\bar{B}^{*}$}
\label{A4}

\begin{figure*}[ht]
     \centering
     \includegraphics[width=1.0\textwidth]{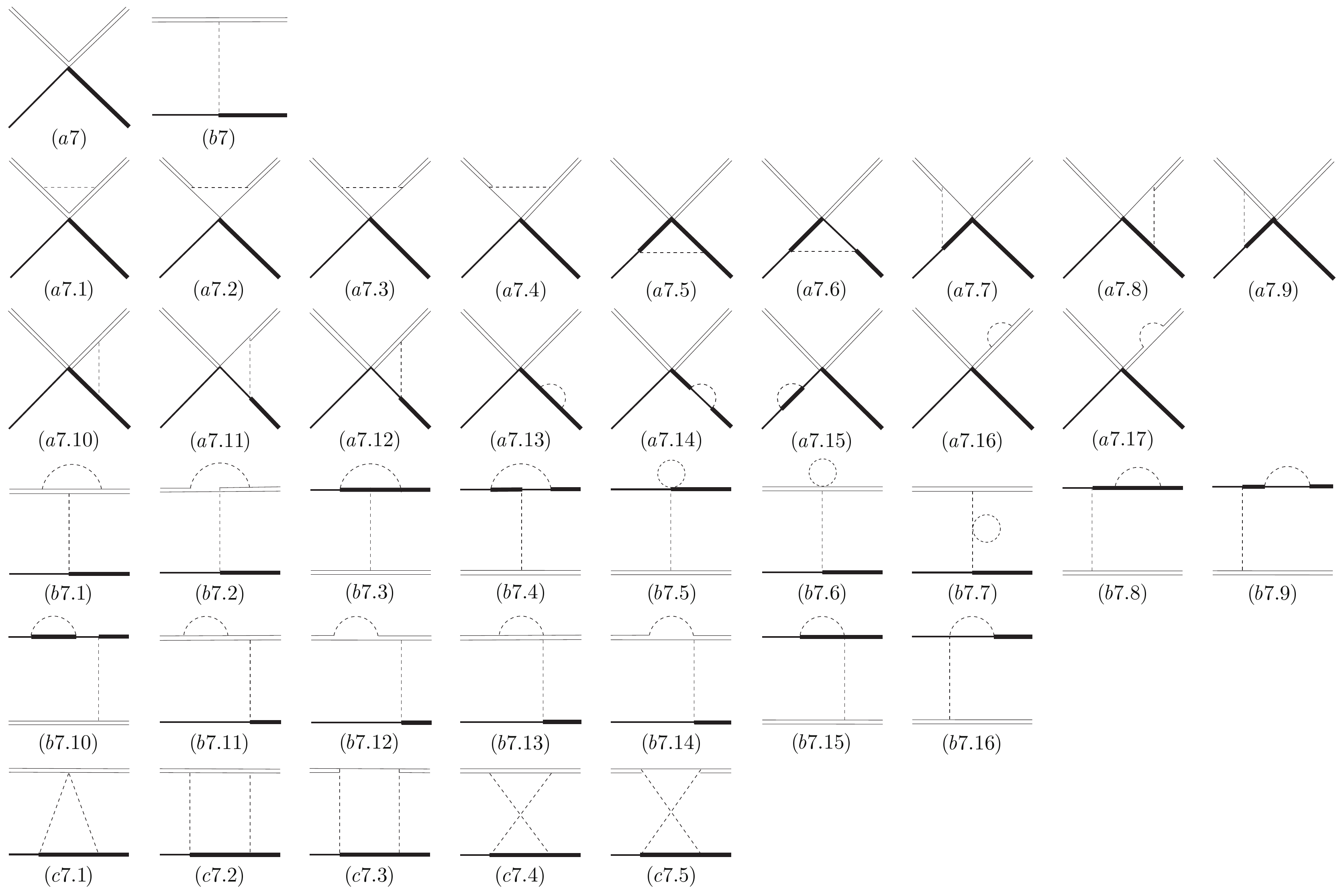}
     \captionsetup{justification=raggedright, singlelinecheck=false}
     \caption{LO contact ($a7$), LO OPE ($b7$), and NLO contact ($a7.1-a7.17$), NLO OPE ($b7.1-b7.16$), and NLO TPE ($c7.1-c7.5$) diagrams of the process $D\bar{B}^{*} \to D^{*}\bar{B}^{*}$.}\label{BsD-BsDs}
\end{figure*}

For the process $D(p_1)\bar{B}^{*}(p_2) \to D^{*}(p_3) \bar{B}^{*}(p_4) $, the contact, OPE, and TPE diagrams are shown in Fig.~\ref{BsD-BsDs}. We can write the scattering amplitudes of LO diagrams as
\begin{align}
    \mathcal{M}_{a7}^{(0)}=&~ 4(D_{b}+E_{b}) \epsilon^{*\mu}_{3}(\epsilon_{2}\times \epsilon^{*}_{4})_{\mu},
\end{align}

\begin{align}
    \mathcal{M}_{b7}^{(0)}=&~ \frac{g_{1}g_{2}}{f^2}\frac{q_{\mu}q_{\nu}}{q^{2}-m^{2}} \epsilon^{\mu}_{2}(\epsilon^{*}_{4}\times \epsilon^{*}_{3})^{\nu},
\end{align}
with isospin $I=1$, and
\begin{align}
    \mathcal{M}_{a7}^{(0)}=&~ 4(D_{b}-3E_{b}) \epsilon^{*\mu}_{3}(\epsilon_{2}\times \epsilon^{*}_{4})_{\mu},
\end{align}

\begin{align}
    \mathcal{M}_{b7}^{(0)}=& -\frac{3g_{1}g_{2}}{f^2}\frac{q_{\mu}q_{\nu}}{q^{2}-m^{2}} \epsilon^{\mu}_{2}(\epsilon^{*}_{4}\times \epsilon^{*}_{3})^{\nu},
\end{align}
with isospin $I=0$.

The one-loop corrections to the contact amplitudes are listed as
\begin{align}
    \mathcal{M}_{a7.1}^{(2)}=&~ 4\frac{g_{2}^{2}}{f^{2}}(d-3)A J_{22}^{g} \epsilon^{*\mu}_{3}(\epsilon_{2}\times \epsilon^{*}_{4})_{\mu},
\end{align}

\begin{align}
    \mathcal{M}_{a7.2}^{(2)}=&~ 0, 
\end{align}

\begin{align}
    \mathcal{M}_{a7.3}^{(2)}=& -4\frac{g_{2}^{2}}{f^{2}}A J_{22}^{g} \epsilon^{*\mu}_{3}(\epsilon_{2}\times \epsilon^{*}_{4})_{\mu}, 
\end{align}

\begin{align}
    \mathcal{M}_{a7.4}^{(2)}=& -4\frac{g_{2}^{2}}{f^{2}}A J_{22}^{g} \epsilon^{*\mu}_{3}(\epsilon_{2}\times \epsilon^{*}_{4})_{\mu}, 
\end{align}

\begin{align}
    \mathcal{M}_{a7.5}^{(2)}=& -8\frac{g_{1}^{2}}{f^{2}}A J_{22}^{g} \epsilon^{*\mu}_{3}(\epsilon_{2}\times \epsilon^{*}_{4})_{\mu}, 
\end{align}

\begin{align}
    \mathcal{M}_{a7.6}^{(2)}=&~ 4\frac{g_{1}^{2}}{f^{2}}A J_{22}^{g} \epsilon^{*\mu}_{3}(\epsilon_{2}\times \epsilon^{*}_{4})_{\mu}, 
\end{align}

\begin{align}
    \mathcal{M}_{a7.7}^{(2)}=& -4\frac{g_{1}g_{2}}{f^{2}}A J_{22}^{h} \epsilon^{*\mu}_{3}(\epsilon_{2}\times \epsilon^{*}_{4})_{\mu}, 
\end{align}

\begin{align}
    \mathcal{M}_{a7.8}^{(2)}=& -4\frac{g_{1}g_{2}}{f^{2}}A J_{22}^{h} \epsilon^{*\mu}_{3}(\epsilon_{2}\times \epsilon^{*}_{4})_{\mu}, 
\end{align}

\begin{align}
    \mathcal{M}_{a7.9}^{(2)}=& -4\frac{g_{1}g_{2}}{f^{2}}A J_{22}^{h} \epsilon^{*\mu}_{3}(\epsilon_{2}\times \epsilon^{*}_{4})_{\mu}, 
\end{align}
    
\begin{align}
    \mathcal{M}_{a7.10}^{(2)}=& -4\frac{g_{1}g_{2}}{f^{2}}(d-3)A J_{22}^{h} \epsilon^{*\mu}_{3}(\epsilon_{2}\times \epsilon^{*}_{4})_{\mu}, 
\end{align}

\begin{align}
    \mathcal{M}_{a7.11}^{(2)}=&~ 0, 
\end{align}

\begin{align}
    \mathcal{M}_{a7.12}^{(2)}=& -4\frac{g_{1}g_{2}}{f^{2}}A J_{22}^{h} \epsilon^{*\mu}_{3}(\epsilon_{2}\times \epsilon^{*}_{4})_{\mu},
\end{align}

\begin{align}
    \mathcal{M}_{a7.(13+14)}^{(2)}=&~ \frac{3g_{2}^{2}}{2f^{2}}A [(d-2) \partial_{\omega} J_{22}^{b} (\omega_{1})+\partial_{\omega} J_{22}^{b}(\omega_{2})] \notag \\ & \times [\epsilon^{*\mu}_{3}(\epsilon_{2}\times \epsilon^{*}_{4})_{\mu}], 
\end{align}

\begin{align}
    \mathcal{M}_{a7.15}^{(2)}=&~ \frac{3g_{1}^{2}}{2f^{2}}(d-1)A \partial_{\omega} J_{22}^{b} \epsilon^{*\mu}_{3}(\epsilon_{2}\times \epsilon^{*}_{4})_{\mu}, 
\end{align}

\begin{align}
    \mathcal{M}_{a7.(16+17)}^{(2)}=&~ \frac{3g_{2}^{2}}{2f^{2}}A [(d-2) \partial_{\omega} J_{22}^{b} (\omega_{1})+\partial_{\omega} J_{22}^{b}(\omega_{2})] \notag \\ & \times [\epsilon^{*\mu}_{3}(\epsilon_{2}\times \epsilon^{*}_{4})_{\mu}].
\end{align}

For the one-loop corrections to the OPE amplitudes, we have
\begin{align}
    \mathcal{M}_{b7.1}^{(2)}=& -4\frac{g_{1}g_{2}^{3}}{f^{2}}(d-3)A J_{22}^{g}\frac{q_{\mu}q_{\nu}}{q^{2}-m^{2}} \epsilon^{\mu}_{2}(\epsilon^{*}_{4}\times \epsilon^{*}_{3})^{\nu}, 
\end{align}

\begin{align}
    \mathcal{M}_{b7.2}^{(2)}=&~ 4\frac{g_{1}g_{2}^{3}}{f^{2}}A J_{22}^{g} \frac{q_{\mu}q_{\nu}}{q^{2}-m^{2}} \epsilon^{\mu}_{2}(\epsilon^{*}_{4}\times \epsilon^{*}_{3})^{\nu}, 
\end{align}

\begin{align}
    \mathcal{M}_{b7.3}^{(2)}=& -8\frac{g_{1}^{3}g_{2}}{f^{2}}A J_{22}^{g}\frac{1}{q^{2}-m^{2}} [q^{2}\epsilon^{*\mu}_{3}(\epsilon_{2}\times \epsilon^{*}_{4})_{\mu} \notag \\ &  +q_{\mu}q_{\nu}\epsilon^{*\mu}_{3}(\epsilon_{2}\times \epsilon^{*}_{4})^{\nu} - q_{\mu}q_{\nu}\epsilon^{*\mu}_{3}(\epsilon^{*}_{4}\times \epsilon_{2})^{\nu}],
\end{align}

\begin{align}
    \mathcal{M}_{b7.4}^{(2)}=& -4\frac{g_{1}^{3}g_{2}}{f^{2}}A J_{22}^{g} \frac{q_{\mu}q_{\nu}}{q^{2}-m^{2}} \epsilon^{\mu}_{2}(\epsilon^{*}_{4}\times \epsilon^{*}_{3})^{\nu}, 
\end{align}

\begin{align}
    \mathcal{M}_{b7.5}^{(2)}=&~ 4\frac{g_{1}g_{2}}{f^{2}}A J_{0}^{c}\frac{q_{\mu}q_{\nu}}{q^{2}-m^{2}} \epsilon^{\mu}_{2}(\epsilon^{*}_{4}\times \epsilon^{*}_{3})^{\nu}, 
\end{align}

\begin{align}
    \mathcal{M}_{b7.6}^{(2)}=&~ 4\frac{g_{1}g_{2}}{f^{2}}A J_{0}^{c}\frac{q_{\mu}q_{\nu}}{q^{2}-m^{2}} \epsilon^{\mu}_{2}(\epsilon^{*}_{4}\times \epsilon^{*}_{3})^{\nu},
\end{align}

\begin{align}
   \mathcal{M}_{b7.7}^{(2)}=&~ \frac{8g_{1}g_{2}}{3f^{2}}A [2m^{2}L+\frac{m^{2}}{8\pi^{2}}{\rm{log}} (\frac{m}{\mu})]\frac{q_{\mu}q_{\nu}}{q^{2}-m^{2}} \notag \\ & \times [\epsilon^{\mu}_{2}(\epsilon^{*}_{4}\times \epsilon^{*}_{3})^{\nu}], 
\end{align}

\begin{align}
    \mathcal{M}_{b7.(8+9)}^{(2)}=& -\frac{3g_{1}^{3}g_{2}}{2f^{2}}A [(d-2)\partial_{\omega} J_{22}^{b} (\omega_{1})+\partial_{\omega} J_{22}^{b}(\omega_{2})] \notag \\ & \times \frac{q_{\mu}q_{\nu}}{q^{2}-m^{2}} \epsilon^{\mu}_{2}(\epsilon^{*}_{4}\times \epsilon^{*}_{3})^{\nu},  
\end{align}

\begin{align}
    \mathcal{M}_{b7.10}^{(2)}=& -\frac{3g_{1}^{3}g_{2}}{2f^{2}}(d-1)A \partial_{\omega} J_{22}^{b} \frac{q_{\mu}q_{\nu}}{q^{2}-m^{2}} \notag \\ & \times [\epsilon^{\mu}_{2}(\epsilon^{*}_{4}\times \epsilon^{*}_{3})^{\nu}], 
\end{align}

\begin{align}
    \mathcal{M}_{b7.(11+12)}^{(2)}=& -\frac{3g_{1}g_{2}^{3}}{2f^{2}}A [(d-2)\partial_{\omega} J_{22}^{b} (\omega_{1})+\partial_{\omega} J_{22}^{b}(\omega_{2})] \notag \\ & \times \frac{q_{\mu}q_{\nu}}{q^{2}-m^{2}} \epsilon^{\mu}_{2}(\epsilon^{*}_{4}\times \epsilon^{*}_{3})^{\nu}, 
\end{align}

\begin{align}
    \mathcal{M}_{b7.13(14)}^{(2)}=&~ \frac{8g_{1}g_{2}}{f^{2}}A [2m^{2}L+\frac{m^{2}}{8\pi^{2}}{\rm{log}} (\frac{m}{\mu})]\frac{q_{\mu}q_{\nu}}{q^{2}-m^{2}} \notag \\ & \times [\epsilon^{\mu}_{2}(\epsilon^{*}_{4}\times \epsilon^{*}_{3})^{\nu}], 
\end{align}

\begin{align}
    \mathcal{M}_{b7.15(16)}^{(2)}=&~ \frac{8g_{1}g_{2}}{f^{2}}A [2m^{2}L+\frac{m^{2}}{8\pi^{2}}{\rm{log}} (\frac{m}{\mu})]\frac{q_{\mu}q_{\nu}}{q^{2}-m^{2}} \notag \\ & \times [\epsilon^{\mu}_{2}(\epsilon^{*}_{4}\times \epsilon^{*}_{3})^{\nu}].
\end{align}

In Tables~\ref{19} and \ref{20}, we list the coefficients of the above contact amplitudes at NLO and OPE amplitudes at NLO, respectively. The TPE amplitudes are given by

\begin{align}
    \mathcal{M}_{c7.1}^{(2)}=&~ 0,
\end{align}

\begin{align}
    \mathcal{M}_{c7.2}^{(2)}=& -4\frac{g_{1}^2g_{2}^2}{f^{4}}A \{(J^{B}_{21}+J^{B}_{31})q^{2}\epsilon^{*\mu}_{3}(\epsilon_{2}\times \epsilon^{*}_{4})_{\mu} (J^{B}_{21}+ \notag \\ & +J^{B}_{31})q_{\mu}q_{\nu}\epsilon^{\mu}_{2}(\epsilon^{*}_{4}\times \epsilon^{*}_{3})^{\nu} +[J^{B}_{21}+(d-2)J^{B}_{31}]\notag \\ &  \times q_{\mu}q_{\nu}\epsilon^{*\mu}_{3}( \epsilon_{2} \times \epsilon^{*}_{4})^{\nu} \},
\end{align}

\begin{align}
     \mathcal{M}_{c7.3}^{(2)}=&~ 4\frac{g_{1}^2g_{2}^2}{f^{4}}A [ (J^{R}_{21}+J^{R}_{31})q_{\mu}q_{\nu}\epsilon^{\mu*}_{3}(\epsilon_{2} \times \epsilon^{*}_{4})^{\nu}   \notag \\ &+J^{R}_{31}q_{\mu}q_{\nu}\epsilon^{*\mu}_{3}(\epsilon^{*}_{4} \times \epsilon_{2})^{\nu} ],
\end{align}

\begin{align}
     \mathcal{M}_{c7.4}^{(2)}=&~ 4\frac{g_{1}^2g_{2}^2}{f^{4}}A \{(J^{R}_{21}+J^{R}_{31})q^{2}\epsilon^{*\mu}_{3}(\epsilon_{2}\times \epsilon^{*}_{4})_{\mu} + (J^{R}_{21} \notag \\ &+J^{R}_{31})q_{\mu}q_{\nu}\epsilon^{\mu}_{2}(\epsilon^{*}_{4}\times \epsilon^{*}_{3})^{\nu}  -[J^{R}_{21}+(d-2)J^{R}_{31}]\notag \\ & \times q_{\mu}q_{\nu}\epsilon^{*\mu}_{3}(\epsilon^{*}_{4} \times \epsilon_{2})^{\nu} \},
\end{align}

\begin{align}
     \mathcal{M}_{c7.5}^{(2)}=&~ 4\frac{g_{1}^2g_{2}^2}{f^{4}}A [ (J^{R}_{21}+J^{R}_{31})q_{\mu}q_{\nu}\epsilon^{*\mu}_{3}(\epsilon^{*}_{4} \times \epsilon_{2})^{\nu}  \notag \\ &+J^{R}_{31}q_{\mu}q_{\nu}\epsilon^{\mu*}_{3}(\epsilon_{2} \times \epsilon^{*}_{4})^{\nu} ].
\end{align}

The coefficients in the above TPE amplitudes are listed in Table~\ref{21}.

\renewcommand{\arraystretch}{1.5}
\begin{center}
    \begin{table}[hb]
          \centering
          \captionsetup{justification=raggedright, singlelinecheck=false}
          \caption{The coefficients in the NLO contact amplitudes for the $D\bar{B}^{*} \to D^{*}\bar{B}^{*}$ process.}\label{19}
    	\setlength{\tabcolsep}{2.2mm}{
    	\begin{tabular}{ccccrr}
    		\toprule[1pt]
    		& { $I=1$  } & { $I=0$ } &    & \\
    		&  $A$     &    $A$    &   $\omega_1$   & $\omega_2$ \\
    		\midrule[1pt]
    		$A_{a7.1}$ & $0$  & $0$  & $0$ & $0$ \\
    		$A_{a7.3(4)}$ & $0$  & $0$  & $0(\delta_{2})$ & $\delta_{2}(0)$ \\
    		$A_{a7.5(6)}$ & $0$  & $0$  & $-\delta_{1}$ & $0(\delta_{1})$ \\
    		$A_{a7.7}$ & $\frac{1}{4}(D_{a}+E_{a})$  & $\frac{-3}{4}(D_{a}-3E_{a})$  & $\delta_{2}$ & $-\delta_{1}$ \\
    		$A_{a7.8}$ & $\frac{1}{4}(D_{b}+E_{b})$  & $\frac{-3}{4}(D_{b}-3E_{b})$  & $0$ & $\delta_{2}$ \\
    		$A_{a7.9}$ & \makecell[c]{$\frac{1}{4}(D_{a}+E_{a}$\\$+D_{b}+E_{b})$}  & \makecell[c]{$\frac{-3}{4}(D_{a}-3E_{a}$\\$+D_{b}-3E_{b})$}  & $0$ & $-\delta_{1}$ \\
                $A_{a7.10}$ & $\frac{1}{4}(D_{b}+E_{b})$  & $\frac{-3}{4}(D_{b}-3E_{b})$  & $0$ & $0$ \\
    	    $A_{a7.12}$ & $\frac{1}{4}(D_{a}+E_{a})$  & $\frac{-3}{4}(D_{a}-3E_{a})$  & $0$ & $\delta_{1}$ \\
                $A_{a7.(13+14)}$ & $D_{b}+E_{b}$  & $D_{b}-3E_{b}$  & $0$ & $\delta_{1}$  \\
                $A_{a7.15}$ & $D_{b}+E_{b}$  & $D_{b}-3E_{b}$  & $-\delta_{1}$ & $0$ \\
                $A_{a7.(16+17)}$ & $D_{b}+E_{b}$  & $D_{b}-3E_{b}$  & $0$ & $\delta_{2}$  \\
    		\bottomrule[1pt]
    	\end{tabular}}
    \end{table}
\end{center}

\renewcommand{\arraystretch}{1.5}
\begin{center}
    \begin{table}[ht]
          \centering
          \captionsetup{justification=raggedright, singlelinecheck=false}
          \caption{The coefficients in the NLO OPE amplitudes for the $D\bar{B}^{*} \to D^{*}\bar{B}^{*}$ process.}\label{20}
    	\setlength{\tabcolsep}{3.7mm}{
    	\begin{tabular}{ccccrr}
    		\toprule[1pt]
    		& { $I=1$  } & { $I=0$ } &    & \\
    		&  $A$     &    $A$    &   $\omega_1$   & $\omega_2$ \\
    		\midrule[1pt]
    		$A_{b7.1(2)}$ & $-\frac{1}{16}$  & $\frac{3}{16}$  & $0(\delta_{2})$ & $0$ \\
    		$A_{b7.3(4)}$ & $-\frac{1}{16}$  & $\frac{3}{16}$  & $-\delta_{1}$ & $0(\delta_{1})$ \\
    		$A_{b7.5(6)}$ & $-\frac{1}{12}$  & $\frac{1}{4}$  & $0$ & $0$ \\
    		$A_{b7.7}$ & $\frac{1}{4}$  & $-\frac{3}{4}$  & $0$ & $0$ \\
    		$A_{b7.(8+9)}$ & $\frac{1}{4}$  & $-\frac{3}{4}$  & $0$ & $\delta_{1}$ \\
    		$A_{b7.10}$ & $\frac{1}{4}$  & $-\frac{3}{4}$  & $-\delta_{1}$ & $0$ \\
                $A_{b7.(11+12)}$ & $\frac{1}{4}$  & $-\frac{3}{4}$  & $0$ & $\delta_{2}$ \\
    	    $A_{b7.13(14)}$ & $\frac{1}{4}$  & $-\frac{3}{4}$  & $0$ & $0$ \\
                $A_{b7.15(16)}$ & $\frac{1}{4}$  & $-\frac{3}{4}$  & $0$ & $0$ \\
    		\bottomrule[1pt]
    	\end{tabular}}
    \end{table}
\end{center}

\renewcommand{\arraystretch}{1.5}
\begin{center}
    \begin{table}[ht]
          \centering
          \captionsetup{justification=raggedright, singlelinecheck=false}
          \caption{The coefficients in the TPE amplitudes for the $D\bar{B}^{*} \to D^{*}\bar{B}^{*}$ process.}\label{21}
    	\setlength{\tabcolsep}{4.8mm}{
    	\begin{tabular}{ccccrr}
    		\toprule[1pt]
    		& { $I=1$  } & { $I=0$ } &    & \\
    		&  $A$     &    $A$    &   $\omega_1$   & $\omega_2$ \\
    		\midrule[1pt]
    		$A_{c7.2}$ & $\frac{1}{16}$  & $\frac{5}{16}$  & $-\delta_{1}$ & $0$ \\
    		$A_{c7.3}$ & $\frac{1}{16}$  & $\frac{5}{16}$  & $-\delta_{1}$ & $\delta_{2}$ \\
    		$A_{c7.4}$ & $\frac{9}{16}$   & $-\frac{3}{16}$ & $-\delta_{1}$ & $0$ \\
    		$A_{c7.5}$ & $\frac{9}{16}$   & $-\frac{3}{16}$ & $-\delta_{1}$ & $\delta_{2}$ \\
    		\bottomrule[1pt]
    	\end{tabular}}
    \end{table}
\end{center}
     
\subsection{$D^{*}\bar{B} \to D^{*}\bar{B}^{*}$}
\label{A5}

\begin{figure*}[ht]
     \centering
     \includegraphics[width=1.0\textwidth]{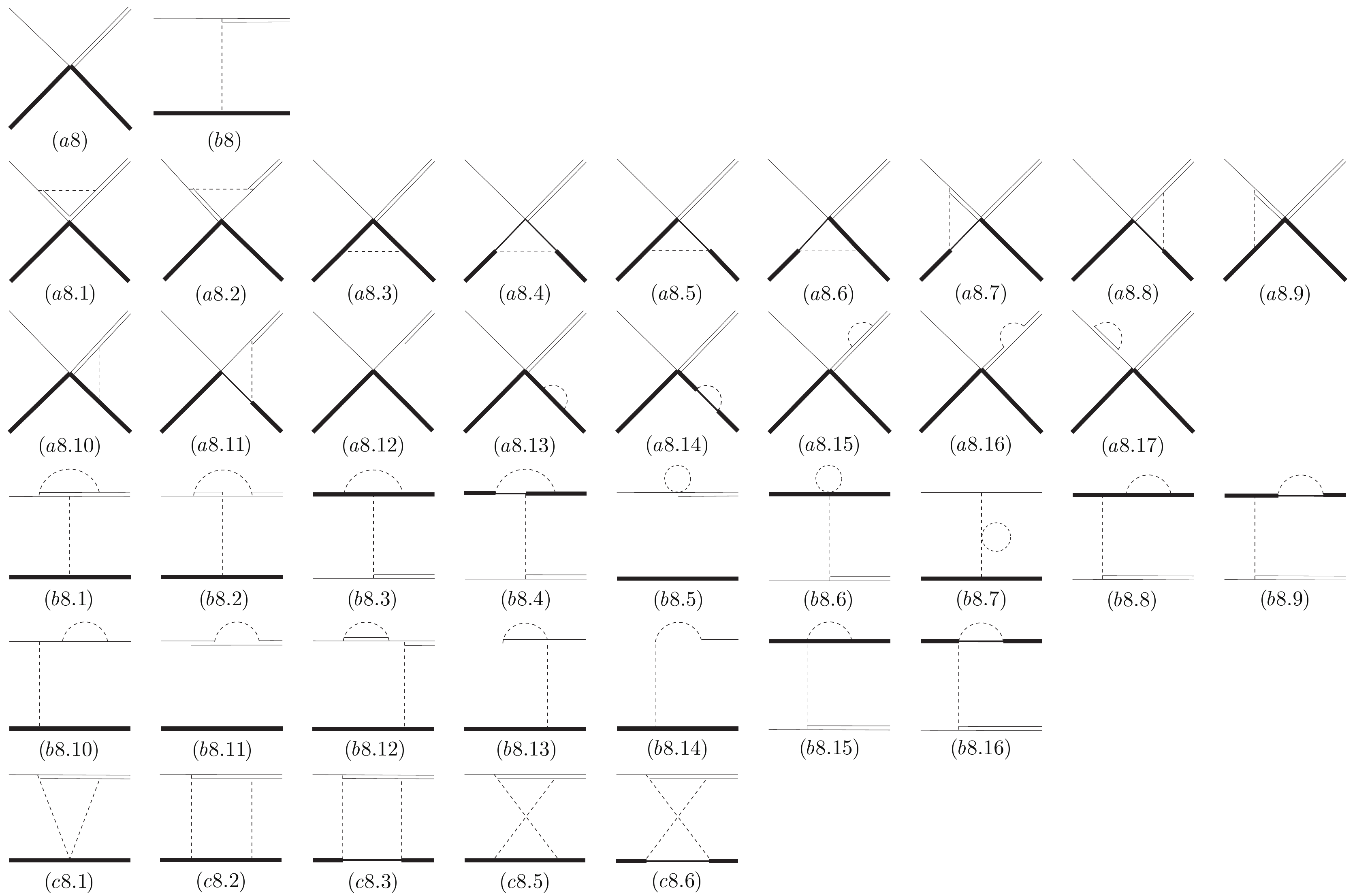}
     \captionsetup{justification=raggedright, singlelinecheck=false}
     \caption{LO contact ($a8$), LO OPE ($b8$), and NLO contact ($a8.1-a8.17$), NLO OPE ($b8.1-b8.16$), and NLO TPE ($c8.1-c8.5$) diagrams of the process $D^{*}\bar{B} \to D^{*}\bar{B}^{*}$.}\label{BDs-BsDs}
\end{figure*}

For the process $ D^{*}(p_1)\bar{B}(p_2) \to D^{*}(p_3) \bar{B}^{*}(p_4) $, the contact, OPE, and TPE diagrams are shown in Fig.~\ref{BDs-BsDs}. The scattering amplitudes of LO diagrams are given by
\begin{align}
    \mathcal{M}_{a8}^{(0)}=& -4(D_{b}+E_{b})\epsilon^{\mu}_{1}(\epsilon^{*}_{3} \times \epsilon^{*}_{4})_{\mu},
\end{align}

\begin{align}
    \mathcal{M}_{b8}^{(0)}=&~ \frac{g_{1}g_{2}}{f^2}\frac{q_{\mu}q_{\nu}}{q^{2}-m^{2}} \epsilon^{\mu}_{1}(\epsilon^{*}_{4}\times \epsilon^{*}_{3})^{\nu},
\end{align}
with isospin $I=1$, and
\begin{align}
    \mathcal{M}_{a8}^{(0)}=& -4(D_{b}-3E_{b})\epsilon^{\mu}_{1}(\epsilon^{*}_{3} \times \epsilon^{*}_{4})_{\mu},
\end{align}

\begin{align}
    \mathcal{M}_{b8}^{(0)}=& -\frac{3g_{1}g_{2}}{f^2}\frac{q_{\mu}q_{\nu}}{q^{2}-m^{2}} \epsilon^{\mu}_{1}(\epsilon^{*}_{4}\times \epsilon^{*}_{3})^{\nu},
\end{align}
with isospin $I=0$.

The one-loop corrections to the contact amplitudes are listed as follows:
\begin{align}
    \mathcal{M}_{a8.1}^{(2)}=&~ 4\frac{g_{2}^{2}}{f^{2}}(d-3)A J_{22}^{g} \epsilon^{\mu}_{1}(\epsilon^{*}_{3} \times \epsilon^{*}_{4})_{\mu},
\end{align}

\begin{align}
    \mathcal{M}_{a8.2}^{(2)}=&~ 0, 
\end{align}

\begin{align}
    \mathcal{M}_{a8.3}^{(2)}=& -4\frac{g_{2}^{2}}{f^{2}}A J_{22}^{g} \epsilon^{\mu}_{1}(\epsilon^{*}_{3} \times \epsilon^{*}_{4})_{\mu}, 
\end{align}

\begin{align}
    \mathcal{M}_{a8.4}^{(2)}=& -4\frac{g_{2}^{2}}{f^{2}}A J_{22}^{g} \epsilon^{\mu}_{1}(\epsilon^{*}_{3} \times \epsilon^{*}_{4})_{\mu}, 
\end{align}

\begin{align}
   \mathcal{M}_{a8.5}^{(2)}=& -8\frac{g_{1}^{2}}{f^{2}}A J_{22}^{g} \epsilon^{\mu}_{1}(\epsilon^{*}_{3} \times \epsilon^{*}_{4})_{\mu}, 
\end{align}

\begin{align}
    \mathcal{M}_{a8.6}^{(2)}=&~ 4\frac{g_{1}^{2}}{f^{2}}A J_{22}^{g} \epsilon^{\mu}_{1}(\epsilon^{*}_{3} \times \epsilon^{*}_{4})_{\mu},
\end{align}

\begin{align}
    \mathcal{M}_{a8.7}^{(2)}=& -4\frac{g_{1}g_{2}}{f^{2}}A J_{22}^{h} \epsilon^{\mu}_{1}(\epsilon^{*}_{3} \times \epsilon^{*}_{4})_{\mu}, 
\end{align}

\begin{align}
    \mathcal{M}_{a8.8}^{(2)}=& -4\frac{g_{1}g_{2}}{f^{2}}A J_{22}^{h} \epsilon^{\mu}_{1}(\epsilon^{*}_{3} \times \epsilon^{*}_{4})_{\mu}, 
\end{align}

\begin{align}
    \mathcal{M}_{a8.9}^{(2)}=& -4\frac{g_{1}g_{2}}{f^{2}}A J_{22}^{h} \epsilon^{\mu}_{1}(\epsilon^{*}_{3} \times \epsilon^{*}_{4})_{\mu},
\end{align}

\begin{align}
    \mathcal{M}_{a8.10}^{(2)}=& -4\frac{g_{1}g_{2}}{f^{2}}(d-3)A J_{22}^{h} \epsilon^{\mu}_{1}(\epsilon^{*}_{3} \times \epsilon^{*}_{4})_{\mu}, 
\end{align}

\begin{align}
    \mathcal{M}_{a8.11}^{(2)}=&~ 0,
\end{align}

\begin{align}
    \mathcal{M}_{a8.12}^{(2)}=& -4\frac{g_{1}g_{2}}{f^{2}}A J_{22}^{h} \epsilon^{\mu}_{1}(\epsilon^{*}_{3} \times \epsilon^{*}_{4})_{\mu}, 
\end{align}

\begin{align}
    \mathcal{M}_{a8.(13+14)}^{(2)}=&~ \frac{3g_{2}^{2}}{2f^{2}}A [(d-2) \partial_{\omega} J_{22}^{b} (\omega_{1})+\partial_{\omega} J_{22}^{b}(\omega_{2})] \notag \\ & \times [\epsilon^{\mu}_{1}(\epsilon^{*}_{3} \times \epsilon^{*}_{4})_{\mu}], 
\end{align}

\begin{align}
    \mathcal{M}_{a8.15}^{(2)}=&~ \frac{3g_{1}^{2}}{2f^{2}}(d-1)A \partial_{\omega} J_{22}^{b} \epsilon^{\mu}_{1}(\epsilon^{*}_{3} \times \epsilon^{*}_{4})_{\mu}, 
\end{align}

\begin{align}
    \mathcal{M}_{a8.(16+17)}^{(2)}=&~ \frac{3g_{2}^{2}}{2f^{2}}A [(d-2) \partial_{\omega} J_{22}^{b} (\omega_{1})+\partial_{\omega} J_{22}^{b}(\omega_{2})] \notag \\ & \times [\epsilon^{\mu}_{1}(\epsilon^{*}_{3} \times \epsilon^{*}_{4})_{\mu}].
\end{align}

The one-loop corrections to the OPE amplitudes are given by
    
\begin{align}
    \mathcal{M}_{b8.1}^{(2)}=& -8\frac{g_{1}g_{2}^{3}}{f^{2}}A J_{22}^{g}\frac{1}{q^{2}-m^{2}} [q^{2}\epsilon^{\mu}_{1}(\epsilon^{*}_{3} \times \epsilon^{*}_{4})_{\mu} \notag \\ &  - q_{\mu}q_{\nu}\epsilon^{\mu}_{1}(\epsilon^{*}_{4} \times \epsilon^{*}_{3})^{\nu} + q_{\mu}q_{\nu}\epsilon^{*\mu}_{3}(\epsilon^{*}_{4}\times \epsilon_{1})^{\nu}], 
\end{align}

\begin{align}
    \mathcal{M}_{b8.2}^{(2)}=& -4\frac{g_{1}g_{2}^{3}}{f^{2}}A J_{22}^{g}\frac{q_{\mu}q_{\nu}}{q^{2}-m^{2}} \epsilon^{\mu}_{1}(\epsilon^{*}_{3} \times \epsilon^{*}_{4})^{\nu}, 
\end{align}

\begin{align}
    \mathcal{M}_{b8.3}^{(2)}=& -4\frac{g_{1}^{3}g_{2}}{f^{2}}(d-3)A J_{22}^{g} \frac{q_{\mu}q_{\nu}}{q^{2}-m^{2}} \epsilon^{\mu}_{1}(\epsilon^{*}_{3} \times \epsilon^{*}_{4})^{\nu}, 
\end{align}

\begin{align}
    \mathcal{M}_{b8.4}^{(2)}=&~ 4\frac{g_{1}^{3}g_{2}}{f^{2}}A J_{22}^{g} \frac{q_{\mu}q_{\nu}}{q^{2}-m^{2}} \epsilon^{\mu}_{1}(\epsilon^{*}_{3} \times \epsilon^{*}_{4})^{\nu}, 
\end{align}

\begin{align}
    \mathcal{M}_{b8.5}^{(2)}=&~ 4\frac{g_{1}g_{2}}{f^{2}}A J_{0}^{c}\frac{q_{\mu}q_{\nu}}{q^{2}-m^{2}} \epsilon^{\mu}_{1}(\epsilon^{*}_{3} \times \epsilon^{*}_{4})^{\nu}, 
\end{align}

\begin{align}
    \mathcal{M}_{b8.6}^{(2)}=&~ 4\frac{g_{1}g_{2}}{f^{2}}A J_{0}^{c}\frac{q_{\mu}q_{\nu}}{q^{2}-m^{2}} \epsilon^{\mu}_{1}(\epsilon^{*}_{3} \times \epsilon^{*}_{4})^{\nu}, 
\end{align}

\begin{align}
    \mathcal{M}_{b8.7}^{(2)}=&~ \frac{8g_{1}g_{2}}{3f^{2}}A_{1}[2m^{2}L+\frac{m^{2}}{8\pi^{2}}{\rm{log}} (\frac{m}{\mu})]\frac{q_{\mu}q_{\nu}}{q^{2}-m^{2}} \notag \\ & \times [\epsilon^{\mu}_{1}(\epsilon^{*}_{3} \times \epsilon^{*}_{4})^{\nu}], 
\end{align}

\begin{align}
    \mathcal{M}_{b8.(8+9)}^{(2)}=& -\frac{3g_{1}^{3}g_{2}}{2f^{2}}A [(d-2)\partial_{\omega} J_{22}^{b} (\omega_{1})+\partial_{\omega} J_{22}^{b}(\omega_{2})] \notag \\ & \times \frac{q_{\mu}q_{\nu}}{q^{2}-m^{2}} \epsilon^{\mu}_{1}(\epsilon^{*}_{3} \times \epsilon^{*}_{4})^{\nu}, 
\end{align}

\begin{align}
    \mathcal{M}_{b8.(10+11)}^{(2)}=& -\frac{3g_{1}g_{2}^{3}}{2f^{2}}A [(d-2)\partial_{\omega} J_{22}^{b} (\omega_{1})+\partial_{\omega} J_{22}^{b}(\omega_{2})] \notag \\ & \times \frac{q_{\mu}q_{\nu}}{q^{2}-m^{2}} \epsilon^{\mu}_{1}(\epsilon^{*}_{3} \times \epsilon^{*}_{4})^{\nu}, 
\end{align}

\begin{align}
    \mathcal{M}_{b8.12}^{(2)}=& -\frac{3g_{1}g_{2}^{3}}{2f^{2}}(d-1)A \partial_{\omega} J_{22}^{b} \frac{q_{\mu}q_{\nu}}{q^{2}-m^{2}} \notag \\ & \times [\epsilon^{\mu}_{1}(\epsilon^{*}_{3} \times \epsilon^{*}_{4})^{\nu}], 
\end{align}

\begin{align}
    \mathcal{M}_{b8.13(14)}^{(2)}=&~ \frac{8g_{1}g_{2}}{f^{2}}A [2m^{2}L+\frac{m^{2}}{8\pi^{2}}{\rm{log}} (\frac{m}{\mu})]\frac{q_{\mu}q_{\nu}}{q^{2}-m^{2}} \notag \\ & \times [\epsilon^{\mu}_{1}(\epsilon^{*}_{3} \times \epsilon^{*}_{4})^{\nu}], 
\end{align}

\begin{align}
    \mathcal{M}_{b8.15(16)}^{(2)}=&~ \frac{8g_{1}g_{2}}{f^{2}}A [2m^{2}L+\frac{m^{2}}{8\pi^{2}}{\rm{log}} (\frac{m}{\mu})]\frac{q_{\mu}q_{\nu}}{q^{2}-m^{2}} \notag \\ & \times [\epsilon^{\mu}_{1}(\epsilon^{*}_{3} \times \epsilon^{*}_{4})^{\nu}].
\end{align}

The coefficients appearing in the NLO contact and OPE amplitudes are listed in Tables~\ref{22} and \ref{23}, respectively, and the TPE amplitudes

\begin{align}
     \mathcal{M}_{c8.1}^{(2)}=&~ 0,
\end{align}

\begin{align}
     \mathcal{M}_{c8.2}^{(2)}=& -4\frac{g_{1}^2g_{2}^2}{f^{4}}A \{ (J^{B}_{21}+J^{B}_{31})q^{2}\epsilon^{\mu}_{1}(\epsilon^{*}_{3} \times \epsilon^{*}_{4})_{\mu} \notag \\ & + (J^{B}_{21}+J^{B}_{31})q_{\mu}q_{\nu}\epsilon^{\mu}_{1}(\epsilon^{*}_{3} \times \epsilon^{*}_{4})^{\nu} +[J^{B}_{21}\notag \\ & +(d-2)J^{B}_{31}]q_{\mu}q_{\nu}\epsilon^{\mu}_{1}(\epsilon^{*}_{4} \times \epsilon^{*}_{3})^{\nu} \}, 
\end{align}

\begin{align}
     \mathcal{M}_{c8.3}^{(2)}=& -4\frac{g_{1}^2g_{2}^2}{f^{4}}A [ (J^{B}_{21}+J^{B}_{31})q_{\mu}q_{\nu}\epsilon^{\mu}_{1}(\epsilon^{*}_{4} \times \epsilon^{*}_{3})^{\nu} \notag \\ &   +J^{B}_{31}q_{\mu}q_{\nu}\epsilon^{\mu*}_{3}(\epsilon^{*}_{4} \times \epsilon_{1})^{\nu} ], 
\end{align}

\begin{align}
     \mathcal{M}_{c8.4}^{(2)}=&~ 4\frac{g_{1}^2g_{2}^2}{f^{4}}A \{ (J^{R}_{21}+J^{R}_{31})q^{2}\epsilon^{\mu}_{1}(\epsilon^{*}_{3} \times \epsilon^{*}_{4})_{\mu} \notag \\ &  + (J^{R}_{21}+J^{R}_{31})q_{\mu}q_{\nu}\epsilon^{\mu}_{1}(\epsilon^{*}_{3} \times \epsilon^{*}_{4})^{\nu} +[J^{R}_{21}\notag \\ & +(d-2)J^{R}_{31}]q_{\mu}q_{\nu}\epsilon^{*\mu}_{3}(\epsilon^{*}_{4} \times \epsilon_{1})^{\nu} \}, 
\end{align}

\begin{align}
     \mathcal{M}_{c8.5}^{(2)}=& -4\frac{g_{1}^2g_{2}^2}{f^{4}}A [(J^{R}_{21}+J^{R}_{31})q_{\mu}q_{\nu}\epsilon^{\mu*}_{3}(\epsilon^{*}_{4} \times \epsilon_{1})^{\nu} \notag \\ &  + J^{R}_{31} q_{\mu}q_{\nu}\epsilon^{\mu}_{1}(\epsilon^{*}_{4} \times \epsilon^{*}_{3})^{\nu}].
\end{align}

The coefficients in the above TPE amplitudes are listed in Table~\ref{24}.

\renewcommand{\arraystretch}{1.5}
\begin{center}
    \begin{table}[ht]
          \centering
          \captionsetup{justification=raggedright, singlelinecheck=false}
          \caption{The coefficients in the NLO contact amplitudes for the process $D^{*}\bar{B} \to D^{*}\bar{B}^{*}$.}\label{22}
    	\setlength{\tabcolsep}{1.8mm}{
    	\begin{tabular}{ccccrr}
    		\toprule[1pt]
    		& { $I=1$  } & { $I=0$ } &    & \\
    		&  $A$     &    $A$    &   $\omega_1$   & $\omega_2$ \\
    		\midrule[1pt]
    		$A_{a8.1}$ & $0$  & $0$  & $-\delta_{2}$ & $0$ \\
    		$A_{a8.2(3)}$ & $0$  & $0$  & $-\delta_{2}(0)$ & $\delta_{2}(0)$ \\
    		$A_{a8.5(6)}$ & $0$  & $0$  & $0(\delta_{1})$ & $\delta_{1}$ \\
    		$A_{a8.7}$ & $\frac{1}{4}(D_{a}+E_{a})$  & $\frac{-3}{4}(D_{a}-3E_{a})$  & $\delta_{}$ & $-\delta_{2}$ \\
    		$A_{a8.8}$ & $\frac{1}{4}(D_{b}+E_{b})$  & $\frac{-3}{4}(D_{b}-3E_{b})$  & $0$ & $\delta_{1}$ \\
    		$A_{a8.9}$ & \makecell[c]{$\frac{1}{4}(D_{a}+E_{a}$\\$+D_{b}+E_{b})$}  & \makecell[c]{$\frac{-3}{4}(D_{a}-3E_{a}$\\$+D_{b}-3E_{b})$}  & $0$ & $-\delta_{2}$ \\
                $A_{a8.10}$ & $\frac{1}{4}(D_{b}+E_{b})$  & $\frac{-3}{4}(D_{b}-3E_{b})$  & $0$ & $0$ \\
    	    $A_{a8.12}$ & $\frac{1}{4}(D_{a}+E_{a})$  & $\frac{-3}{4}(D_{a}-3E_{a})$  & $\delta_{2}$ & $0$ \\
                $A_{a8.(13+14)}$ & $D_{b}+E_{b}$  & $D_{b}-3E_{b}$  & $0$ & $\delta_{1}$  \\
                $A_{a8.(15+16)}$ & $D_{b}+E_{b}$  & $D_{b}-3E_{b}$  & $0$ & $\delta_{2}$ \\
                $A_{a8.(16+17)}$ & $D_{b}+E_{b}$  & $D_{b}-3E_{b}$  & $-\delta_{2}$ & $0$  \\
    		\bottomrule[1pt]
    	\end{tabular}}
    \end{table}
\end{center}

\renewcommand{\arraystretch}{1.4}
\begin{center}
    \begin{table}[ht]
          \centering
          \captionsetup{justification=raggedright, singlelinecheck=false}
          \caption{The coefficients in the NLO OPE amplitudes for the process $D^{*}\bar{B} \to D^{*}\bar{B}^{*}$.}\label{23}
    	\setlength{\tabcolsep}{3.8mm}{
    	\begin{tabular}{ccccrr}
    		\toprule[1pt]
    		& { $I=1$  } & { $I=0$ } &    & \\
    		&  $A$     &    $A$    &   $\omega_1$   & $\omega_2$ \\
    		\midrule[1pt]
    		$A_{b8.1(2)}$ & $-\frac{1}{16}$  & $\frac{3}{16}$  & $-\delta_{2}$ & $0(\delta_{2})$ \\
    		$A_{b8.3(4)}$ & $-\frac{1}{16}$  & $\frac{3}{16}$  & $0(\delta_{1})$ & $0$ \\
    		$A_{b8.5(6)}$ & $-\frac{1}{12}$  & $\frac{1}{4}$  & $0$ & $0$ \\
    		$A_{b8.7}$ & $\frac{1}{4}$  & $-\frac{3}{4}$  & $0$ & $0$ \\
    		$A_{b8.(8+9)}$ & $\frac{1}{4}$  & $-\frac{3}{4}$  & $0$ & $\delta_{1}$ \\
    		$A_{b8.(10+11)}$ & $\frac{1}{4}$  & $-\frac{3}{4}$  & $0$ & $\delta_{2}$ \\
                $A_{b8.12}$ & $\frac{1}{4}$  & $-\frac{3}{4}$  & $-\delta_{2}$ & $0$  \\
    	    $A_{b8.13}$ & $\frac{1}{4}$  & $-\frac{3}{4}$  & $0$ & $0$ \\
                $A_{b8.14}$ & $\frac{1}{4}$  & $-\frac{3}{4}$  & $0$ & $0$ \\
                $A_{b8.15}$ & $\frac{1}{4}$  & $-\frac{3}{4}$  & $0$ & $0$ \\
                $A_{b8.16}$ & $\frac{1}{4}$  & $-\frac{3}{4}$  & $0$ & $0$ \\
    		\bottomrule[1pt]
    	\end{tabular}}
    \end{table}
\end{center}

\renewcommand{\arraystretch}{1.4}
\begin{center}
    \begin{table}[ht]
          \centering
          \captionsetup{justification=raggedright, singlelinecheck=false}
          \caption{The coefficients in the NLO TPE amplitudes for the process $D^{*}\bar{B} \to D^{*}\bar{B}^{*}$.}\label{24}
    	\setlength{\tabcolsep}{5.0mm}{
    	\begin{tabular}{ccccrr}
    		\toprule[1pt]
    		& { $I=1$  } & { $I=0$ } &    & \\
    		&  $A$     &    $A$    &   $\omega_1$   & $\omega_2$ \\
    		\midrule[1pt]
    		$A_{c8.2}$ & $\frac{1}{16}$  & $\frac{5}{16}$  & $-\delta_{2}$ & $0$ \\
    		$A_{c8.3}$ & $\frac{1}{16}$  & $\frac{5}{16}$  & $-\delta_{2}$ & $\delta_{1}$ \\
    		$A_{c8.4}$ & $\frac{9}{16}$   & $-\frac{3}{16}$ & $-\delta_{2}$ & $0$ \\
    		$A_{c8.5}$ & $\frac{9}{16}$   & $-\frac{3}{16}$ & $-\delta_{2}$ & $\delta_{1}$ \\
    		\bottomrule[1pt]
    	\end{tabular}}
    \end{table}
\end{center}

\section{Definitions of some loop functions} \label{SecAppB}

The various loop functions used in amplitudes are defined as

\begin{widetext}

\begin{align}
&~ i\int \frac{d^D l \mu^{4-D} }{ {(2\pi)}^D } 
\frac{\{1,~l^\alpha,~ l^\alpha l^\beta,~ l^\alpha l^\beta l^\gamma\} }
{[(+/-)v\cdot l+\omega+i\varepsilon](l^2-m^2+i\varepsilon)} 
\notag \\ &  \equiv \{J^{a/b}_0,~ v^\alpha J^{a/b}_{11},~
v^\alpha v^\beta J^{a/b}_{21}+g^{\alpha\beta}J^{a/b}_{22},~
(g\vee v)J^{a/b}_{31}+v^\alpha v^\beta v^\gamma J^{a/b}_{32} \} (m,\omega), 
\end{align}

\begin{align}
&~ i\int\frac{d^D l \mu^{4-D} }{ {(2\pi)}^D } 
\frac{ \{1,~l^\alpha,~ l^\alpha l^\beta,~ l^\alpha l^\beta l^\gamma\} }
{(v\cdot l+\omega_1+i\varepsilon)[(+/-)v \cdot l +\omega_2+i\varepsilon](l^2-m^2+i\varepsilon) } 
\notag \\ & \equiv  \{J^{g/h}_0,~ v^\alpha J^{g/h}_{11},
~ v^\alpha v^\beta J^{g/h}_{21}+g^{\alpha\beta}J^{g/h}_{22}, 
(g\vee v)J^{g/h}_{31}+v^\alpha v^\beta v^\gamma J^{g/h}_{32} \}(m,\omega_1,\omega_2),
\end{align}

\begin{align}
&~ i\int\frac{d^D l \mu^{4-D} }{ {(2\pi)}^D } 
\frac{ \{1,~l^\alpha,~ l^\alpha l^\beta,~ l^\alpha l^\beta l^\gamma\} }
{(l^2-m_1^2+i\varepsilon)[(q+l)^2-m_2^2+i\varepsilon] } \notag \\  
& \equiv \{ J^F_0,~ q^\alpha J^F_{11},~ q^\alpha q^\beta J^F_{21}+g^{\alpha\beta}J^F_{22},~ (g\vee q)J^F_{31}+q^\alpha q^\beta q^\gamma J^F_{32} \}(m_1,m_2,q),
\end{align}

\begin{align}
&~i \int \frac{d^D l \mu^{4-D} }{ {(2\pi)}^D } 
\frac{ \{1,~l^\alpha,~ l^\alpha l^\beta,~ l^\alpha l^\beta l^\gamma,~l^\alpha l^\beta l^\gamma l^\delta\} } 
{[(+/-)v \cdot l+\omega+i\varepsilon](l^2-m_1^2+i\varepsilon)[(q+l)^2-m_2^2+i\varepsilon]} 
\notag \\ 
& \equiv  \{ J^{T/S}_0,~ q^\alpha
J^{T/S}_{11}+v^\alpha J^{T/S}_{12},~ g^{\alpha \beta}
J^{T/S}_{21}+q^\alpha q^\beta J^{T/S}_{22}+v^\alpha v^\beta
J^{T/S}_{23}+(q\vee v)J^{T/S}_{24}, (g\vee
q)J^{T/S}_{31}+q^\alpha q^\beta q^\gamma J^{T/S}_{32}  
\notag \\  
&~ +(q^2\vee v)J^{T/S}_{33}
+(g\vee v)J^{T/S}_{34}+(q\vee v^2)J^{T/S}_{35}+v^\alpha v^\beta
v^\gamma J^T_{36},~ (g\vee g)J^{T/S}_{41}+(g\vee
q^2)J^{T/S}_{42}+q^\alpha q^\beta q^\gamma q^\delta J^{T/S}_{43} 
\notag \\
&~ +(g\vee v^2)J^{T/S}_{44} + v^\alpha v^\beta v^\gamma v^\delta J^{T/S}_{45} 
+(q^3\vee v) J^{T/S}_{46} +(q^2\vee v^2)J^{T/S}_{47} +(q\vee v^3) J^{T/S}_{48}
\notag \\   
&~  +(g\vee q\vee v) J^{T/S}_{49} \}(m_1,m_2,\omega,q), 
\end{align}

\begin{align}
&~i\int
\frac{d^D l \mu^{4-D} }{ {(2\pi)}^D } \frac{ \{1,~
l^\alpha,~ l^\alpha l^\beta,~ l^\alpha l^\beta l^\gamma,~
l^\alpha l^\beta l^\gamma l^\delta\} } {(v\cdot
l+\omega_1+i\varepsilon)[(+/-)v\cdot
l+\omega_2+i\varepsilon](l^2-m_1^2+i\varepsilon)[(q+l)^2-m_2^2+i\varepsilon]} 
\notag \\  
& \equiv \{ J^{R/B}_0,~ q^\alpha
J^{R/B}_{11}+v^\alpha J^{R/B}_{12},~ g^{\alpha \beta}
J^{R/B}_{21}+q^\alpha q^\beta J^{R/B}_{22}+v^\alpha v^\beta
J^{R/B}_{23}+(q\vee v)J^{R/B}_{24},  (g \vee
q ) J^{R/B}_{31}+q^\alpha q^\beta q^\gamma J^{R/B}_{32} 
\notag \\ 
&~ +( q^2\vee v)J^{R/B}_{33} +(g\vee v)J^{R/B}_{34} +(q\vee
v^2)J^{R/B}_{35} +v^\alpha v^\beta v^\gamma J^{R/B}_{36},
(g\vee g)J^{R/B}_{41}+(g\vee q^2)J^{R/B}_{42}+q^\alpha q^\beta q^\gamma q^\delta J^{R/B}_{43}
\notag \\   
&~ +(g\vee v^2)J^{R/B}_{44} + v^\alpha v^\beta v^\gamma v^\delta J^{R/B}_{45}  
+(q^3\vee v) J^{R/B}_{46}+(q^2\vee v^2)J^{R/B}_{47} +(q\vee v^3) J^{R/B}_{48}
\notag \\   
&~ +(g\vee q\vee v)
J^{R/B}_{49} \}(m_1,m_2,\omega_1,\omega_2,q),  \label{LoopFunction2}
\end{align}
with
\begin{flalign}
q \vee v &~\equiv q^\alpha v^\beta+q^\beta v^\alpha, \quad g \vee
q \equiv
g^{\alpha\beta}q^\gamma+g^{\alpha\gamma}q^\beta+g^{\gamma\beta}q^\alpha,
\quad g \vee v \equiv
g^{\alpha\beta}v^\gamma+g^{\alpha\gamma}v^\beta+g^{\gamma\beta}v^\alpha,
\quad 
\notag\\  
q^2 \vee v &~\equiv q^{\beta } q^{\gamma }
v^{\alpha }+q^{\alpha }
q^{\gamma } v^{\beta }+q^{\alpha } q^{\beta } v^{\gamma }, \quad
q \vee v^2 \equiv q^{\gamma } v^{\alpha }
v^{\beta }+q^{\beta } v^{\alpha } v^{\gamma }+q^{\alpha } v^{\beta } v^{\gamma },
\notag\\ 
g \vee g &~ \equiv g^{\alpha \beta } g^{\gamma \delta
}+g^{\alpha \delta } g^{\beta \gamma }+g^{\alpha \gamma } g^{\beta
\delta }, \quad g \vee q^2 \equiv q^{\alpha } q^{\beta } g^{\gamma
\delta }+q^{\alpha } q^{\delta } g^{\beta \gamma } +q^{\alpha}
q^{\gamma } g^{\beta \delta }+q^{\gamma } q^{\delta } g^{\alpha
\beta } +q^{\beta } q^{\delta } g^{\alpha \gamma } +q^{\beta }
q^{\gamma } g^{\alpha \delta }, 
\notag\\  
g \vee v^2 &~\equiv v^{\alpha } v^{\beta } g^{\gamma \delta } +v^{\alpha }
v^{\delta } g^{\beta \gamma }+v^{\alpha } v^{\gamma } g^{\beta
\delta }
+v^{\gamma } v^{\delta } g^{\alpha \beta }+v^{\beta } v^{\delta   } g^{\alpha \gamma }
+v^{\beta } v^{\gamma } g^{\alpha \delta }, \quad
\notag\\  
q^3\vee v &~\equiv q^{\beta } q^{\gamma } q^{\delta }
v^{\alpha }+q^{\alpha } q^{\gamma } q^{\delta} v^{\beta }
+q^{\alpha } q^{\beta } q^{\delta } v^{\gamma }+q^{\alpha }
q^{\beta } q^{\gamma } v^{\delta } ,\quad q\vee v^3 \equiv
q^{\delta } v^{\alpha } v^{\beta } v^{\gamma }+q^{\gamma }
v^{\alpha } v^{\beta } v^{\delta } +q^{\beta } v^{\alpha }
v^{\gamma } v^{\delta }+q^{\alpha } v^{\beta } v^{\gamma }
v^{\delta }, 
\notag \\  
q^2 \vee v^2 &~\equiv q^{\gamma }
q^{\delta } v^{\alpha } v^{\beta }+q^{\beta } q^{\delta }
v^{\alpha } v^{\gamma } +q^{\alpha } q^{\delta } v^{\beta }
v^{\gamma }+q^{\beta } q^{\gamma } v^{\alpha } v^{\delta }
+q^{\alpha } q^{\gamma } v^{\beta }  v^{\delta }+q^{\alpha }
q^{\beta } v^{\gamma } v^{\delta }, 
\notag \\  
g\vee q \vee v
&~\equiv q^{\beta } v^{\alpha } g^{\gamma \delta }+q^{\alpha }
v^{\beta } g^{\gamma \delta } +q^{\delta } v^{\alpha } g^{\beta
\gamma }+q^{\gamma } v^{\alpha } g^{\beta \delta }+q^{\alpha }
v^{\delta } g^{\beta \gamma } +q^{\alpha } v^{\gamma } g^{\beta
\delta}+q^{\delta } v^{\gamma } g^{\alpha \beta }+q^{\delta }
v^{\beta } g^{\alpha \gamma } +q^{\gamma } v^{\delta } g^{\alpha
\beta  } 
\notag \\ 
&~\quad +q^{\gamma } v^{\beta }
g^{\alpha \delta }+q^{\beta } v^{\delta } g^{\alpha \gamma }
+q^{\beta } v^{\gamma } g^{\alpha \delta }.
\end{flalign}

\end{widetext}

$J^b$ is related to $J^a$:
\begin{align}
  J^b_0 & =J^a_0, \quad J^b_{11}=-J^a_{11}, \quad J^b_{21}=J^a_{21}, \notag \\
  J^b_{22} & =J^a_{22}, \quad J^b_{31}=-J^a_{31}, \quad J^b_{32}=-J^a_{32}.
\end{align}
$J^g$ and $J^h$ can be deduced to
\begin{align}
J^g(\omega_1,\omega_2) &=  \frac{1}{\omega_2-\omega_1} \left[ J^a(\omega_1)-J^a(\omega_2)\right],  
\end{align}

\begin{align}
J^h(\omega_1,\omega_2) &=  \frac{1}{\omega_2+\omega_1} \left[ J^a(\omega_1)+J^b(\omega_2)\right].
\end{align}
$J^S$ is related to $J^T$:
\begin{align}
 J^S_0(v \cdot q) &= J^T_0(-v \cdot q) , \quad J^S_{11}(v\cdot q)=J^T_{11}(-v\cdot q), \notag \\
 J^S_{12}(v \cdot q) & =-J^T_{12}(-v\cdot q), \quad J^S_{21}=J^T_{21}(-v\cdot q), \notag \\
 J^S_{22}(v \cdot q) & =J^T_{22}(-v\cdot q), \quad J^S_{23}(v \cdot q)=J^T_{23}(-v \cdot q ). \notag \\
 J^S_{24}(v \cdot q) & =-J^T_{24}(-v\cdot q), \quad J^S_{31}(v \cdot q)=J^T_{31}(-v \cdot q ). \notag \\
 J^S_{32}(v \cdot q) & =J^T_{32}(-v\cdot q), \quad J^S_{33}(v \cdot q)=-J^T_{33}(-v \cdot q ). \notag \\
 J^S_{34}(v \cdot q) & =-J^T_{34}(-v\cdot q), \quad J^S_{35}(v \cdot q)=J^T_{35}(-v \cdot q ). \notag \\
 J^S_{36}(v \cdot q) & =-J^T_{34}(-v\cdot q), \quad J^S_{41}(v \cdot q)=J^T_{41}(-v \cdot q ). \notag \\
 J^S_{42}(v \cdot q) & =J^T_{42}(-v\cdot q), \quad J^S_{43}(v \cdot q)=J^T_{43}(-v \cdot q ). \notag \\
 J^S_{44}(v \cdot q) & =J^T_{44}(-v\cdot q), \quad J^S_{45}(v \cdot q)=J^T_{45}(-v \cdot q ). \notag \\
 J^S_{46}(v \cdot q) & =-J^T_{46}(-v\cdot q), \quad J^S_{47}(v \cdot q)=J^T_{47}(-v \cdot q ). \notag \\
 J^S_{48}(v \cdot q) & =-J^T_{48}(-v\cdot q), \quad J^S_{49}(v \cdot q)=-J^T_{49}(-v \cdot q ). 
\end{align}

$J^R$ and $J^B$ can be deduced to
\begin{align}
 J^R(\omega_1,\omega_2) =  \frac{1}{\omega_2-\omega_1} \left[ J^T(\omega_1)-J^T(\omega_2)\right],  
\end{align}

\begin{align}
 J^B(\omega_1,\omega_2) =  \frac{1}{\omega_2+\omega_1} \left[ J^T(\omega_1)+J^S(\omega_2)\right].
\end{align}



\begin{thebibliography}{300}


\bibitem{LHCb:2020bls}
R.~Aaij \textit{et al.} [LHCb], A model-independent study of resonant structure in $B^+\to D^+D^-K^+$ decays, Phys. Rev. Lett. \textbf{125}, 242001 (2020).

\bibitem{LHCb:2020pxc}
R.~Aaij \textit{et al.} [LHCb], Amplitude analysis of the $B^+\to D^+D^-K^+$ decay,
Phys. Rev. D \textbf{102}, 112003 (2020).

\bibitem{LHCb:2021vvq}
R.~Aaij \textit{et al.} [LHCb], Observation of an exotic narrow doubly charmed tetraquark, Nature Phys. \textbf{18}, 751-754 (2022).

\bibitem{LHCb:2021auc}
R.~Aaij \textit{et al.} [LHCb], Study of the doubly charmed tetraquark $T_{cc}^{+}$, Nature Commun. \textbf{13}, 3351 (2022).

\bibitem{LHCb:2022sfr}
R.~Aaij \textit{et al.} [LHCb], First Observation of a Doubly Charged Tetraquark and Its Neutral Partner,
Phys. Rev. Lett. \textbf{131}, 041902 (2023).

\bibitem{LHCb:2022lzp}
R.~Aaij \textit{et al.} [LHCb], Amplitude analysis of $B^{+}\to D^{-}D^{+}_{s}\pi^{+}$ and $B^{0}\to \bar{D}^{0}D^{+}_{s}\pi^{+}$ decays, Phys. Rev. D \textbf{108}, 012017 (2023).

\bibitem{Cui:2006mp}
Y.~Cui, X.~L.~Chen, W.~Z.~Deng and S.~L.~Zhu,
The Possible Heavy Tetraquarks $qQ \bar q \bar Q$, $qq \bar Q \bar Q$ and $qQ \bar Q \bar Q$,
HEPNP \textbf{31}, 7-13 (2007).

\bibitem{Detmold:2007wk}
W.~Detmold, K.~Orginos and M.~J.~Savage,
$BB$ Potentials in Quenched Lattice QCD,
Phys. Rev. D \textbf{76} (2007), 114503.

\bibitem{Carlucci:2007um}
M.~V.~Carlucci, F.~Giannuzzi, G.~Nardulli, M.~Pellicoro and S.~Stramaglia,
AdS-QCD quark-antiquark potential, meson spectrum and tetraquarks,
Eur. Phys. J. C \textbf{57}, 569-578 (2008).

\bibitem{Navarra:2007yw}
F.~S.~Navarra, M.~Nielsen and S.~H.~Lee,
QCD sum rules study of $QQ - \bar{u}\bar{d}$ mesons,
Phys. Lett. B \textbf{649}, 166-172 (2007).

\bibitem{Yang:2009zzp}
Y.~Yang, C.~Deng, J.~Ping and T.~Goldman,
$S$-wave $Q Q \bar q \bar q$ state in the constituent quark model,
Phys. Rev. D \textbf{80}, 114023 (2009).

\bibitem{Carames:2011zz}
T.~F.~Carames, A.~Valcarce and J.~Vijande,
Doubly charmed exotic mesons: A gift of nature?,
Phys. Lett. B \textbf{699}, 291-295 (2011).

\bibitem{Molina:2010tx}
R.~Molina, T.~Branz and E.~Oset, A new interpretation for the $D^*_{s2}(2573)$ and the prediction of novel exotic charmed mesons, Phys. Rev. D \textbf{82}, 014010 (2010).

\bibitem{Du:2012wp}
M.~L.~Du, W.~Chen, X.~L.~Chen and S.~L.~Zhu, Exotic $QQ\bar{q}\bar{q}$, $QQ\bar{q}\bar{s}$ and $QQ\bar{s}\bar{s}$ states, Phys. Rev. D \textbf{87}, 014003 (2013).

\bibitem{Hyodo:2012pm}
T.~Hyodo, Y.~R.~Liu, M.~Oka, K.~Sudoh and S.~Yasui, 
Production of doubly charmed tetraquarks with exotic color configurations in electron-positron collisions,
Phys. Lett. B \textbf{721}, 56-60 (2013).

\bibitem{Ohkoda:2012hv}
S.~Ohkoda, Y.~Yamaguchi, S.~Yasui, K.~Sudoh and A.~Hosaka,
Exotic mesons with double charm and bottom flavor,
Phys. Rev. D \textbf{86}, 034019 (2012).


\bibitem{Vijande:2013qr}
J.~Vijande, A.~Valcarce and J.~M.~Richard,
Adiabaticity and color mixing in tetraquark spectroscopy,
Phys. Rev. D \textbf{87}, 034040 (2013).

\bibitem{Ikeda:2013vwa}
Y.~Ikeda, B.~Charron, S.~Aoki, T.~Doi, T.~Hatsuda, T.~Inoue, N.~Ishii, K.~Murano, H.~Nemura and K.~Sasaki,
Charmed tetraquarks $T_{cc}$ and $T_{cs}$ from dynamical lattice QCD simulations,
Phys. Lett. B \textbf{729}, 85-90 (2014).


\bibitem{Bicudo:2015kna}
P.~Bicudo, K.~Cichy, A.~Peters and M.~Wagner,
$BB$ interactions with static bottom quarks from Lattice QCD,
Phys. Rev. D \textbf{93}, 034501 (2016).

\bibitem{Bicudo:2015vta}
P.~Bicudo, K.~Cichy, A.~Peters, B.~Wagenbach and M.~Wagner,
Evidence for the existence of $u d \bar{b} \bar{b}$ and the non-existence of $s s \bar{b} \bar{b}$ and $c c \bar{b} \bar{b}$ tetraquarks from lattice QCD,
Phys. Rev. D \textbf{92}, 014507 (2015).

\bibitem{Peters:2015tra}
A.~Peters, P.~Bicudo, K.~Cichy, B.~Wagenbach and M.~Wagner,
Exploring possibly existing $q q \bar b \bar b$ tetraquark states with $q q = ud, ss, cc$,
PoS \textbf{LATTICE2015}, 095 (2016).

\bibitem{Peters:2016isf}
A.~Peters, P.~Bicudo, L.~Leskovec, S.~Meinel and M.~Wagner,
Lattice QCD study of heavy-heavy-light-light tetraquark candidates,
PoS \textbf{LATTICE2016}, 104 (2016).

\bibitem{Francis:2016hui}
A.~Francis, R.~J.~Hudspith, R.~Lewis and K.~Maltman, Lattice Prediction for Deeply Bound Doubly Heavy Tetraquarks, Phys. Rev. Lett. \textbf{118}, 142001 (2017).

\bibitem{Bicudo:2017szl}
P.~Bicudo, M.~Cardoso, A.~Peters, M.~Pflaumer and M.~Wagner,
$ud \bar{b} \bar{b}$ tetraquark resonances with lattice QCD potentials and the Born-Oppenheimer approximation,
Phys. Rev. D \textbf{96}, 054510 (2017).

\bibitem{Chen:2017rhl}
W.~Chen, H.~X.~Chen, X.~Liu, T.~G.~Steele and S.~L.~Zhu, Open-flavor charm and bottom $sq\bar q\bar Q$ and $qq\bar q\bar Q$ tetraquark states, Phys. Rev. D \textbf{95}, 114005 (2017). 

\bibitem{Azizi:2018mte}
K.~Azizi and U.~\"Ozdem, The electromagnetic multipole moments of the charged open-flavor $Z_{\bar cq}$ states, J. Phys. G \textbf{45}, 055003 (2018).

\bibitem{Bicudo:2019mny}
P.~Bicudo, M.~Cardoso, A.~Peters, M.~Pflaumer and M.~Wagner, Doubly heavy tetraquark resonances in lattice QCD, J. Phys. Conf. Ser. \textbf{1137}, 012039 (2019).

\bibitem{Wang:2019xzt}
Z.~G.~Wang,
Analysis of the axialvector $B_c$-like tetraquark states with the QCD sum rules, EPL \textbf{128}, 11001 (2019).

\bibitem{Lu:2020qmp}
Q.~F.~L\"u, D.~Y.~Chen and Y.~B.~Dong, Open charm and bottom tetraquarks in an extended relativized quark model, Phys. Rev. D \textbf{102}, 074021 (2020).

\bibitem{He:2020jna}
X.~G.~He, W.~Wang and R.~Zhu, Open-charm tetraquark $X_c$ and open-bottom tetraquark $X_b$, Eur. Phys. J. C \textbf{80}, 1026 (2020).

\bibitem{Cheng:2020nho}
J.~B.~Cheng, S.~Y.~Li, Y.~R.~Liu, Y.~N.~Liu, Z.~G.~Si and T.~Yao, Spectrum and rearrangement decays of tetraquark states with four different flavors, Phys. Rev. D \textbf{101}, 114017 (2020).

\bibitem{Albuquerque:2020ugi}
R.~M.~Albuquerque, S.~Narison, D.~Rabetiarivony and G.~Randriamanatrika, $X_{0,1}$(2900) and $(D^-K^+)$ invariant mass from QCD Laplace sum rules at NLO, Nucl. Phys. A \textbf{1007}, 1221Phys. Rev. D 105, 054018
13 (2021).

\bibitem{Guo:2021mja}
T.~Guo, J.~Li, J.~Zhao and L.~He, Mass spectra and decays of open-heavy tetraquark states, Phys. Rev. D \textbf{105}, 054018 (2022).

\bibitem{Chen:2022asf}
H.~X.~Chen, W.~Chen, X.~Liu, Y.~R.~Liu and S.~L.~Zhu, An updated review of the new hadron states, Rept. Prog. Phys. \textbf{86}, 026201 (2023).

\bibitem{Li:2012ss}
N.~Li, Z.~F.~Sun, X.~Liu and S.~L.~Zhu,
Coupled-channel analysis of the possible $D^{(*)}D^{(*)}, \bar{B}^{(*)}\bar{B}^{(*)}$ and $D^{(*)}\bar{B}^{(*)}$ molecular states,
Phys. Rev. D \textbf{88}, 114008 (2013).

\bibitem{Chen:2013aba}
W.~Chen, T.~G.~Steele and S.~L.~Zhu, Exotic open-flavor $bc\bar{q}\bar{q}$, $bc\bar{s}\bar{s}$ and $qc\bar{q}\bar{b}$, $sc\bar{s}\bar{b}$ tetraquark states, Phys. Rev. D \textbf{89}, 054037 (2014).

\bibitem{Luo:2017eub}
S.~Q.~Luo, K.~Chen, X.~Liu, Y.~R.~Liu and S.~L.~Zhu, Exotic tetraquark states with the $qq\bar{Q}\bar{Q}$ configuration, Eur. Phys. J. C \textbf{77}, 709 (2017).

\bibitem{Deng:2021gnb}
C.~Deng and S.~L.~Zhu, $T_{cc}^+$ and its partners, Phys. Rev. D \textbf{105}, 054015 (2022).

\bibitem{Wu:2024zbx}
W.~L.~Wu, Y.~Ma, Y.~K.~Chen, L.~Meng and S.~L.~Zhu, Doubly heavy tetraquark bound and resonant states, [arXiv:2409.03373 [hep-ph]].


\bibitem{Eichten:2017ffp}
E.~J.~Eichten and C.~Quigg, Heavy-quark symmetry implies stable heavy tetraquark mesons $Q_iQ_j \bar q_k \bar q_l$, Phys. Rev. Lett. \textbf{119}, 202002 (2017).


\bibitem{Silvestre-Brac:1993zem}
B.~Silvestre-Brac and C.~Semay, Systematics of $L=0$ $q^{2}\bar{q}^{2}$ systems, 
Z. Phys. C \textbf{57}, 273-282 (1993).

\bibitem{Ebert:2007rn}
D.~Ebert, R.~N.~Faustov, V.~O.~Galkin and W.~Lucha, Masses of tetraquarks with two heavy quarks in the relativistic quark model, Phys. Rev. D \textbf{76}, 114015 (2007).

\bibitem{Park:2018wjk}
W.~Park, S.~Noh and S.~H.~Lee, Masses of the doubly heavy tetraquarks in a constituent quark model, Nucl. Phys. A \textbf{983}, 1-19 (2019)

\bibitem{Lu:2020rog}
Q.~F.~L$\ddot{\rm{u}}$, D.~Y.~Chen and Y.~B.~Dong, Masses of doubly heavy tetraquarks $T_{QQ^\prime}$ in a relativized quark model, Phys. Rev. D \textbf{102}, 034012 (2020).

\bibitem{Braaten:2020nwp}
E.~Braaten, L.~P.~He and A.~Mohapatra, Masses of doubly heavy tetraquarks with error bars, 
Phys. Rev. D \textbf{103}, 016001 (2021).

\bibitem{Kim:2022mpa}
Y.~Kim, M.~Oka and K.~Suzuki, Doubly heavy tetraquarks in a chiral-diquark picture, Phys. Rev. D \textbf{105}, 074021 (2022).

\bibitem{Song:2023izj}
Y.~Song and D.~Jia, Mass spectra of doubly heavy tetraquarks in diquark\ensuremath{-}antidiquark picture, Commun. Theor. Phys. \textbf{75}, 055201 (2023).


\bibitem{Ali:2018xfq}
A.~Ali, Q.~Qin and W.~Wang, Discovery potential of stable and near-threshold doubly heavy tetraquarks at the LHC, Phys. Lett. B \textbf{785}, 605-609 (2018).


\bibitem{Meinel:2022lzo}
S.~Meinel, M.~Pflaumer and M.~Wagner, Search for $\bar{b}\bar{b}us$ and $\bar{b}\bar{c}ud$ tetraquark bound states using lattice QCD, Phys. Rev. D \textbf{106}, 034507 (2022).

\bibitem{Alexandrou:2023cqg}
C.~Alexandrou, J.~Finkenrath, T.~Leontiou, S.~Meinel, M.~Pflaumer and M.~Wagner,
Shallow Bound States and Hints for Broad Resonances with Quark Content $\bar{b}\bar{c}ud$ in $B-\bar{D}$ and $B^{*}-\bar{D}$ Scattering from Lattice QCD,
Phys. Rev. Lett. \textbf{132}, 151902 (2024).

\bibitem{Padmanath:2023rdu}
M.~Padmanath, A.~Radhakrishnan and N.~Mathur, Bound Isoscalar Axial-Vector $bc\bar{u}\bar{d}$ Tetraquark $T_{bc}$ from Lattice QCD Using Two-Meson and Diquark-Antidiquark Variational Basis, Phys. Rev. Lett. \textbf{132}, 20 (2024).

\bibitem{Radhakrishnan:2024ihu}
A.~Radhakrishnan, M.~Padmanath and N.~Mathur, Study of the isoscalar scalar $bc\bar{u}\bar{d}$ tetraquark $T_{bc}$ with lattice QCD, Phys. Rev. D \textbf{110}, 034506 (2024).


\bibitem{Liu:2024lmv}
M.~Z.~Liu and L.~S.~Geng,
Investigations of the weak decays of $D\bar B$ molecules,
Phys. Rev. D \textbf{110}, 053002 (2024).

\bibitem{Machleidt:2011zz}
R.~Machleidt and D.~R.~Entem,
Chiral effective field theory and nuclear forces,
Phys. Rept. \textbf{503}, 1-75 (2011).

\bibitem{Hammer:2019poc}
H.~W.~Hammer, S.~K\"onig and U.~van Kolck,
Nuclear effective field theory: status and perspectives,
Rev. Mod. Phys. \textbf{92} (2020), 025004.


\bibitem{Xu:2017tsr}
H.~Xu, B.~Wang, Z.~W.~Liu and X.~Liu,
$DD^{*}$ potentials in chiral perturbation theory and possible molecular states,
Phys. Rev. D \textbf{99}, 014027 (2019)
[erratum: Phys. Rev. D \textbf{104}, 119903 (2021)].

\bibitem{Wang:2022jop}
B.~Wang and L.~Meng,
Revisiting the $DD^{*}$ chiral interactions with the local momentum-space regularization up to the third order and the nature of $T_{cc}^+$,
Phys. Rev. D \textbf{107}, 094002 (2023).


\bibitem{Wang:2018atz}
B.~Wang, Z.~W.~Liu and X.~Liu,
$\bar{B}^{(*)} \bar{B}^{*)}$ interactions in chiral effective field theory, Phys. Rev. D \textbf{99}, 036007 (2019).

\bibitem{Xu:2021vsi}
H.~Xu,
Study of the hidden charm $D\bar{D}^{*}$ interactions in chiral effective field theory, Phys. Rev. D \textbf{105}, 034013 (2022).

\bibitem{Abreu:2022sra}
L.~M.~Abreu,
A note on the possible bound $D^{(*)}D^{(*)}$,$\bar B^{(*)}\bar B^{(*)}$ and $D^{(*)}\bar B^{(*)}$ states,
Nucl. Phys. B \textbf{985}, 115994 (2022).

\bibitem{Meng:2019ilv}
L.~Meng, B.~Wang, G.~J.~Wang and S.~L.~Zhu,
The hidden charm pentaquark states and $\Sigma_c\bar{D}^{(*)}$ interaction in chiral perturbation theory,
Phys. Rev. D \textbf{100}, 014031 (2019).


\bibitem{Wang:2019ato}
B.~Wang, L.~Meng and S.~L.~Zhu,
Hidden-charm and hidden-bottom molecular pentaquarks in chiral effective field theory,
JHEP \textbf{11} (2019), 108.

\bibitem{Wang:2022ztm}
B.~Wang, L.~Meng and S.~L.~Zhu,
Molecular tetraquarks and pentaquarks in chiral effective field theory,
Nucl. Part. Phys. Proc. \textbf{324-329}, 45-48 (2023).

\bibitem{Du:2016tgp}
M.~L.~Du, F.~K.~Guo, U.~G.~Mei{\ss}ner and D.~L.~Yao,
Aspects of the low-energy constants in the chiral Lagrangian for charmed mesons,
Phys. Rev. D \textbf{94}, 094037 (2016).

      
\bibitem{Ecker:1988te}
G.~Ecker, J.~Gasser, A.~Pich and E.~de Rafael,
The Role of Resonances in Chiral Perturbation Theory,
Nucl. Phys. B \textbf{321}, 311-342 (1989).


\bibitem{Ecker:1989yg}
G.~Ecker, J.~Gasser, H.~Leutwyler, A.~Pich and E.~de Rafael,
Chiral Lagrangians for Massive Spin $1$ Fields,
Phys. Lett. B \textbf{223}, 425-432 (1989).

\bibitem{Donoghue:1988ed}
J.~F.~Donoghue, C.~Ramirez and G.~Valencia,
The Spectrum of QCD and Chiral Lagrangians of the Strong and Weak Interactions,
Phys. Rev. D \textbf{39}, 1947 (1989).

\bibitem{Epelbaum:2014efa}
E.~Epelbaum, H.~Krebs and U.~G.~Mei{\ss}ner,
Improved chiral nucleon-nucleon potential up to next-to-next-to-next-to-leading order, Eur. Phys. J. A \textbf{51}, 53 (2015).


\bibitem{Entem:2003ft}
D.~R.~Entem and R.~Machleidt,
Accurate charge-dependent nucleon-nucleon potential at fourth order of chiral perturbation theory,
Phys. Rev. C \textbf{68}, 041001 (2003).

\bibitem{Entem:2017gor}
D.~R.~Entem, R.~Machleidt and Y.~Nosyk,
High-quality two-nucleon potentials up to fifth order of the chiral expansion,
Phys. Rev. C \textbf{96}, 024004 (2017).

\bibitem{Thomson:2013zua}
M.~Thomson, Modern particle physics, Cambridge University Press, 2013.

\bibitem{Zhai:2021uap}
H. Zhai, Ultracold Atomic Physics, Cambridge University Press, 2021.

\bibitem{Taylor:1972pty}
J.~R.~Taylor, Scattering Theory: The Quantum Theory of Nonrelativistic Collisions, John Wiley \& Sons, Inc., 1972.


\bibitem{Zhe:2025dat}
Z. Liu, H. Xu and X. Liu, $S$-wave potentials and scattering rates of $D^{(*)}\bar{B}^{(*)}$ systems in chiral effective field theory, 10.5281/zenodo.16628036 (2025).
        
\end{thebibliography}
\end{document}